\documentclass[a4paper,unpublished,onecolumn,11pt]{quantumarticle}
\pdfoutput=1
\usepackage[utf8]{inputenc}
\usepackage[english]{babel}
\usepackage[T1]{fontenc}
\usepackage{amsmath}
\usepackage{amsfonts}
\usepackage{hyperref}
\usepackage{subcaption}
\usepackage{multirow}

\usepackage{tikz}
\usepackage{lipsum}
\usepackage{quantikz}
\usepackage{adjustbox}

\usepackage{xcolor}
\usepackage{cleveref}
\usepackage{stmaryrd}

\makeatletter
\AddToHook{cmd/appendix/before}{\def\cref@section@alias{appendix}}
\makeatother

\definecolor{floquet-red}{RGB}{205,41,41}
\definecolor{floquet-green}{RGB}{0,143,144}
\definecolor{floquet-blue}{RGB}{32,74,211}

\begin{document}

\newcommand{\todo}[1]{\textcolor{red}{\uppercase{TODO: #1 }}}
\newcommand{\etal}{\textit{et al.\ }}

\title{Logical gates on Floquet codes via folds and twists}

\author{Alexandra E. Moylett}
\affiliation{Nu Quantum Ltd., Cambridge, United Kingdom}
\email{alex.moylett@nu-quantum.com}
\orcid{0000-0003-0163-5262}
\author{Bhargavi Jonnadula}
\affiliation{Nu Quantum Ltd., Cambridge, United Kingdom}
\email{bhargavi.jonnadula@nu-quantum.com}
\orcid{0009-0001-2919-7746}
\maketitle

\begin{abstract}
Floquet codes have recently emerged as a new family of error-correcting codes, and have drawn significant interest across both theoretical and practical quantum computing. A central open question has been how to implement logical operations on these codes. In this work, we show how two techniques from static quantum error-correcting codes can also be implemented on Floquet codes. First, we present a way of implementing fold-transversal operations on Floquet codes in order to yield logical Hadamard and $\operatorname{S}$ gates. And second, we present a way of implementing logical $\operatorname{CNOT}$ gates on Floquet codes via Dehn twists. We discuss the requirements for these techniques, and show that they are applicable to a wide family of Floquet codes defined on colour code lattices. Through numerical benchmarking of the logical operations on the CCS Floquet code, we establish a logical-gate threshold of $0.25$-$0.35\%$ and verify sub-threshold exponential error suppression. Our results show that these logical operations are robust, featuring a performance that is close to the baseline set by a quantum memory benchmark. Finally, we explain in detail how to implement logical gates on Floquet codes by operating on the embedded codes.
\end{abstract}

\section{Introduction}

Quantum error correction (``QEC'') has developed significantly over recent years, from the identification of high-performance high-rate QEC codes \cite{Panteleev2022GoodQLDPCCodes, Bravyi2024BBCodes} to the experimental demonstration of QEC codes on real-world hardware \cite{Chen2024TransversalLogic, Google2025SurfaceCodeThreshold}. One area of particular interest has been in identifying what logical gates can be implemented on a given QEC code. A wide variety of techniques have been developed for fault-tolerantly implementing logical gates on different codes, including transversal gates \cite{Bombin2006ColourCode, Chen2024TransversalLogic}, twist defects \cite{Brown2017PunchingHolesCuttingCorners}, code surgery \cite{Horsman2012LatticeSurgery, Breuckmann2017HyperbolicSurfaceCodes, Baspin2025FastSurgeryQuantumLDPC}, and gates which are induced by exchanging qubits \cite{Breuckmann2024FoldTransversal, Sayginel2025AutQEC, Malcolm2025SHYPS}.

A recent shift in quantum error correction has been the move towards thinking about quantum error correction as a dynamic process \cite{McEwen2023RelaxingHardware}. One of the earliest examples of this shift was the introduction of Floquet codes \cite{Hastings2021FloquetCodes}. Unlike traditional QEC codes, where logical information is static throughout the duration of the code, Floquet codes allow the logical information to evolve dynamically, whilst still protecting the logical information from errors. Floquet codes also have some desirable hardware properties: all the measurements on a Floquet code are between pairs of qubits, making them suitable for hardware with native two-qubit measurements \cite{Microsoft2025Roadmap} as well as for hardware distributed across a quantum network \cite{Sutcliffe2025DistributedQEC}.

Since their initial proposal by Hastings and Haah, the question of what logical gates can be implemented on Floquet codes has prompted some interesting results \cite{Hastings2021FloquetCodes}. In Hastings and Haah, it was shown that the process of measuring the stabilisers in the original Floquet code induces a logical operation similar to the Hadamard gate. This in turn led to the development of dynamic automorphism codes \cite{Davydova2024DynamicAutomorphism}, which are able to implement the full Clifford group across $k$ logical qubits, but requires these logical qubits to be arranged as a vertical stack of $k+1$ layers of qubits. It has also been shown that some techniques from static codes can be extended to Floquet codes, such as lattice surgery \cite{Haah2022BoundaryFloquet, Gidney2021FloquetBenchmark} and twist defects \cite{Ellison2023FloquetCodesTwist}.

In Higgott and Breuckmann \cite{Higgott2024HyperbolicFloquet}, it was suggested that two further techniques from the study of logical gates on static codes could be adapted to Floquet codes. The first is fold-transversal gates \cite{Moussa2016FoldTransversal, Breuckmann2024FoldTransversal}, which utilise symmetries between the $\operatorname{Z}$ and $\operatorname{X}$ stabilisers of a code to implement logical gates. And the second is Dehn twists \cite{Zhu2020DehnTwist}, which distort the lattice of a code to implement logical $\operatorname{CNOT}$ gates. Higgott and Breuckmann leave the details of adapting these techniques to Floquet codes as an open question.

It is not necessarily clear how to adapt these techniques from static codes to their Floquet counterparts. In principle one can use the fact that a Floquet code during each moment in time embeds a static code to implement logical gates, by performing logical gates on the static code. However there are key technical details which are missing from this proposal, particularly around ensuring that the embedded code features good logical gates, and ensuring that the embedded code can still be mapped back to the Floquet code in a fault-tolerant manner.

\begin{figure}
    \centering
    \begin{subfigure}{0.3\linewidth}
        \includegraphics[width=\linewidth]{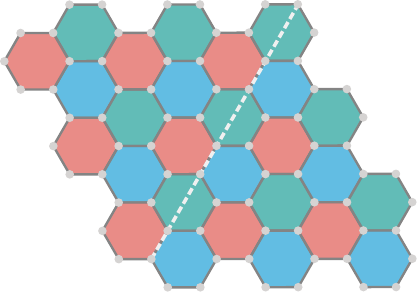}
        \caption{}
    \end{subfigure}
    \begin{subfigure}{0.3\linewidth}
        \includegraphics[width=\linewidth]{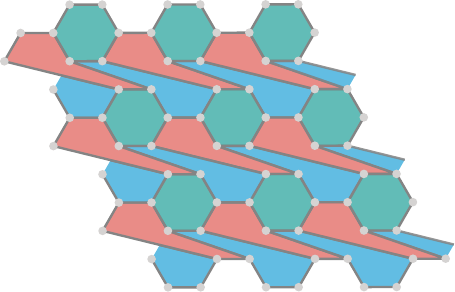}
        \caption{}
    \end{subfigure}
    \begin{subfigure}{0.3\linewidth}
        \includegraphics[width=\linewidth]{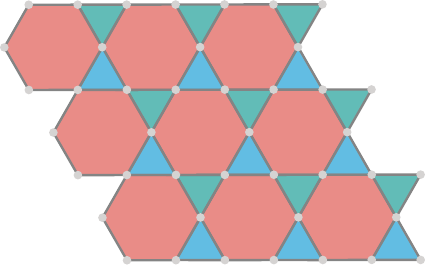}
        \caption{}
    \end{subfigure}
    \caption{Logical-gate techniques adapted and applied to Floquet codes. (a) In \Cref{sec:fold-transversal-floquet}, we extend the concept of fold-transversal logical gates to Floquet codes, and show how this can be utilised to implement logical Hadamard and $\operatorname{S}$ gates. (b) In \Cref{sec:dehn-twist-floquet}, we show how distortion of the Floquet code lattice can be used to implement $\operatorname{CNOT}$ gates via Dehn twists. (c) In \Cref{sec:embedded-codes}, we investigate how to implement logical gates on the embedded stabiliser codes.}
    \label{fig:summary}
\end{figure}

In this work, we introduce novel techniques for fault-tolerantly implementing both fold-transversal logical gates and Dehn twists on Floquet codes. We apply these techniques to a variety of Floquet codes using different measurement schedules and lattices. We benchmark these logical gates in Stim, and through Monte Carlo sampling show that under circuit-level noise they demonstrate a fault-tolerant threshold of $0.25$-$0.35\%$. At a physical error rate of $0.05\%$, simulations which can be decoding using PyMatching \cite{Higgott2025SparseBlossom} reach logical error rates of $\sim10^{-6}$ with 294 physical qubits, whereas circuits which require belief propagation with ordered statistics decoding \cite{Hillmann2025BPLSD} achieve a logical error rate of $\sim10^{-5}$. These results are close to the results found for quantum memory experiments under the same decoder, noise model, and number of QEC rounds. Finally, while these new techniques do not utilise the embedded stabiliser code, we provide a technical discussion on how logical gates on the embedded codes can lead to logical gates on Floquet codes. We summarise these developments in \Cref{fig:summary}.

The rest of this paper is laid out as follows. In \Cref{sec:background} we provide some background material on fold-transversal gates, Dehn twists, and Floquet codes. In \Cref{sec:fold-transversal-floquet}, we identify emergent $\operatorname{ZX}$-dualities which are exploited to implement fold-transversal logical gates on Floquet codes. In \Cref{sec:dehn-twist-floquet} we introduce an \textit{edge-swapping gadget} which is utilised to implement Dehn twists on Floquet codes. We benchmark both of these techniques against quantum memory experiments in \Cref{sec:benchmarking}. In \Cref{sec:embedded-codes} we provide a technical discussion of how to implement logical gates on Floquet codes via their embedded stabiliser codes. Finally, we conclude with some open questions in \Cref{sec:conclusion}.

\section{Background}
\label{sec:background}

\subsection{Fold-transversal gates on static codes}
\label{ssec:fold-transversal-static}

Fold-transversal gates are a way of implementing logical Hadamard and $\operatorname{S}$ gates through symmetries available in stabiliser codes \cite{Moussa2016FoldTransversal, Breuckmann2024FoldTransversal}. They originated from understanding the colour code as a surface code which has been folded along its diagonal \cite{Kubica2015UnfoldingColourCode, Moussa2016FoldTransversal}. The triangular colour code features transversal logical Clifford gates, and by equivalence, the folded surface code also features transversal logical gates. Subsequent work showed that fold-transversal logical gates are also implementable on more general CSS codes, including hyperbolic surface codes \cite{Breuckmann2024FoldTransversal}.

\begin{figure}
    \centering
    \begin{subfigure}{0.3\linewidth}
        \includegraphics[width=\linewidth]{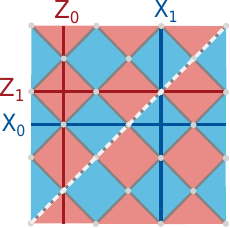}
        \caption{}
        \label{fig:duality-toric-code}
    \end{subfigure}
    \hfill
    \begin{subfigure}{0.3\linewidth}
        \includegraphics[width=\linewidth]{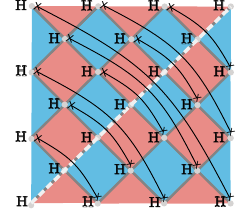}
        \caption{}
        \label{fig:hadamard-toric-code}
    \end{subfigure}
    \hfill
    \begin{subfigure}{0.3\linewidth}
        \includegraphics[width=\linewidth]{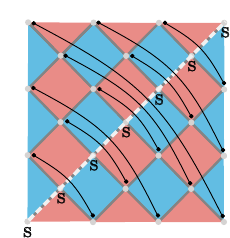}
        \caption{}
        \label{fig:s-toric-code}
    \end{subfigure}
    \caption{Fold-transversal gates on the unrotated toric code. An example of unrotated toric code along with logical operators is shown in (a). Nodes represent physical qubits, and red \& blue plaquettes denote $\operatorname{X}$ \& $\operatorname{Z}$ stabilisers, respectively. The fold line for the $\operatorname{ZX}$-duality is shown as a dashed line. (b) A fold-transversal Hadamard-type gate $\operatorname{H_0}\otimes \operatorname{H_1}$ is implemented by applying transversal Hadamard gates to every qubit and $\operatorname{SWAP}$ gates between pairs of qubits across the fold line. (c) A fold-transversal $\operatorname{S}$-type gate $\operatorname{S_0}\otimes \operatorname{S_1}$ is implemented by applying $\operatorname{S}$ gates along the fold line and $\operatorname{CZ}$ gates between pairs of qubits across the fold line.}
    \label{fig:fold-surface-code}
\end{figure}

The main ingredient that gives rise to fold-transversal gates in a CSS code is a $\operatorname{ZX}$-duality \cite{Breuckmann2024FoldTransversal}. Let $C$ be a CSS code. A $\operatorname{ZX}$-duality $\tau$ is an isomorphism $C\rightarrow C^{\top}$ of the code such that for any $\operatorname{X} = \bigotimes_i\operatorname{X}_i$ (resp.\ $\operatorname{Z} = \bigotimes_i\operatorname{Z}_i$) stabiliser in $C$ there exists a corresponding stabiliser $\operatorname{Z} = \bigotimes_i\operatorname{Z}_{\tau(i)}$ (resp.\ $\operatorname{X} = \bigotimes_i\operatorname{X}_{\tau(i)}$). One example can be seen in the unrotated toric code, shown in \Cref{fig:duality-toric-code}: if we were to fold along the dashed white line then each red face ($\operatorname{X}$ stabiliser) on one side of the line maps to a blue face ($\operatorname{Z}$ stabiliser) on the other side of the line, and vice versa. Another example, the colour code, features a trivial $\operatorname{ZX}$-duality $\tau(i) = i$ for all physical qubits $i$, as all plaquettes of the colour code measure both $\operatorname{X}$ and $\operatorname{Z}$ stabilisers.

A logical operation is then fold-transversal if it can be formed by applying a combination of single-qubit operations and two-qubit operations between each qubit $i$ and its corresponding dual qubit $\tau(i)$ \cite{Breuckmann2024FoldTransversal}. Two common types of such gates are Hadamard-type gates and $\operatorname{S}$-type gates. Hadamard-type gates are implemented as shown in \Cref{fig:hadamard-toric-code}:

\begin{equation}
    \bigotimes_{\substack{i=1\\i<\tau(i)}}^n\operatorname{SWAP}_{i, \tau(i)}\bigotimes_{i=1}^n\operatorname{H}_i,
    \label{eq:hadamard-type}
\end{equation}

\noindent whereas $\operatorname{S}$-type gates are implemented as shown in \Cref{fig:s-toric-code}:

\begin{equation}
    \bigotimes_{\substack{i=1\\i<\tau(i)}}^n\operatorname{CZ}_{i, \tau(i)}\bigotimes_{\substack{i=1\\i=\tau(i)}}^n\operatorname{S}_i.
    \label{eq:s-type}
\end{equation}

The way the logical observables transform as a result of these gate sequences depends on the particular code and $\operatorname{ZX}$-duality \cite{Breuckmann2024FoldTransversal}. For the unrotated toric code, which encodes two logical qubits, the Hadamard-type gate maps the logical $\operatorname{Z}$ (resp.\ $\operatorname{X}$) operator on each qubit to the logical $\operatorname{X}$ (resp.\ $\operatorname{Z}$) operator on the same qubit. Note however that this operation occurs across \textit{both} logical qubits, so rather than performing a logical Hadamard gate across a single logical qubit, the logical operation is instead $\operatorname{H}_0\otimes\operatorname{H_1}$. The $\operatorname{S}$-type gate leaves the logical $\operatorname{Z}$ operator of each qubit unchanged, but maps the logical $\operatorname{X}$ operator of each qubit to the logical $\operatorname{Y}$ operator of the same qubit, therefore implementing $\operatorname{S}_0\otimes\operatorname{S}_1$. Note as with the Hadamard-type gate, the $\operatorname{S}$-type gate does not implement a logical $\operatorname{S}$ gate on a \textit{single} logical qubit, but instead acts across \textit{both} logical qubits. Finally, we emphasise that fold-transversal Hadamard-type gates (resp.\ $\operatorname{S}$-type gates) will not necessarily produce logical operations of the form $\bigotimes_i\operatorname{H}_i$ (resp.\ $\bigotimes_i\operatorname{S}_i$): in \Cref{ssec:folded-4-8-8} we will provide an example of a code which supports a $\operatorname{ZX}$-duality but produces different logical gates.

The fault tolerance of these operations comes from the fact that detecting regions can be formed surrounding the fold-transversal gates. Detectors are sets of measurements whose product is deterministic in the absence of errors \cite{McEwen2023RelaxingHardware}. Each detector has a corresponding detecting region, which is the area in spacetime during which a physical error can cause a detector to flip. For the Hadamard-type gate, detectors can be formed by comparing $\operatorname{X}$ (resp.\ $\operatorname{Z}$) stabilisers before the fold-transversal operation with $\operatorname{Z}$ (resp.\ $\operatorname{X}$) stabilisers after the fold-transversal operation. For the $\operatorname{S}$-type gate, $\operatorname{Z}$ stabilisers are unaffected by the operation and can therefore be directly compared to form detectors, and $\operatorname{X}$ stabilisers before the fold can be compared with a product of $\operatorname{X}$ and $\operatorname{Z}$ stabilisers after the fold. One disadvantage to this approach is that the fold-transversal $\operatorname{S}$-type gate can be harder to decode, as a measurement error on a $\operatorname{Z}$ stabiliser after the fold-transversal operation will now change three detectors rather than two \cite{Chen2024TransversalLogic}. Recent work has explored the challenges of fast decoding across such logical operations for the rotated surface code \cite{Wan2025CNOTDecoder, Cain2025FastCNOTDecoder, Serraperalta2025FastCNOTDecoder, Turner2025FastCNOTDecoder}.

Note that other related constructions have also been investigated in the literature. For example, Chen \etal \cite{Chen2024TransversalLogic} presented similar techniques for implementing logical Hadamard and $\operatorname{S}$ gates on the rotated surface code\footnote{The rotated surface code is based on a square lattice like the unrotated toric code, but rotated by $45^\circ$ and featuring open boundaries \cite{Bombin2007RotatedSurface, Horsman2012LatticeSurgery}.}. It is also known that for both the rotated and unrotated surface code, an error-corrected Hadamard gate can be realised by applying transversal Hadamard gates to each physical qubit and then rotating the code by $90^\circ$ \cite{Blunt2024Compilation, Geher2024LogicalHadamard}.

\subsection{Dehn twists on static codes}
\label{ssec:dehn-twist-static}

\begin{figure}[!th]
     \centering
     \begin{subfigure}[b]{0.3\textwidth}
         \centering
         \includegraphics[width=\textwidth]{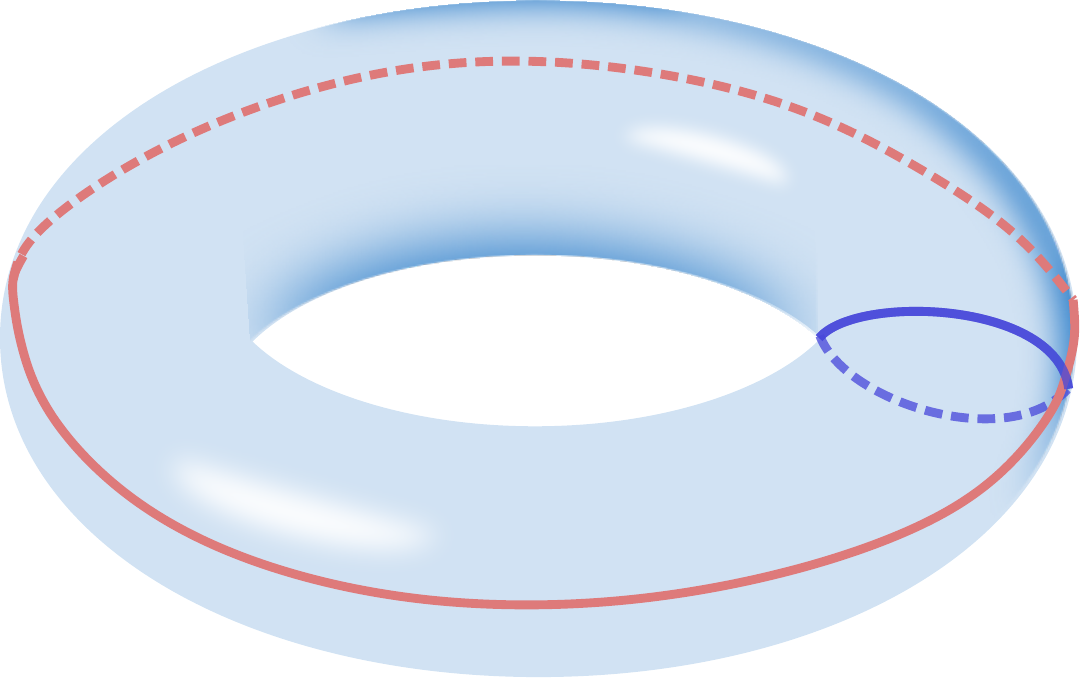}
         \caption{}
         \label{fig:torus}
     \end{subfigure}
     \hfill
     \begin{subfigure}[b]{0.3\textwidth}
         \centering
         \includegraphics[width=\textwidth]{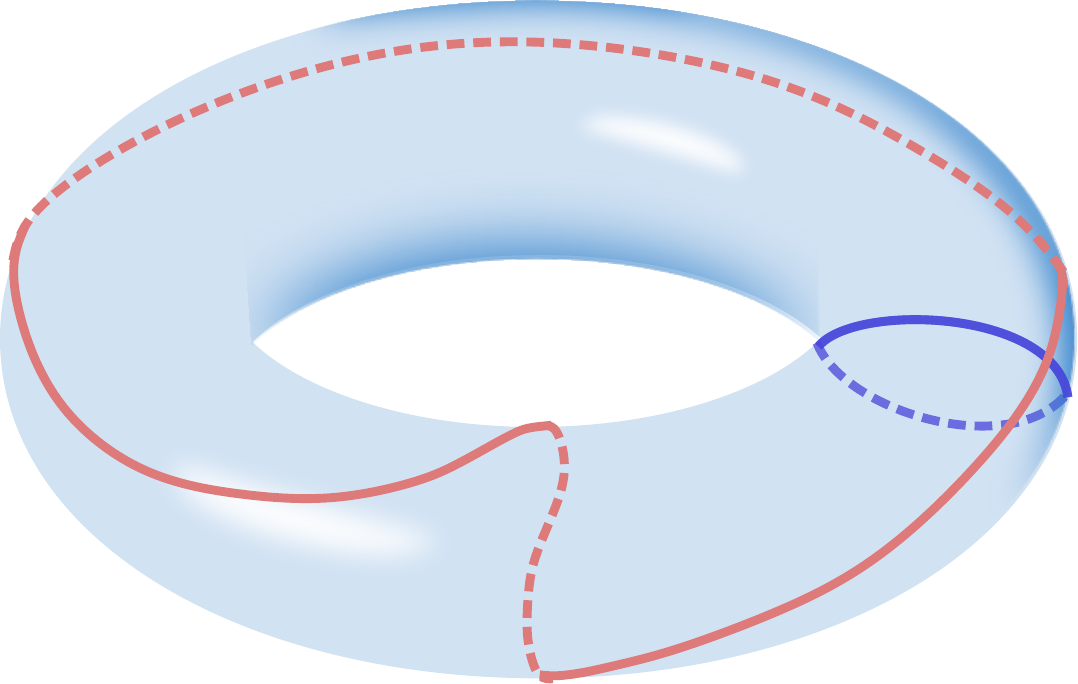}
         \caption{}
         \label{fig:torus_meredian_twist}
     \end{subfigure}
     \hfill
     \begin{subfigure}[b]{0.3\textwidth}
         \centering
         \includegraphics[width=\textwidth]{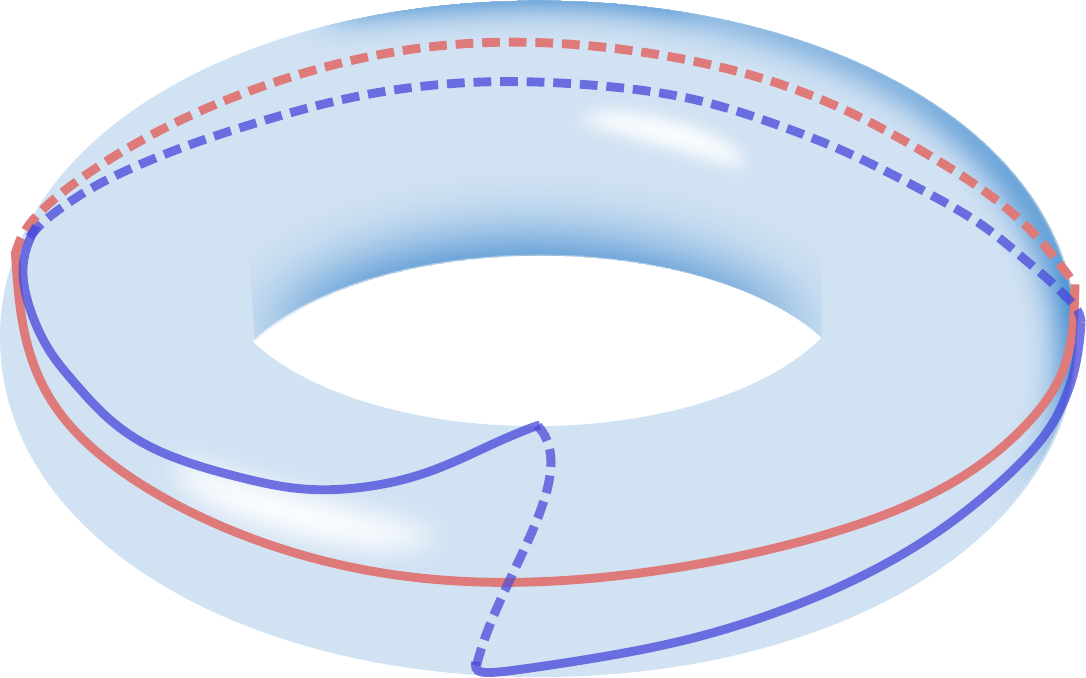}
         \caption{}
         \label{fig:torus_logitudinal_twist}
     \end{subfigure}
        \caption{Dehn twist schematic on a torus. In (a), independent non-trivial loops along which the logical operators lie are shown. The Dehn twist of the toroidal loop (red) along the poloidal loop is shown in (b). The Dehn twist along the toroidal loop is displayed in (c).}
        \label{fig:torus_cartoon_twist}
\end{figure}

Dehn twists on topological surfaces are orientation-preserving self-homeomorphisms \cite{Farb2011primer}. This deformation technique enables the implementation of non-trivial logical operations, such as logical $\operatorname{CNOT}$ gates, in quantum codes by exploiting topological properties. During the protocol, the surface undergoes a twist along a selected non-trivial loop. Consequently, any non-contractable loop that intersects the chosen loop changes in a non-trivial way. \Cref{fig:torus_cartoon_twist} shows the non-trivial loops on a torus and Dehn twists around the loops.

Each logical qubit encoded in the quantum code is associated with two types of logical gates, $\operatorname{X}$ and $\operatorname{Z}$, which are supported on the non-trivial loops of the topological surface. For $\mathbb{Z}_2$ codes, the lattice distortion can be implemented using only the two-qubit $\operatorname{CNOT}$ gate.
This approach works for the toric code, which is a $\mathbb{Z}_2$ code. Therefore, it is useful to know how the Paulis transform under the action of $\operatorname{CNOT}$. Since $\operatorname{CNOT}$ is in the Clifford group, we have
\begin{align}\label{eq:cnot_heisenberg}
\begin{split}
\operatorname{CNOT}(\operatorname{X}_\text{control}\otimes \operatorname{I}_\text{target})\operatorname{CNOT}^\dagger &= \operatorname{X}_\text{control}\otimes \operatorname{X}_\text{target},\\
\operatorname{CNOT}(\operatorname{I}_\text{control}\otimes \operatorname{X}_\text{target})\operatorname{CNOT}^\dagger &= \operatorname{I}_\text{control}\otimes \operatorname{X}_\text{target},\\
\operatorname{CNOT}(\operatorname{Z}_\text{control}\otimes \operatorname{I}_\text{target})\operatorname{CNOT}^\dagger &= \operatorname{Z}_\text{control}\otimes \operatorname{I}_\text{target},\\
\operatorname{CNOT}(\operatorname{I}_\text{control}\otimes \operatorname{Z}_\text{target})\operatorname{CNOT}^\dagger &= \operatorname{Z}_\text{control}\otimes \operatorname{Z}_\text{target}.
\end{split}
\end{align}
Note that a non-trivial transformation occurs on the Pauli $\operatorname{X}$ (resp.\ $\operatorname{Z}$) when the corresponding qubit is the control (resp.\ target) in the $\operatorname{CNOT}$ gate. Using this fact, a Dehn twist can be implemented as given in \cite{Breuckmann2017HyperbolicSurfaceCodes, Guernut2024ToricCliffordGates, Zhu2020DehnTwist}. The procedure goes as follows:
\begin{itemize}
    \item Choose  representatives of the logical operators along non-contractable loops
    \item Identify a specific loop along which Dehn twist needs to be performed. The logical $\operatorname{X}$ (resp. $\operatorname{Z}$) on it serves as the target (resp. control), denoted by  $\operatorname{X}_{\text{target}}$ (resp. $\operatorname{Z}_{\text{control}}$)
    \item Determine a different loop that should be twisted around the first loop. The logical $\operatorname{X}$ (resp. $\operatorname{Z}$) on the second loop serves as $\operatorname{X}_{\text{control}}$ (resp. $\operatorname{Z}_{\text{target}}$).
    \item Apply $\operatorname{CNOT}$s with targets on the qubits in logical $\operatorname{X}_{\text{target}}$ and controls from a parallel loop that intersects with logical $\operatorname{X}_{\text{control}}$
    \item Update stabilisers and perform an error correction round
    \item Repeat the above two steps until the original lattice is recovered
\end{itemize}
The first three points are straightforward, so we now explain the reasoning behind the remaining points. The loop along which the twist happens is unaffected, and the other loop deforms during the transformation.  Thus, \Cref{eq:cnot_heisenberg} suggests applying $\operatorname{CNOT}$s with targets on the qubits in logical $\operatorname{X}_{\text{target}}$. The controls are chosen from a parallel loop that intersects with logical $\operatorname{X}_{\text{control}}$. As a result, logical $\operatorname{X}_{\text{target}}$ remains unaffected. The support of logical $\operatorname{X}_{\text{control}}$ increases by one, where the additional qubit is along the direction of $\operatorname{X}_{\text{target}}$.  Since logicals $\operatorname{Z}_{\text{control}}$ and $\operatorname{Z}_{\text{target}}$ are parallel to $\operatorname{X}_{\text{target}}$ and $\operatorname{X}_{\text{control}}$, respectively, similar reasoning can be used to argue that $\operatorname{Z}_{\text{control}}$ remains unchanged while $\operatorname{Z}_{\text{target}}$ gains an extra qubit. Applying $\operatorname{CNOT}$s deforms the lattice, which requires updating stabilisers according to \Cref{eq:cnot_heisenberg}. Hence, stabilisers should be appropriately compared before and after the deformation while executing a QEC round. In order to recover the original geometry of the lattice, we repeat the process of applying $\operatorname{CNOT}$s by shifting targets along the first loop until the original stabilisers are recovered. At the end, the procedure implements a logical $\operatorname{CNOT}$ as logical $\operatorname{X}_{\text{control}}$ ($\operatorname{Z}_{\text{target}}$) acquires the component of $\operatorname{X}_{\text{target}}$ ($\operatorname{Z}_{\text{control}}$) while $\operatorname{X}_{\text{target}}$ ($\operatorname{Z}_{\text{control}}$) remains the same.

\begin{figure}[!th]
     \centering
     \begin{subfigure}[b]{0.24\textwidth}
         \centering
         \includegraphics[width=\textwidth]{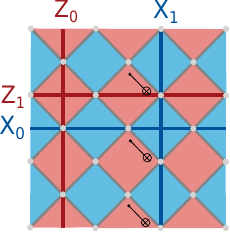}
         \caption{}
         \label{fig:toric_linear_intial}
     \end{subfigure}
     \hfill
     \begin{subfigure}[b]{0.24\textwidth}
         \centering
         \includegraphics[width=\textwidth]{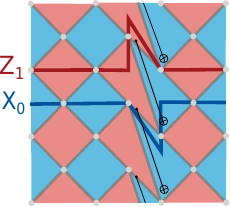}
         \caption{}
         \label{fig:toric_linear_layer1}
     \end{subfigure}
     \hfill
     \begin{subfigure}[b]{0.24\textwidth}
         \centering
         \includegraphics[width=\textwidth]{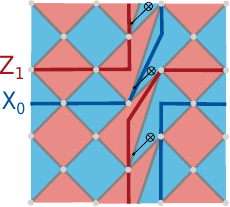}
         \caption{}
         \label{fig:toric_linear_layer2}
     \end{subfigure}
    \hfill
     \begin{subfigure}[b]{0.24\textwidth}
         \centering
         \includegraphics[width=\textwidth]{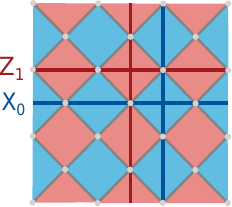}
         \caption{}
         \label{fig:toric_linear_final}
     \end{subfigure}
        \caption{Linear Dehn twist implementing logical $\operatorname{CNOT}_{0,1}$. The selection of logical operators and the initial layer of $\operatorname{CNOT}$ gates are illustrated in (a). Note that the support of $\operatorname{Z}_0$ and $\operatorname{X}_1$ remains the same in subsequent figures, hence omitted. The deformed lattice, modifications to logical operators, and the $\operatorname{CNOT}$ sequence for subsequent layers are presented in (b) and (c). After the application of $\operatorname{CNOT}$ gates in (c), the original lattice structure is restored in (d), and operators $\operatorname{X}_0$ and $\operatorname{Z}_1$ acquire a component along the vertical loop.}
        \label{fig:toric_linear_dehn}
\end{figure}

The Dehn twist protocol along the vertical loop is explained in \Cref{fig:toric_linear_dehn} for the unrotated toric code of size $d=3$. The initial logical strings are shown in \Cref{fig:toric_linear_intial}. We apply $\operatorname{CNOT}$ gates with targets on qubits in $\operatorname{X}_1$. The control lies along a line parallel to the vertical loop, which also intersects with $\operatorname{X}_0$. The stabilisers transform obeying \Cref{eq:cnot_heisenberg}. In particular, the support of each $\operatorname{X}$  and $\operatorname{Z}$ stabilizer changes by one qubit as shown in \Cref{fig:toric_linear_layer1}. The support of logicals along the horizontal loop also changes, which can be tracked using \Cref{eq:cnot_heisenberg}. After repeating the process $d-1$ times, we obtain the logical transformation $\operatorname{X}_0\rightarrow \operatorname{X}_0\operatorname{X}_1$, $\operatorname{Z}_0\rightarrow\operatorname{Z}_0$, $\operatorname{X}_1\rightarrow\operatorname{X}_1$, and $\operatorname{Z}_1\rightarrow\operatorname{Z}_0\operatorname{Z}_1$ thus implementing a logical $\operatorname{CNOT}_{0,1}$. A similar technique can be used to perform logical $\operatorname{CNOT}_{10}$ with the role of horizontal and vertical loops interchanged.

\begin{figure}[!th]
     \centering
     \begin{subfigure}[b]{0.3\textwidth}
         \centering
         \includegraphics[width=\textwidth]{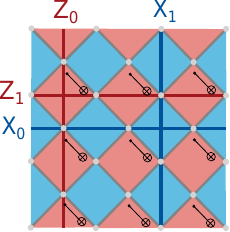}
         \caption{}
         \label{fig:toric_instantaneous_initial}
     \end{subfigure}
     \hfill
     \begin{subfigure}[b]{0.3\textwidth}
         \centering
         \includegraphics[width=\textwidth]{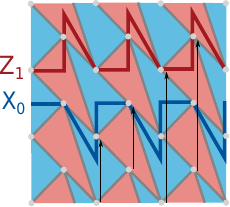}
         \caption{}
         \label{fig:toric_instantaneous_swap}
     \end{subfigure}
     \hfill
     \begin{subfigure}[b]{0.3\textwidth}
         \centering
         \includegraphics[width=\textwidth]{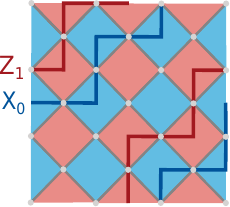}
         \caption{}
         \label{fig:toric_instantaneous_final}
     \end{subfigure}
        \caption{Instantaneous Dehn twist implementing logical $\operatorname{CNOT}_{0,1}$. The $\operatorname{CNOT}$ sequence along with logicals is indicated in (a). The support of $\operatorname{Z}_0$ and $\operatorname{X}_1$ remains the same throughout the process. The transformed stabilisers and logicals are presented in (b). An indicative cyclic shift is also shown in (b). Note that the range increases from left to right. When the lattice is restored in (c), the operators $\operatorname{X}_0$ and $\operatorname{Z}_1$ gain the vertical component, thus finishing the protocol.}
        \label{fig:toric_instantaneous_dehn}
\end{figure}

The above implementation takes $O(d)$ time steps for lattice distortion using $\operatorname{CNOT}$s. Thus, it is termed linear-time Dehn twist \cite{Breuckmann2017HyperbolicSurfaceCodes}. A different implementation, called instantaneous Dehn twist \cite{Zhu2020DehnTwist}, can be used to reduce time overhead while deforming the lattice by applying $\operatorname{CNOT}$s in parallel across a wider spatial region. \Cref{fig:toric_instantaneous_dehn} presents the instantaneous Dehn twist that implements the logical $\operatorname{CNOT}_{0,1}$ operation. This approach works as follows. First, choose a loop along which the Dehn twist should be realised. Perform the first layer of $\operatorname{CNOT}$s similar to the linear version, but now simultaneously along all loops parallel to the specified loop.  This step can be executed in a constant amount of time. All the stabilisers are deformed, and the logicals are changed according to \Cref{eq:cnot_heisenberg}. The lattice can be restored by applying cyclic shift operations along the chosen loop, which can be done by long-range $\operatorname{SWAP}$s or qubit shuttling. Here, the code size determines the support of long-range $\operatorname{SWAP}$s. While the $\operatorname{CNOT}$ step requires $O(1)$ time, the $\operatorname{SWAP}$ operations may require up to $O(d-1)$ time steps. The instantaneous version is therefore potentially more suitable for architectures that realise long-range $\operatorname{SWAP}$ operations, as enabled by qubit shuttling or optical interconnects.

We will now discuss how to implement these Dehn twists in a fault-tolerant way. As discussed, the lattice is deformed using $\operatorname{CNOT}$s between the data qubits. Each of these $\operatorname{CNOT}$'s can introduce further errors. Hence, to maintain fault tolerance, it is essential to perform an error correction round by comparing stabilizers with those from the previous layer. 
One can also perform an error correction round, say after applying $n$ layers of $\operatorname{CNOT}$s at the cost of reducing the distance by $O(n)$.

Note that in the instantaneous case, long-range $\operatorname{SWAP}$s are required to restore the code space. Since qubits separated by a length of $O(d)$ can interact, two-qubit errors from applying the SWAP gates can reduce the code distance by half. It may be possible to restore code distance by inserting QEC rounds during the SWAP layer. 
Two rounds of error correction are sufficient if the interaction range during $\operatorname{SWAP}$s is $O(1)$ compared to the code distance; such an implementation is discussed in hyperbolic codes, see \cite{Lavasani2019universallogical}.

Using either of the above methods, one can implement logical $\operatorname{SWAP}$ by performing three sequential logical $\operatorname{CNOT}$s in the order $\operatorname{CNOT}_{0,1}$, $\operatorname{CNOT}_{1,0}$, $\operatorname{CNOT}_{0,1}$. So far, we have discussed executing one Dehn twist at a time. Multiple Dehn twists can also be implemented at once if we allow long-range interactions for $\operatorname{CNOT}$s, as demonstrated in \cite{Zhu2020DehnTwist}.

\subsection{Floquet codes}
\label{ssec:floquet-codes}

Floquet codes are a variant of dynamic codes which allow the encoded logical information to evolve over time \cite{Hastings2021FloquetCodes, Davydova2023CSSFloquet, Kesselring2024AnyonFloquet, Bombin2024Unifying}. They can be defined on a colour code lattice, meaning a 2-D lattice which is trivalent and three-face-colourable \cite{Vuillot2021PlanarFloquetCodes}. Each vertex in the lattice corresponds to a qubit, and each edge corresponds to a two-qubit measurement. The ability to decompose all operations on Floquet codes into two-qubit measurements has made them particularly relevant for QEC on Majorana devices \cite{Microsoft2025Roadmap} as well as for networked architectures based on distributed Bell pairs \cite{Higgott2024HyperbolicFloquet, Dessertaine2025PhotonicFloquet, Sutcliffe2025DistributedQEC}.

\begin{figure}
    \centering
    \begin{subfigure}{0.4\linewidth}
        \includegraphics[width=\linewidth]{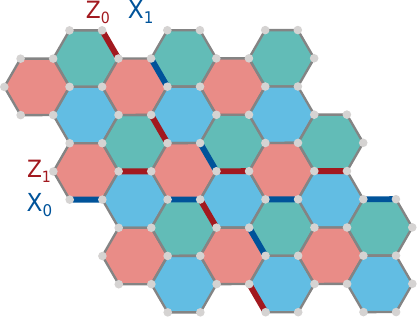}
        \caption{}
        \label{fig:honeycomb}
    \end{subfigure}
    \hspace{2cm}
    \begin{subfigure}{0.32\linewidth}
        \includegraphics[width=\linewidth]{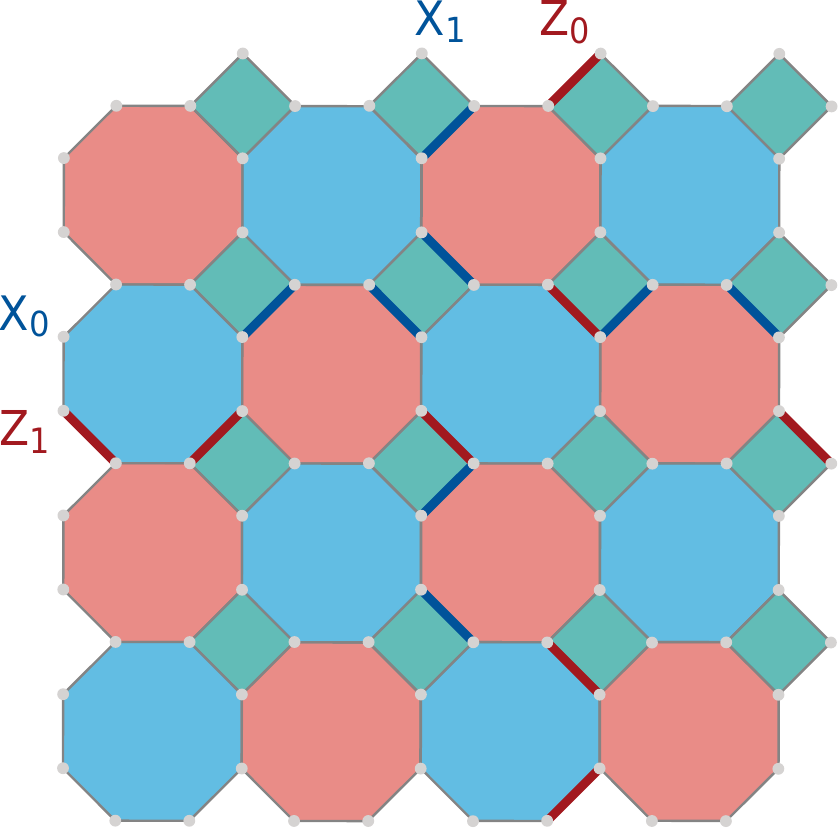}
        \caption{}
        \label{fig:4-8-8}
    \end{subfigure}
    \caption{Floquet codes can be defined on a colour code lattice, such as the (a) $6.6.6$ (``honeycomb'') and (b) $4.8.8$ (``square-octagon'') lattices. Nodes represent physical qubits, edges represent two-qubit Pauli measurements, and plaquettes represent stabilisers. The specific measurements, stabilisers, and logical operators depend on the measurement schedule, see Tables \ref{tab:floquet-gidney-schedule} and \ref{tab:floquet-css-schedule} for some example schedules. For both measurement schedules, logical operators between the $\color{floquet-blue}\operatorname{bZZ}$ and $\color{floquet-red}\operatorname{rXX}$ measurement rounds are presented in both figures as red and blue lines. In both figures there are four logical operators representing two logical qubits.}
    \label{fig:example-lattices}
\end{figure}

The structure of the lattice allows us to assign colourings to edges based on the colour of the faces that a given edge connects. In this work we will primarily focus on the $6.6.6$\footnote{Throughout this work we shall refer to lattices using Cundy \& Rollett’s notation for Euclidean tilings \cite{CundyNotation}. Under this notation, each vertex in a $i_0.i_1...i_{n-1}.i_n$ lattice is surrounded by $n$ polygons arranged clockwise such that the $j$-th polygon has $i_j$ sides. For example, a $6.6.6$ lattice means that each vertex is surrounded by three hexagons, a $4.8.8$ lattice means that each vertex is surrounded by one square and two octagons, and a $3.6.3.6$ lattice means that each vertex is surrounded clockwise by one triangle, one hexagon, a second triangle, and a second hexagon, in that order.} (``honeycomb'') lattice shown in \Cref{fig:honeycomb} \cite{Hastings2021FloquetCodes}, but in general these codes can be defined on any colour code lattice, including the $4.8.8$ (``square-octagon'') lattice shown in \Cref{fig:4-8-8} \cite{Paetznick2023PlanarFloquetPerformance}, (semi-)hyperbolic lattices such as the 8.8.8 lattice \cite{Higgott2024HyperbolicFloquet, Fahimniya2025Hyperbolic}, and lattices which are non-uniform due to hardware imperfections such as defective qubits \cite{Aasen2023DeadQubits, McLauchlan2024FabDefects}.

The other important component to a Floquet code is the measurement schedule. The measurement schedule is a sequence of Pauli pair measurements which repeats with some period\footnote{Note that aperiodic dynamic codes can also be defined \cite{Davydova2023CSSFloquet}.}. We define a QEC round as implementing a full iteration of the measurement schedule, and each layer of measurements in the schedule as a sub-round. In each sub-round of the measurement schedule, a colour $C\in\{\textrm{red}, \textrm{green}, \textrm{blue}\}$ and pair of Pauli bases $\operatorname{P_0}, \operatorname{P_1}$ are specified. Pairs of qubits connected by an edge of a given colour are then measured in the given Pauli bases\footnote{Note that some Floquet codes might apply different Pauli measurements to different pairs of qubits in the same measurement round, such as the original Floquet code described in Hastings and Haah \cite{Hastings2021FloquetCodes}. The Hastings and Haah measurement schedule is equivalent to the measurement schedule presented in Gidney \etal up to single-qubit Clifford gates \cite{Gidney2021FloquetBenchmark}.}. For example, during an $\color{floquet-red}{\operatorname{rXX}}$ sub-round, qubits connected by red edges will be measured in the joint Pauli $\operatorname{XX}$ basis. Two common measurement schedules are the period-three schedule ${\color{floquet-red}{\operatorname{rXX}}} \rightarrow {\color{floquet-green}{\operatorname{gYY}}} \rightarrow {\color{floquet-blue}{\operatorname{bZZ}}}$ \cite{Hastings2021FloquetCodes, Gidney2021FloquetBenchmark}, and the period-six schedule ${\color{floquet-red}{\operatorname{rXX}}} \rightarrow {\color{floquet-green}{\operatorname{gZZ}}} \rightarrow {\color{floquet-blue}{\operatorname{bXX}}} \rightarrow {\color{floquet-red}{\operatorname{rZZ}}} \rightarrow {\color{floquet-green}{\operatorname{gXX}}} \rightarrow {\color{floquet-blue}{\operatorname{bZZ}}}$, the latter of which gives rise to CSS Floquet codes \cite{Davydova2023CSSFloquet, Kesselring2024AnyonFloquet, Bombin2024Unifying}. For this paper, we will primarily focus on the latter schedule, which exhibits a structure comparable to CSS codes \cite{Davydova2023CSSFloquet}, but we will offer some comments about how our work can be generalised to the period-three schedule as well as others \cite{Haah2022BoundaryFloquet, Dua2024FloquetRewinding}.

For each sub-round we can define an instantaneous stabiliser group (ISG) to describe how the code evolves over time \cite{Hastings2021FloquetCodes}. For a given sub-round, the ISG consists of Pauli operators which were measured during that sub-round, plus any Pauli operators from the previous ISG which commute with the measurements from this sub-round. If a given Pauli in the previous ISG can be formed by a product of measurements in the current sub-round, then a detector can be formed and used to detect errors in the code. Logical observables can be preserved as products of non-trivial Pauli operators which commute with each ISG. Measurement outcomes from each sub-round must be multiplied into the observables to ensure the observables continue to commute with each ISG, this is what distinguishes Floquet codes from their static counterparts.

We give examples of how detectors and observables evolve for the period-three and period-six schedules in Tables \ref{tab:floquet-gidney-schedule} and \ref{tab:floquet-css-schedule}, respectively. Each row specifies a Pauli string, with each column showing how that Pauli string evolves during each sub-round. A row starting with a red, green, or blue plaquette refers to a Pauli product $\bigotimes_i\operatorname{P}_i$ of the qubits around that plaquette. Take for example a blue plaquette in \Cref{tab:floquet-gidney-schedule}. There is a $\bigotimes_i\operatorname{Z}_i$ stabiliser on this blue plaquette which persists throughout the measurement schedule, which we show in the first row of \Cref{tab:floquet-gidney-schedule}, which is indexed by a blue plaquette. There is also a time-dependent detecting region which exists on this plaquette, which we present in the fourth row, which is also indexed by a blue plaquette: this detecting region starts during the $\color{floquet-red}\operatorname{rXX}$ sub-round as $\bigotimes_i\operatorname{X_i}$, evolves to $\bigotimes_i\operatorname{Z_i}$ during the $\color{floquet-green}\operatorname{gYY}$ sub-round, remains unchanged during the $\color{floquet-blue}\operatorname{bZZ}$, evolves again to $\bigotimes_i\operatorname{Y_i}$ during the second $\color{floquet-red}\operatorname{rXX}$, before being fully measured out during the second $\color{floquet-green}\operatorname{gYY}$ sub-round. The last two rows correspond to Pauli strings along non-trivial loops in the lattice: each loop defines two logical operators, which correspond to Pauli strings along edges of a given colour in the loop.

We can use these tables to see what Pauli information exists during different sub-rounds. For example, by looking at the third column of \Cref{tab:floquet-gidney-schedule}, which is indexed by $\color{floquet-blue}\operatorname{bZZ}$, we can see reading down that column that the following Pauli information exists after the $\color{floquet-blue}\operatorname{bZZ}$ sub-round:

\begin{itemize}
    \item a $\operatorname{Z}$ stabiliser on each blue plaquette;
    \item a $\operatorname{X}$ stabiliser on each red plaquette;
    \item a $\operatorname{Y}$ stabiliser on each green plaquette;
    \item a $\operatorname{Z}$ detecting region on each blue plaquette;
    \item a $\operatorname{X}$ detecting region on each red plaquette;
    \item a $\operatorname{Z}$ detecting region on each green plaquette;
    \item logical observables formed by a product of Pauli $\operatorname{Z}$ operators on red edges along each non-trivial loop; and
    \item logical observables formed by a product of Pauli $\operatorname{X}$ operators on blue edges along each non-trivial loop.
\end{itemize}

\begin{table}[]
    \centering
    \caption{Detecting regions and logical observables for each sub-round in the bulk of the period-three honeycomb Floquet code \cite{Hastings2021FloquetCodes, Gidney2021FloquetBenchmark}. $\operatorname{P}_i$ refers to a Pauli operator $\operatorname{P}$ on the qubit indexed by $i$ for all qubits in a plaquette. $\operatorname{X}$, $\operatorname{Y}$, and $\operatorname{Z}$ stabilisers persist across the red, green, and blue plaquettes for the full duration of the code. $\operatorname{I}$ appearing in a cell indicates that detectors on plaquettes of that colour can be measured during that sub-round. $\operatorname{X_L}$ and $\operatorname{Z_L}$ correspond to the operators for a logical qubit, for each sub-round these correspond to Pauli operators along edges of a given colour in a non-trivial loop. Note that the period-three schedule exchanges $\operatorname{X}$ and $\operatorname{Z}$ logical Paulis in the same homology class every three sub-rounds. This is not the same as a logical Hadamard gate, which exchanges $\operatorname{X}$ and $\operatorname{Z}$ logical Paulis across different homology classes.}
    \label{tab:floquet-gidney-schedule}
    \begin{tabular}{|c|c|c|c|c|c|c|c|}
        \hline
        \multirow{2}{2.9em}{Pauli string} & \multicolumn{7}{|c|}{Sub-round}\\\cline{2-8}
        & $\color{floquet-red}\operatorname{rXX}$ & $\color{floquet-green}\operatorname{gYY}$ & $\color{floquet-blue}\operatorname{bZZ}$ & $\color{floquet-red}\operatorname{rXX}$ & $\color{floquet-green}\operatorname{gYY}$ & $\color{floquet-blue}\operatorname{bZZ}$ & $\color{floquet-red}\operatorname{rXX}$ \\\hline
        \includegraphics[width=1.5em]{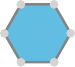} & $\color{floquet-blue}\operatorname{Z}_i$ & $\color{floquet-blue}\operatorname{Z}_i$ & $\color{floquet-blue}\operatorname{Z}_i$ & $\color{floquet-blue}\operatorname{Z}_i$ & $\color{floquet-blue}\operatorname{Z}_i$ & $\color{floquet-blue}\operatorname{Z}_i$ & $\color{floquet-blue}\operatorname{Z}_i$ \\\hline
        \includegraphics[width=1.5em]{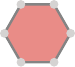} & $\color{floquet-red}\operatorname{X}_i$ & $\color{floquet-red}\operatorname{X}_i$ & $\color{floquet-red}\operatorname{X}_i$ & $\color{floquet-red}\operatorname{X}_i$ & $\color{floquet-red}\operatorname{X}_i$ & $\color{floquet-red}\operatorname{X}_i$ & $\color{floquet-red}\operatorname{X}_i$ \\\hline
        \includegraphics[width=1.5em]{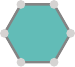} & $\color{floquet-green}\operatorname{Y}_i$ & $\color{floquet-green}\operatorname{Y}_i$ & $\color{floquet-green}\operatorname{Y}_i$ & $\color{floquet-green}\operatorname{Y}_i$ & $\color{floquet-green}\operatorname{Y}_i$ & $\color{floquet-green}\operatorname{Y}_i$ & $\color{floquet-green}\operatorname{Y}_i$ \\\hline
        \includegraphics[width=1.5em]{blue-plaquette.pdf} & $\color{floquet-blue}\operatorname{X}_i$ & $\color{floquet-blue}\operatorname{Z}_i$ & $\color{floquet-blue}\operatorname{Z}_i$ & $\color{floquet-blue}\operatorname{Y}_i$ & $\color{floquet-blue}\operatorname{I}_i$ & & \\\hline
        \includegraphics[width=1.5em]{red-plaquette.pdf} &  & $\color{floquet-red}\operatorname{Y}_i$ & $\color{floquet-red}\operatorname{X}_i$ & $\color{floquet-red}\operatorname{X}_i$ & $\color{floquet-red}\operatorname{Z}_i$ & $\color{floquet-red}\operatorname{I}_i$ & \\\hline
        \includegraphics[width=1.5em]{green-plaquette.pdf} &  &  & $\color{floquet-green}\operatorname{Z}_i$ & $\color{floquet-green}\operatorname{Y}_i$ & $\color{floquet-green}\operatorname{Y}_i$ & $\color{floquet-green}\operatorname{X}_i$ & $\color{floquet-green}\operatorname{I}_i$ \\\hline
        $\operatorname{X_L}$ & $\color{floquet-green}\operatorname{gXX}$ & $\color{floquet-green}\operatorname{gZZ}$ & $\color{floquet-red}\operatorname{rZZ}$ & $\color{floquet-red}\operatorname{rYY}$ & $\color{floquet-blue}\operatorname{bYY}$ & $\color{floquet-blue}\operatorname{bXX}$ & $\color{floquet-green}\operatorname{gXX}$ \\\hline
        $\operatorname{Z_L}$ & $\color{floquet-red}\operatorname{rYY}$ & $\color{floquet-blue}\operatorname{bYY}$ & $\color{floquet-blue}\operatorname{bXX}$ & $\color{floquet-green}\operatorname{gXX}$ & $\color{floquet-green}\operatorname{gZZ}$ & $\color{floquet-red}\operatorname{rZZ}$ & $\color{floquet-red}\operatorname{rYY}$ \\\hline
    \end{tabular}
\end{table}

\begin{table}[]
    \centering
    \caption{Detecting regions and logical observables for each sub-round in the bulk of the period-six CSS honeycomb Floquet code \cite{Davydova2023CSSFloquet, Kesselring2024AnyonFloquet, Bombin2024Unifying}. Each plaquette measures an $\operatorname{X}$ and $\operatorname{Z}$ detector, though the detectors are measured at different times. Unlike the period-three schedule, no detecting regions persist across the full QEC round, and the $\operatorname{X}$ and $\operatorname{Z}$ Pauli strings which form the logical operators are never exchanged.}
    \label{tab:floquet-css-schedule}
    \begin{tabular}{|c|c|c|c|c|c|c|c|c|c|c|}
        \hline
        \multirow{2}{2.9em}{Pauli string} & \multicolumn{10}{|c|}{Sub-round}\\\cline{2-11}
        & $\color{floquet-red}\operatorname{rXX}$ & $\color{floquet-green}\operatorname{gZZ}$ & $\color{floquet-blue}\operatorname{bXX}$ & $\color{floquet-red}\operatorname{rZZ}$ & $\color{floquet-green}\operatorname{gXX}$ & $\color{floquet-blue}\operatorname{bZZ}$ & $\color{floquet-red}\operatorname{rXX}$ & $\color{floquet-green}\operatorname{gZZ}$ & $\color{floquet-blue}\operatorname{bXX}$ & $\color{floquet-red}\operatorname{rZZ}$ \\\hline
        \includegraphics[width=1.5em]{blue-plaquette.pdf} & $\color{floquet-blue}\operatorname{X}_i$ & $\color{floquet-blue}\operatorname{X}_i$ & $\color{floquet-blue}\operatorname{X}_i$ & $\color{floquet-blue}\operatorname{X}_i$ & $\color{floquet-blue}\operatorname{I}_i$ & & & & & \\\hline
        \includegraphics[width=1.5em]{red-plaquette.pdf} & & $\color{floquet-red}\operatorname{Z}_i$ & $\color{floquet-red}\operatorname{Z}_i$ & $\color{floquet-red}\operatorname{Z}_i$ & $\color{floquet-red}\operatorname{Z}_i$ & $\color{floquet-red}\operatorname{I}_i$ & & & & \\\hline
        \includegraphics[width=1.5em]{green-plaquette.pdf} &  &  & $\color{floquet-green}\operatorname{X}_i$ & $\color{floquet-green}\operatorname{X}_i$ & $\color{floquet-green}\operatorname{X}_i$ & $\color{floquet-green}\operatorname{X}_i$ & $\color{floquet-green}\operatorname{I}_i$ & & & \\\hline
        \includegraphics[width=1.5em]{blue-plaquette.pdf} & & & & $\color{floquet-blue}\operatorname{Z}_i$ & $\color{floquet-blue}\operatorname{Z}_i$ & $\color{floquet-blue}\operatorname{Z}_i$ & $\color{floquet-blue}\operatorname{Z}_i$ & $\color{floquet-blue}\operatorname{I}_i$ & & \\\hline
        \includegraphics[width=1.5em]{red-plaquette.pdf} & & & &  & $\color{floquet-red}\operatorname{X}_i$ & $\color{floquet-red}\operatorname{X}_i$ & $\color{floquet-red}\operatorname{X}_i$ & $\color{floquet-red}\operatorname{X}_i$ & $\color{floquet-red}\operatorname{I}_i$ & \\\hline
        \includegraphics[width=1.5em]{green-plaquette.pdf} & & & & & & $\color{floquet-green}\operatorname{Z}_i$ & $\color{floquet-green}\operatorname{Z}_i$ & $\color{floquet-green}\operatorname{Z}_i$ & $\color{floquet-green}\operatorname{Z}_i$ & $\color{floquet-green}\operatorname{I}_i$ \\\hline
        $\operatorname{X_L}$ & $\color{floquet-green}\operatorname{gXX}$ & $\color{floquet-green}\operatorname{gXX}$ & $\color{floquet-red}\operatorname{rXX}$ & $\color{floquet-red}\operatorname{rXX}$ & $\color{floquet-blue}\operatorname{bXX}$ & $\color{floquet-blue}\operatorname{bXX}$ & $\color{floquet-green}\operatorname{gXX}$ & $\color{floquet-green}\operatorname{gXX}$ & $\color{floquet-red}\operatorname{rXX}$ & $\color{floquet-red}\operatorname{rXX}$ \\\hline
        $\operatorname{Z_L}$ & $\color{floquet-red}\operatorname{rZZ}$ & $\color{floquet-blue}\operatorname{bZZ}$ & $\color{floquet-blue}\operatorname{bZZ}$ & $\color{floquet-green}\operatorname{gZZ}$ & $\color{floquet-green}\operatorname{gZZ}$ & $\color{floquet-red}\operatorname{rZZ}$ & $\color{floquet-red}\operatorname{rZZ}$ & $\color{floquet-blue}\operatorname{bZZ}$ & $\color{floquet-blue}\operatorname{bZZ}$ & $\color{floquet-green}\operatorname{gZZ}$ \\\hline
    \end{tabular}
\end{table}

For Floquet codes defined on a 2-D colour code lattice, the instantaneous stabiliser group in each sub-round corresponds to a surface code \cite{Hastings2021FloquetCodes, Davydova2023CSSFloquet}. This can be particularly relevant when searching for logical gates that can be implemented on these codes, as one can implement logical gates on the embedded code \cite{Davydova2024DynamicAutomorphism}. We offer some comments about this approach in \Cref{sec:embedded-codes}.

Finally, we note that certain Floquet codes introduce non-trivial automorphisms through their measurement schedules. For example, the period-three ${\color{floquet-red}{\operatorname{rXX}}} \rightarrow {\color{floquet-green}{\operatorname{gYY}}} \rightarrow {\color{floquet-blue}{\operatorname{bZZ}}}$ schedule induces an exchange between $\operatorname{X}$ and $\operatorname{Z}$ logical information\footnote{Note that Pauli information is only exchanged between strings of the same homology, i.e.\ a horizontal (resp.\ vertical) $\operatorname{X}$ string is exchanged with a horizontal (resp.\ vertical) $\operatorname{Z}$ string.}, as can be seen in \Cref{tab:floquet-gidney-schedule} by comparing the logical operators during the two $\color{floquet-blue}\operatorname{bZZ}$ sub-rounds. This property subsequently inspired the concept of dynamic automorphism codes, which encode $k$ logical qubits by utilising $k$ stacks of 2-D lattices and are able to implement the full $k$-qubit Clifford group using pairwise measurements \cite{Davydova2024DynamicAutomorphism}. Note that while a full set of $k$-qubit Clifford gates has been identified for the dynamic automorphism codes, it is yet to be confirmed that this full set of Clifford gates is fault-tolerant.

\section{Fold-transversal gates on Floquet codes}
\label{sec:fold-transversal-floquet}

We will now discuss how the fold-transversal gates described in \Cref{ssec:fold-transversal-static} can be extended to Floquet codes. Our approach is to use emergent $\operatorname{ZX}$-dualities in the instantaneous stabiliser group to identify Hadamard-type and $\operatorname{S}$-type logical gates.

Our primary example of this emergent symmetry and the resulting logical gates is using the CSS Floquet code defined on a honeycomb lattice with periodic boundaries. On this code the Hadamard-type and $\operatorname{S}$-type logical gates produced are $\operatorname{H_0}\otimes \operatorname{H_1}$ and $\operatorname{S_0} \otimes \operatorname{S_1}$, respectively. However, we will show how this work can also be generalised beyond this example: in \Cref{ssec:folded-gidney-schedule}, we show how this approach can generalise to the period-3 measurement schedule, producing logical gates $\operatorname{H_0}\otimes \operatorname{H_1}$ and $\operatorname{S_0} \otimes \operatorname{S_1}$; and in \Cref{ssec:folded-4-8-8} we show how lattice growth techniques similar to those used by Moussa \cite{Moussa2016FoldTransversal} can be used to implement fold-transversal gates on the planar $4.8.8$ code, thus producing logical $\operatorname{H}$ and $\operatorname{S}$ gates.

\subsection{Emergent $\operatorname{ZX}$-dualities in CSS Floquet codes}
\label{ssec:floquet-zx-duality}

We shall start by looking further at the instantaneous stabiliser group of CSS Floquet codes and what $\operatorname{ZX}$-dualities emerge across the measurement schedule.

Suppose we are between the ${\color{floquet-blue}\operatorname{bZZ}}$ and ${\color{floquet-red}\operatorname{rXX}}$ measurement sub-rounds, using the period-six CSS measurement schedule. Referring back to \Cref{tab:floquet-css-schedule}, we can see that at this point the instantaneous stabiliser group consists of the following Pauli operations:
\begin{itemize}
    \item $\operatorname{X}$ checks on the red plaquettes;
    \item $\operatorname{Z}$ checks on the blue plaquettes;
    \item $\operatorname{X}$ checks on the green plaquettes;
    \item $\operatorname{Z}$ checks on the green plaquettes; and
    \item $\operatorname{Z}$ checks on the blue edges.
\end{itemize}

\begin{figure}
    \centering
    \includegraphics[width=0.4\linewidth]{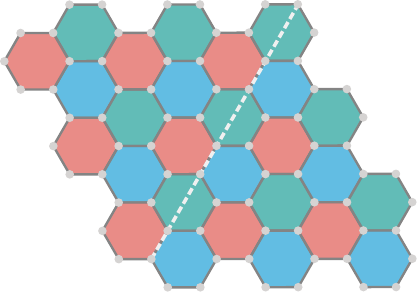}
    \caption{The honeycomb Floquet code with a $\operatorname{ZX}$-duality. The fold lies along the dashed line. Each qubit on one side of the dashed line maps to a qubit on the other side of the line, with qubits on the line mapping to themselves. Red and blue plaquettes map to blue and red plaquettes, respectively. Green plaquettes map to green plaquettes.}
    \label{fig:honeycomb-zx-duality}
\end{figure}

Based on these stabilisers, we argue that a $\operatorname{ZX}$-duality during this sub-round is a mapping of qubits such that each red (resp.\ blue) plaquette is mapped to a corresponding blue (resp.\ red) plaquette, and each green plaquette is mapped to a corresponding green plaquette. An example of such a duality is shown for the honeycomb lattice in \Cref{fig:honeycomb-zx-duality}.

It can be verified that if a qubit mapping correctly maps the plaquette colours as described above, it will also correctly map between the $\operatorname{Z}$ and $\operatorname{X}$ stabilisers on the plaquettes of the lattice between the $\color{floquet-blue}\operatorname{bZZ}$ and $\color{floquet-red}\operatorname{rXX}$ sub-rounds. To see this, note that each $\operatorname{X}$ (resp.\ $\operatorname{Z}$) stabiliser on a red (resp.\ blue) plaquette will map to a corresponding $\operatorname{Z}$ stabiliser on a blue plaquettes, and each $\operatorname{X}$ (resp.\ $\operatorname{Z}$) stabiliser on a green plaquettes will map to a corresponding $\operatorname{Z}$ (resp.\ $\operatorname{X}$) stabiliser on a green plaquette. Note that this is not true for the instantaneous stabilisers defined on the \textit{edges} of the lattice: the $\operatorname{Z}$ checks on the blue edges would have to map to $\operatorname{X}$ checks on the red edges of the lattice, but such checks are not in the instantaneous stabiliser group at this point in time. We will discuss the impact of this on the logical gates for the codes we consider in Sections \ref{ssec:logical-hadamard-type-gate} and \ref{ssec:logical-s-type-gate}.

Note that multiple $\operatorname{ZX}$-dualities can be defined depending on which sub-rounds we are between. For instance, another $\operatorname{ZX}$-duality for the honeycomb lattice can be defined between the ${\color{floquet-green}\operatorname{gXX}}$ and ${\color{floquet-blue}\operatorname{bZZ}}$ measurement sub-rounds, this time mapping blue (resp.\ green) plaquettes to green (resp.\ blue) plaquettes and red plaquettes to red plaquettes. This could potentially allow us to implement multiple logical gates in a single QEC round, thus speeding up code performance. However, as we will see later the fold-transversal operations produce longer detecting regions. It could be that an adversarily-chosen sequence of logical gates produces arbitrarily long detecting regions, which might in turn impact the performance of the code. For this work we will therefore only focus on the single $\operatorname{ZX}$-duality shown in \Cref{fig:honeycomb-zx-duality} implemented between the $\color{floquet-blue}\operatorname{bZZ}$ and $\color{floquet-red}\operatorname{rXX}$ sub-rounds.

\subsection{Logical Hadamard-type gates}
\label{ssec:logical-hadamard-type-gate}

The logical Hadamard-type gate is implemented in the same way as in \Cref{eq:hadamard-type}. Referring to \Cref{tab:hadamard-css-floquet}, we can see that after applying transversal Hadamard gates to all physical qubits, the instantaneous stabiliser group consists of the following:
\begin{itemize}
    \item $\operatorname{Z}$ checks on the red plaquettes;
    \item $\operatorname{X}$ checks on the blue plaquettes;
    \item $\operatorname{Z}$ checks on the green plaquettes;
    \item $\operatorname{X}$ checks on the green plaquettes; and
    \item $\operatorname{X}$ checks on the blue edges.
\end{itemize}

\begin{table}[]
    \centering
    \caption{Evolution of the detecting regions and observables when implementing a logical Hadamard-type gate via $\operatorname{ZX}$-duality $\tau$ on the CSS Floquet code schedule. Note that this logical operation affects the size of the detecting regions: in \Cref{tab:floquet-css-schedule} it can be seen that all detecting regions span five sub-rounds, whereas now the shortest detecting region is just two sub-rounds, and the longest is eight sub-rounds. For notational convenience we assume that $\tau = \tau^{-1}$.}
    \label{tab:hadamard-css-floquet}
    \begin{tabular}{|c|c|c|c|c|c|c|c|c|c|c|}
        \hline
        \multirow{2}{2.9em}{Pauli string} & \multicolumn{10}{|c|}{Sub-round}\\\cline{2-11}
        & $\color{floquet-blue}\operatorname{bXX}$ & $\color{floquet-red}\operatorname{rZZ}$ & $\color{floquet-green}\operatorname{gXX}$ & $\color{floquet-blue}\operatorname{bZZ}$ & $\operatorname{H}$ & $\operatorname{SWAP}$ & $\color{floquet-red}\operatorname{rXX}$ & $\color{floquet-green}\operatorname{gZZ}$ & $\color{floquet-blue}\operatorname{bXX}$ & $\color{floquet-red}\operatorname{rZZ}$ \\\hline
        \includegraphics[width=1.5em]{green-plaquette.pdf} & $\color{floquet-green}\operatorname{X}_i$ & $\color{floquet-green}\operatorname{X}_i$ & $\color{floquet-green}\operatorname{X}_i$ & $\color{floquet-green}\operatorname{X}_i$ & $\color{floquet-green}\operatorname{Z}_i$ & $\color{floquet-green}\operatorname{Z_{\tau(i)}}$ & $\color{floquet-green}\operatorname{Z_{\tau(i)}}$ & $\color{floquet-green}\operatorname{Z}_{\tau(i)}$ & $\color{floquet-green}\operatorname{Z}_{\tau(i)}$ & $\color{floquet-green}\operatorname{I}_{\tau(i)}$ \\\hline
        \includegraphics[width=1.5em]{blue-plaquette.pdf} & & $\color{floquet-blue}\operatorname{Z}_i$ & $\color{floquet-blue}\operatorname{Z}_i$ & $\color{floquet-blue}\operatorname{Z}_i$ & $\color{floquet-blue}\operatorname{X}_i$ & $\color{floquet-blue}\color{floquet-red}\operatorname{X}_{\tau(i)}$ & $\color{floquet-blue}\color{floquet-red}\operatorname{X}_{\tau(i)}$ & $\color{floquet-blue}\color{floquet-red}\operatorname{X}_{\tau(i)}$ & $\color{floquet-blue}\color{floquet-red}\operatorname{I}_{\tau(i)}$ & \\\hline
        \includegraphics[width=1.5em]{red-plaquette.pdf} & & & $\color{floquet-red}\operatorname{X}_i$ & $\color{floquet-red}\operatorname{X}_i$ & $\color{floquet-red}\operatorname{Z}_i$ & $\color{floquet-red}\color{floquet-blue}\operatorname{Z}_{\tau(i)}$ & $\color{floquet-red}\color{floquet-blue}\operatorname{Z}_{\tau(i)}$ & $\color{floquet-red}\color{floquet-blue}\operatorname{I}_{\tau(i)}$ & & \\\hline
        \includegraphics[width=1.5em]{green-plaquette.pdf} & & & & $\color{floquet-green}\operatorname{Z}_i$ & $\color{floquet-green}\operatorname{X}_i$ & $\color{floquet-green}\operatorname{X}_{\tau(i)}$ & $\color{floquet-green}\operatorname{I}_{\tau(i)}$ & & & \\\hline
        $\operatorname{X_L}$ & $\color{floquet-red}\operatorname{rXX}$ & $\color{floquet-red}\operatorname{rXX}$ & $\color{floquet-blue}\operatorname{bXX}$ & $\color{floquet-blue}\operatorname{bXX}$ & $\color{floquet-blue}\operatorname{bZZ}$ & $\color{floquet-red}\operatorname{rZZ}$ & $\color{floquet-red}\operatorname{rZZ}$ & $\color{floquet-blue}\operatorname{bZZ}$ & $\color{floquet-blue}\operatorname{bZZ}$ & $\color{floquet-green}\operatorname{gZZ}$ \\\hline
        $\operatorname{Z_L}$ & $\color{floquet-blue}\operatorname{bZZ}$ & $\color{floquet-green}\operatorname{gZZ}$ & $\color{floquet-green}\operatorname{gZZ}$ & $\color{floquet-red}\operatorname{rZZ}$ & $\color{floquet-red}\operatorname{rXX}$ & $\color{floquet-blue}\operatorname{bXX}$ & $\color{floquet-green}\operatorname{gXX}$ & $\color{floquet-green}\operatorname{gXX}$ & $\color{floquet-red}\operatorname{rXX}$ & $\color{floquet-red}\operatorname{rXX}$ \\\hline
    \end{tabular}
\end{table}

Next, we apply $\operatorname{SWAP}$ gates between pairs of qubits which map to each other via the $\operatorname{ZX}$-duality. After this step, the instantaneous stabiliser group looks similar to the one described in \Cref{ssec:floquet-zx-duality}, with the only difference being that there are now $\operatorname{X}$ checks on red edges.

Detectors can be formed by comparing plaquettes before and after the fold operation, as shown in \Cref{tab:hadamard-css-floquet}. In this case, $\operatorname{X}$ (resp.\ $\operatorname{Z}$) checks on red (resp.\ blue) plaquettes before the logical gate can be compared to $\operatorname{Z}$ (resp.\ $\operatorname{X}$) checks on blue (resp.\ red) plaquettes after the fold, and $\operatorname{Z}$ (resp.\ $\operatorname{X}$) checks on green plaquettes can be compared with $\operatorname{X}$ (resp.\ $\operatorname{Z}$) checks on green plaquettes. All of the detecting regions from the original code are preserved, therefore also preserving the fault-tolerance of the code.

The stabilisers on the edges being modified actually gives us an opportunity to take a more fine-grained approach when correcting errors. To see this, note that the next measurement sub-round performed is ${\color{floquet-red}\operatorname{rXX}}$, which is in the instantaneous stabiliser group following the Hadamard-type gate. As a result, we can form additional detectors by comparing edge measurements from the ${\color{floquet-red}\operatorname{rXX}}$ sub-round with corresponding edge measurements from the preceding ${\color{floquet-blue}\operatorname{bZZ}}$ sub-round. Note that there is some redundancy in the detector definitions here, as products of these edge detectors will also form plaquette detectors. When we implement the logical Hadamard-type gate in \Cref{ssec:hadamard-experiment}, we will omit any detectors which can be formed by these edge measurements. After a single QEC round, the instantaneous stabiliser group has returned to its original form.

Finally, we must consider how the logical operators evolve after this gate. We can see from \Cref{tab:hadamard-css-floquet} that the effect of this operation is to swap the $\operatorname{X}$ and $\operatorname{Z}$ logical operators for both logical qubits. Therefore, for this code, the resulting operation is $\operatorname{H_0}\otimes \operatorname{H_1}$.

\subsection{Logical $\operatorname{S}$-type gates}
\label{ssec:logical-s-type-gate}

The logical $\operatorname{S}$-type gate is also implemented using the same sequence as \Cref{eq:s-type}. We can see from \Cref{tab:floquet-s-gate} that the instantaneous stabiliser group after applying both the $\operatorname{S}$ gates and the $\operatorname{CZ}$ gates is formed by the following Pauli checks:
\begin{itemize}
    \item $\operatorname{X}$ checks on the red plaquettes combined with $\operatorname{Z}$ checks on the blue plaquettes;
    \item $\operatorname{Z}$ checks on the blue plaquettes;
    \item $\operatorname{X}$ checks on the green plaquettes combined with $\operatorname{Z}$ checks on the green plaquettes;
    \item $\operatorname{Z}$ checks on the green plaquettes; and
    \item $\operatorname{Z}$ checks on the blue edges.
\end{itemize}

All of the plaquette stabilisers can be measured using the same measurement schedule as before, as we show in \Cref{tab:floquet-s-gate}. The detectors for the $\operatorname{X}$ checks have been made slightly larger by the addition of $\operatorname{Z}$ checks, but only by a constant amount. As a result, this operation preserves the fault tolerance of the original code.

The evolution of the logical operators is also shown in \Cref{tab:floquet-s-gate}. The logical $\operatorname{Z}$ operators commute with the $\operatorname{S}$-type operation and are therefore unaffected. The logical $\operatorname{X}$ operators on the other hand evolve to become a product of the logical $\operatorname{X}$ and logical $\operatorname{Z}$ operators, giving an overall logical operation of $\operatorname{S_0} \otimes \operatorname{S_1}$.

\begin{table}[]
    \centering
    \caption{Evolution of the detecting regions and observables when implementing a logical $\operatorname{S}$-type gate via $\operatorname{ZX}$-duality $\tau$ on a CSS Floquet code schedule. Note that $X$ detecting regions now have a $\operatorname{Z}$ component included.}
    \label{tab:floquet-s-gate}
    \begin{tabular}{|c|c|c|c|c|c|c|c|c|c|}
        \hline
        \multirow{2}{2.9em}{Pauli string} & \multicolumn{9}{|c|}{Sub-round}\\\cline{2-10}
        & $\color{floquet-blue}\operatorname{bXX}$ & $\color{floquet-red}\operatorname{rZZ}$ & $\color{floquet-green}\operatorname{gXX}$ & $\color{floquet-blue}\operatorname{bZZ}$ & $\operatorname{S}$/$\operatorname{CZ}$ & $\color{floquet-red}\operatorname{rXX}$ & $\color{floquet-green}\operatorname{gZZ}$ & $\color{floquet-blue}\operatorname{bXX}$ & $\color{floquet-red}\operatorname{rZZ}$ \\\hline
        \includegraphics[width=1.5em]{green-plaquette.pdf} & $\color{floquet-green}\operatorname{X}_i$ & $\color{floquet-green}\operatorname{X}_i$ & $\color{floquet-green}\operatorname{X}_i$ & $\color{floquet-green}\operatorname{X}_i$ & $\color{floquet-green}\operatorname{X}_i\operatorname{Z}_{\tau(i)}$ & $\color{floquet-green}\operatorname{I}_i\operatorname{Z}_{\tau(i)}$ & $\color{floquet-green}\operatorname{Z}_{\tau(i)}$ & $\color{floquet-green}\operatorname{Z}_{\tau(i)}$ & $\color{floquet-green}\operatorname{I}_{\tau(i)}$ \\\hline
        \includegraphics[width=1.5em]{blue-plaquette.pdf} & & $\color{floquet-blue}\operatorname{Z}_i$ & $\color{floquet-blue}\operatorname{Z}_i$ & $\color{floquet-blue}\operatorname{Z}_i$ & $\color{floquet-blue}\operatorname{Z}_i$ & $\color{floquet-blue}\operatorname{Z}_i$ & $\color{floquet-blue}\operatorname{I}_i$ & & \\\hline
        \includegraphics[width=1.5em]{red-plaquette.pdf} & &  & $\color{floquet-red}\operatorname{X}_i$ & $\color{floquet-red}\operatorname{X}_i$ & $\color{floquet-red}\operatorname{X}_i\color{floquet-blue}\operatorname{Z}_{\tau(i)}$ & $\color{floquet-red}\operatorname{X}_i\color{floquet-blue}\operatorname{Z}_{\tau(i)}$ & $\color{floquet-red}\operatorname{X}_i\color{floquet-blue}\operatorname{I}_{\tau(i)}$ & $\color{floquet-red}\operatorname{I}_i$ & \\\hline
        \includegraphics[width=1.5em]{green-plaquette.pdf} & & & & $\color{floquet-green}\operatorname{Z}_i$ & $\color{floquet-green}\operatorname{Z}_i$ & $\color{floquet-green}\operatorname{Z}_i$ & $\color{floquet-green}\operatorname{Z}_i$ & $\color{floquet-green}\operatorname{Z}_i$ & $\color{floquet-green}\operatorname{I}_i$ \\\hline
        \multirow{2}{1.5em}{$\operatorname{X_L}$} & \multirow{2}{2.5em}{$\color{floquet-red}\operatorname{rXX}$} & \multirow{2}{2.5em}{$\color{floquet-red}\operatorname{rXX}$} & \multirow{2}{2.5em}{$\color{floquet-blue}\operatorname{bXX}$} & \multirow{2}{2.5em}{$\color{floquet-blue}\operatorname{bXX}$} & $\color{floquet-blue}\operatorname{bXX}$ & $\color{floquet-green}\operatorname{gXX}$ & $\color{floquet-green}\operatorname{gXX}$ & $\color{floquet-red}\operatorname{rXX}$ & $\color{floquet-red}\operatorname{rXX}$ \\
        & & & & & $\color{floquet-red}\operatorname{rZZ}$ & $\color{floquet-red}\operatorname{rZZ}$ & $\color{floquet-blue}\operatorname{bZZ}$ & $\color{floquet-blue}\operatorname{bZZ}$ & $\color{floquet-green}\operatorname{gZZ}$ \\\hline
        $\operatorname{Z_L}$ & $\color{floquet-blue}\operatorname{bZZ}$ & $\color{floquet-green}\operatorname{gZZ}$ & $\color{floquet-green}\operatorname{gZZ}$ & $\color{floquet-red}\operatorname{rZZ}$ & $\color{floquet-red}\operatorname{rZZ}$ & $\color{floquet-red}\operatorname{rZZ}$ & $\color{floquet-blue}\operatorname{bZZ}$ & $\color{floquet-blue}\operatorname{bZZ}$ & $\color{floquet-green}\operatorname{gZZ}$ \\\hline
    \end{tabular}
\end{table}

Under a depolarising noise model, errors in a memory experiment on the CSS Floquet honeycomb code can be decomposed into errors which flip at most two detectors (``graph-like'' errors) \cite{Gidney2021FloquetBenchmark, Kesselring2024AnyonFloquet}. When only graph-like errors exist, high-performance graph-based decoders such as PyMatching \cite{Higgott2025SparseBlossom} can be used to quickly correct errors. As we will show in \Cref{sec:benchmarking}, all other logical operations discussed in this paper also produce graph-like errors. However, the fact that $\operatorname{X}$ checks spread across to $\operatorname{Z}$ checks after the $\operatorname{S}$-type logical gate means that some errors are now weight-three. For example, a measurement error when measuring a $\color{floquet-green}\operatorname{gZZ}$ check can flip $\operatorname{Z}$ detectors on two blue plaquettes and an $\operatorname{X}$ detector on one red plaquette. This error is no longer graph-like, similarly to the errors we saw when initially talking about fold-transversal gates on static codes in \Cref{ssec:fold-transversal-static}. This will impact the decoders we are available to use when benchmarking this protocol in \Cref{ssec:s-gate-experiment}. We close this section noting that there has been a recent interest in developing high-performance decoders for transversal logical gates on the rotated surface code, including decoders for fold-transversal $S$ gates \cite{Wan2025CNOTDecoder, Cain2025FastCNOTDecoder, Serraperalta2025FastCNOTDecoder, Turner2025FastCNOTDecoder}. We believe it is possible to extend these decoders so that they are also able to correct errors on Floquet codes, but we leave the details of this for future work.

\subsection{Extensions to other measurement schedules}
\label{ssec:folded-gidney-schedule}

We will now discuss how we can extend this work to the period-three measurement schedule considered in Gidney \etal \cite{Gidney2021FloquetBenchmark}. Being able to perform fold-transversal logical gates across this schedule would also allow for logical gates across the original Hastings and Haah schedule \cite{Hastings2021FloquetCodes}, which is equivalent up to single-qubit Clifford gates.

As before, we will consider the instantaneous stabiliser group between the $\color{floquet-blue}{\operatorname{bZZ}}$ and $\color{floquet-red}\operatorname{rXX}$ sub-rounds. We will use the same $\operatorname{ZX}$-duality and the same operations as in Equations (\ref{eq:hadamard-type}) and (\ref{eq:s-type}). Referring back to \Cref{tab:floquet-gidney-schedule}, we can see that between the $\color{floquet-blue}{\operatorname{bZZ}}$ and $\color{floquet-red}\operatorname{rXX}$ sub-rounds, the instantaneous stabiliser group consists of the following Pauli operations:
\begin{itemize}
    \item $\operatorname{X}$ checks on all red plaquettes;
    \item $\operatorname{Y}$ checks on all green plaquettes;
    \item $\operatorname{Z}$ checks on all blue plaquettes; and
    \item $\operatorname{Z}$ checks on all blue edges.
\end{itemize}

Note that the stabilisers on the red and blue plaquettes are the same as in the period-six schedule, so our operations will perform the same way as before for those stabilisers. Similarly, the logical operators at this point are also the same as for the period-six schedule, and therefore the action on the logical operators will be the same. Therefore, our main focus will be on the green plaquettes.

Unlike the period-six schedule used by CSS Floquet codes which ensures that all stabilisers are either in the $\operatorname{X}$ basis or $\operatorname{Z}$ basis \cite{Davydova2023CSSFloquet, Kesselring2024AnyonFloquet, Bombin2024Unifying}, the period-three schedule introduces $\operatorname{Y}$ basis stabilisers. These stabilisers do also exist in the instantaneous stabiliser group for the CSS Floquet code, but in that instance they are decomposed into $\operatorname{X}$ and $\operatorname{Z}$ stabilisers which are preserved separately. Here we need to preserve the $\operatorname{Y}$ basis stabiliser itself.

\begin{table}[]
    \centering
    \caption{Evolution of the detecting regions and observables when implementing a fold-transversal Hadamard-type gate via $\operatorname{ZX}$-duality $\tau$ using the period-three measurement schedule. $X$ stabilisers on the red plaquettes are swapped with $Z$ stabilisers on the blue plaquettes, similarly to the swapping of $X$ and $Z$ stabilisers on static codes \cite{Moussa2016FoldTransversal, Breuckmann2024FoldTransversal}. The $Y$ stabilisers on the green plaquettes are mapped to $Y$ stabilisers on other green plaquettes.}
    \label{tab:floquet-hadamard-gate-gidney-schedule}
    \begin{tabular}{|c|c|c|c|c|c|c|c|c|c|c|}
        \hline
        \multirow{2}{2.9em}{Pauli string} & \multicolumn{10}{|c|}{Sub-round}\\\cline{2-11}
        & $\color{floquet-blue}\operatorname{bZZ}$ & $\color{floquet-red}\operatorname{rXX}$ & $\color{floquet-green}\operatorname{gYY}$ & $\color{floquet-blue}\operatorname{bZZ}$ & $\operatorname{H}$ & $\operatorname{SWAP}$ & $\color{floquet-red}\operatorname{rXX}$ & $\color{floquet-green}\operatorname{gYY}$ & $\color{floquet-blue}\operatorname{bZZ}$ & $\color{floquet-red}\operatorname{rXX}$ \\\hline
        \includegraphics[width=1.5em]{green-plaquette.pdf} & $\color{floquet-green}\operatorname{Y}_i$ & $\color{floquet-green}\operatorname{Y}_i$ & $\color{floquet-green}\operatorname{Y}_i$ & $\color{floquet-green}\operatorname{Y}_i$ & $\color{floquet-green}\operatorname{Y}_i$ & $\color{floquet-green}\operatorname{Y}_{\tau(i)}$ & $\color{floquet-green}\operatorname{Y}_{\tau(i)}$ & $\color{floquet-green}\operatorname{Y}_{\tau(i)}$ & $\color{floquet-green}\operatorname{Y}_{\tau(i)}$ & $\color{floquet-green}\operatorname{Y}_{\tau(i)}$ \\\hline
        \includegraphics[width=1.5em]{blue-plaquette.pdf} & $\color{floquet-blue}\operatorname{Z}_i$ & $\color{floquet-blue}\operatorname{Z}_i$ & $\color{floquet-blue}\operatorname{Z}_i$ & $\color{floquet-blue}\operatorname{Z}_i$ & $\color{floquet-blue}\operatorname{X}_i$ & $\color{floquet-red}\operatorname{X}_{\tau(i)}$ & $\color{floquet-red}\operatorname{X}_{\tau(i)}$ & $\color{floquet-red}\operatorname{X}_{\tau(i)}$ & $\color{floquet-red}\operatorname{X}_{\tau(i)}$ & $\color{floquet-red}\operatorname{X}_{\tau(i)}$ \\\hline
        \includegraphics[width=1.5em]{red-plaquette.pdf} & $\color{floquet-red}\operatorname{X}_i$ & $\color{floquet-red}\operatorname{X}_i$ & $\color{floquet-red}\operatorname{X}_i$ & $\color{floquet-red}\operatorname{X}_i$ & $\color{floquet-red}\operatorname{Z}_i$ & $\color{floquet-blue}\operatorname{Z}_{\tau(i)}$ & $\color{floquet-blue}\operatorname{Z}_{\tau(i)}$ & $\color{floquet-blue}\operatorname{Z}_{\tau(i)}$ & $\color{floquet-blue}\operatorname{Z}_{\tau(i)}$ & $\color{floquet-blue}\operatorname{Z}_{\tau(i)}$ \\\hline
        \includegraphics[width=1.5em]{green-plaquette.pdf} & $\color{floquet-green}\operatorname{Z}_i$ & $\color{floquet-green}\operatorname{Y}_i$ & $\color{floquet-green}\operatorname{Y}_i$ & $\color{floquet-green}\operatorname{X}_i$ & $\color{floquet-green}\operatorname{Z}_i$ & $\color{floquet-green}\operatorname{Z}_{\tau(i)}$ & $\color{floquet-green}\operatorname{Y}_{\tau(i)}$ & $\color{floquet-green}\operatorname{Y}_{\tau(i)}$ & $\color{floquet-green}\operatorname{X}_{\tau(i)}$ & $\color{floquet-green}\operatorname{I}_{\tau(i)}$ \\\hline
        \includegraphics[width=1.5em]{blue-plaquette.pdf} & & $\color{floquet-blue}\operatorname{X}_i$ & $\color{floquet-blue}\operatorname{Z}_i$ & $\color{floquet-blue}\operatorname{Z}_i$ & $\color{floquet-blue}\operatorname{X}_i$ & $\color{floquet-red}\operatorname{X}_{\tau(i)}$ & $\color{floquet-red}\operatorname{X}_{\tau(i)}$ & $\color{floquet-red}\operatorname{Z}_{\tau(i)}$ & $\color{floquet-red}\operatorname{I}_{\tau(i)}$ & \\\hline
        \includegraphics[width=1.5em]{red-plaquette.pdf} & & & $\color{floquet-red}\operatorname{Y}_i$ & $\color{floquet-red}\operatorname{X}_i$ & $\color{floquet-red}\operatorname{Z}_i$ & $\color{floquet-blue}\operatorname{Z}_{\tau(i)}$ & $\color{floquet-blue}\operatorname{Y}_{\tau(i)}$ & $\color{floquet-blue}\operatorname{I}_{\tau(i)}$ & & \\\hline
        \includegraphics[width=1.5em]{green-plaquette.pdf} & & & & $\color{floquet-green}\operatorname{Z}_i$ & $\color{floquet-green}\operatorname{X}_i$ & $\color{floquet-green}\operatorname{X}_{\tau(i)}$ & $\color{floquet-green}\operatorname{I}_{\tau(i)}$ & & & \\\hline
        $\operatorname{X_L}$ & $\color{floquet-red}\operatorname{rXX}$ & $\color{floquet-red}\operatorname{rXX}$ & $\color{floquet-blue}\operatorname{bXX}$ & $\color{floquet-blue}\operatorname{bXX}$ & $\color{floquet-blue}\operatorname{bZZ}$ & $\color{floquet-red}\operatorname{rZZ}$ & $\color{floquet-red}\operatorname{rZZ}$ & $\color{floquet-blue}\operatorname{bZZ}$ & $\color{floquet-blue}\operatorname{bZZ}$ & $\color{floquet-green}\operatorname{gZZ}$ \\\hline
        $\operatorname{Z_L}$ & $\color{floquet-blue}\operatorname{bZZ}$ & $\color{floquet-green}\operatorname{gZZ}$ & $\color{floquet-green}\operatorname{gZZ}$ & $\color{floquet-red}\operatorname{rZZ}$ & $\color{floquet-red}\operatorname{rXX}$ & $\color{floquet-blue}\operatorname{bXX}$ & $\color{floquet-green}\operatorname{gXX}$ & $\color{floquet-green}\operatorname{gXX}$ & $\color{floquet-red}\operatorname{rXX}$ & $\color{floquet-red}\operatorname{rXX}$ \\\hline
    \end{tabular}
\end{table}

The simpler logical operation to consider in this case is the Hadamard-type gate. Using the same sequence of Hadamard and $\operatorname{SWAP}$ gates as before, we note that $\operatorname{Y}$ stabilisers are invariant under Hadamard gates, and therefore after the transversal Hadamard gates the stabilisers on the green plaquettes will be left unchanged. The transversal $\operatorname{SWAP}$ gates will map green plaquettes to green plaquettes, allowing for detectors to be formed as before. The action on the logical operators is the same as before, giving a logical $\operatorname{H_0}\otimes \operatorname{H_1}$ gate. The action of this operation on stabilisers and observables can be seen in \Cref{tab:floquet-hadamard-gate-gidney-schedule}.

We emphasise that this operation is different from the period-three schedule's automorphism, which exchanges logical $\operatorname{X}$ and $\operatorname{Z}$ information on the same homology. This is a similar operation to the Hadamard gate, but because the homologies are not exchanged the logical operation on the honeycomb Floquet code is instead $\operatorname{SWAP}_{0, 1}(\operatorname{H}_0\otimes\operatorname{H}_1)$. Applying a fold-transversal layer of $\operatorname{SWAP}$ gates would produce the homology exchange but change the stabilisers: $\operatorname{X}$ stabilisers would now be on blue plaquettes and $\operatorname{Z}$ stabilisers would be on red plaquettes.

The more interesting example to consider is the logical $\operatorname{S}$-type gate. Note that if we apply the same approach as before, the $\operatorname{Y}$ stabilisers on the green plaquettes will become a product of $\operatorname{Y}$ checks on the green plaquettes \textit{and} $\operatorname{Z}$ checks on the green plaquettes. In order to preserve these stabilisers, we multiply in edge checks from the $\color{floquet-red}\operatorname{rXX}$ sub-round. The result will be a $\operatorname{Y}$ basis stabiliser across \textit{two} green plaquettes. Following a full QEC round, this stabiliser can be reduced to just a $\operatorname{Y}$ basis stabiliser across a single green plaquette as before. We demonstrate how this stabiliser evolves and how detecting regions for it can be formed before and after the fold-transversal operation in \Cref{tab:floquet-s-gate-gidney-schedule}.

\begin{table}[]
    \centering
    \caption{Evolution of the stabilisers, detectors, and observables when implementing a fold-transversal $\operatorname{S}$-type gate via $\operatorname{ZX}$-duality $\tau$ using the period-three measurement schedule. The $\operatorname{X}$ stabilisers on the red plaquettes become a product with the $\operatorname{Z}$ stabilisers on the blue plaquettes. The $\operatorname{Y}$ stabilisers on the green plaquettes become a product with $\operatorname{Z}$ stabilisers on other green plaquettes. By multiplying in edge measurements from the $\color{floquet-red} \operatorname{rXX}$ sub-round, these also become $\operatorname{Y}$ stabilisers on the green plaquettes. After four sub-rounds, these components can be removed from the stabilisers and the stabilisers return to their original form.}
    \label{tab:floquet-s-gate-gidney-schedule}
    \begin{tabular}{|c|c|c|c|c|c|c|c|c|c|}
        \hline
        \multirow{2}{2.9em}{Pauli string} & \multicolumn{9}{|c|}{Sub-round}\\\cline{2-10}
        & $\color{floquet-blue}\operatorname{bZZ}$ & $\color{floquet-red}\operatorname{rXX}$ & $\color{floquet-green}\operatorname{gYY}$ & $\color{floquet-blue}\operatorname{bZZ}$ & $\operatorname{S}$/$\operatorname{CZ}$ & $\color{floquet-red}\operatorname{rXX}$ & $\color{floquet-green}\operatorname{gYY}$ & $\color{floquet-blue}\operatorname{bZZ}$ & $\color{floquet-red}\operatorname{rXX}$ \\\hline
        \includegraphics[width=1.5em]{green-plaquette.pdf} & $\color{floquet-green}\operatorname{Y}_i$ & $\color{floquet-green}\operatorname{Y}_i$ & $\color{floquet-green}\operatorname{Y}_i$ & $\color{floquet-green}\operatorname{Y}_i$ & $\color{floquet-green}\operatorname{Y}_i\operatorname{Z}_{\tau(i)}$ & $\color{floquet-green}\operatorname{Y}_i\operatorname{Y}_{\tau(i)}$ & $\color{floquet-green}\operatorname{Y}_i\operatorname{Y}_{\tau(i)}$ & $\color{floquet-green}\operatorname{Y}_i\operatorname{X}_{\tau(i)}$ & $\color{floquet-green}\operatorname{Y}_i\operatorname{I}_{\tau(i)}$ \\\hline
        \includegraphics[width=1.5em]{blue-plaquette.pdf} & $\color{floquet-blue}\operatorname{Z}_i$ & $\color{floquet-blue}\operatorname{Z}_i$ & $\color{floquet-blue}\operatorname{Z}_i$ & $\color{floquet-blue}\operatorname{Z}_i$ & $\color{floquet-blue}\operatorname{Z}_i$ & $\color{floquet-blue}\operatorname{Z}_i$ & $\color{floquet-blue}\operatorname{Z}_i$ & $\color{floquet-blue}\operatorname{Z}_i$ & $\color{floquet-blue}\operatorname{Z}_i$ \\\hline
        \includegraphics[width=1.5em]{red-plaquette.pdf} & $\color{floquet-red}\operatorname{X}_i$ &  $\color{floquet-red}\operatorname{X}_i$ & $\color{floquet-red}\operatorname{X}_i$ & $\color{floquet-red}\operatorname{X}_i$ & $\color{floquet-red}\operatorname{X}_i\color{floquet-blue}\operatorname{Z}_{\tau(i)}$ & $\color{floquet-red}\operatorname{X}_i\color{floquet-blue}\operatorname{Y}_{\tau(i)}$ & $\color{floquet-red}\operatorname{X}_i\color{floquet-blue}\operatorname{I}_{\tau(i)}$ & $\color{floquet-red}\operatorname{X}_i$ & $\color{floquet-red}\operatorname{X}_i$ \\\hline
        \includegraphics[width=1.5em]{green-plaquette.pdf} & $\color{floquet-green}\operatorname{Z}_i$ & $\color{floquet-green}\operatorname{Y}_i$ & $\color{floquet-green}\operatorname{Y}_i$ & $\color{floquet-green}\operatorname{X}_i$ & $\color{floquet-green}\operatorname{X}_i\operatorname{Z}_{\tau(i)}$ & $\color{floquet-green}\operatorname{I}_i\operatorname{Y}_{\tau(i)}$ & $\color{floquet-green}\operatorname{Y}_{\tau(i)}$ & $\color{floquet-green}\operatorname{X}_{\tau(i)}$ & $\color{floquet-green}\operatorname{I}_{\tau(i)}$ \\\hline
        \includegraphics[width=1.5em]{blue-plaquette.pdf} & & $\color{floquet-blue}\operatorname{X}_i$ & $\color{floquet-blue}\operatorname{Z}_i$ & $\color{floquet-blue}\operatorname{Z}_i$ & $\color{floquet-blue}\operatorname{Z}_i$ & $\color{floquet-blue}\operatorname{Y}_i$ & $\color{floquet-blue}\operatorname{I}_i$ & & \\\hline
        \includegraphics[width=1.5em]{red-plaquette.pdf} & &  & $\color{floquet-red}\operatorname{Y}_i$ & $\color{floquet-red}\operatorname{X}_i$ & $\color{floquet-red}\operatorname{X}_i\color{floquet-blue}\operatorname{Z}_{\tau(i)}$ & $\color{floquet-red}\operatorname{X}_i\color{floquet-blue}\operatorname{Y}_{\tau(i)}$ & $\color{floquet-red}\operatorname{Z}_i\color{floquet-blue}\operatorname{I}_{\tau(i)}$ & $\color{floquet-red}\operatorname{I}_i$ & \\\hline
        \includegraphics[width=1.5em]{green-plaquette.pdf} & & & & $\color{floquet-green}\operatorname{Z}_i$ & $\color{floquet-green}\operatorname{Z}_i$ & $\color{floquet-green}\operatorname{Y}_i$ & $\color{floquet-green}\operatorname{Y}_i$ & $\color{floquet-green}\operatorname{X}_i$ & $\color{floquet-green}\operatorname{I}_i$ \\\hline
        \multirow{2}{1.5em}{$\operatorname{X_L}$} & \multirow{2}{2.5em}{$\color{floquet-red}\operatorname{rZZ}$} & \multirow{2}{2.5em}{$\color{floquet-red}\operatorname{rYY}$} & \multirow{2}{2.5em}{$\color{floquet-blue}\operatorname{bYY}$} & \multirow{2}{2.5em}{$\color{floquet-blue}\operatorname{bXX}$} & $\color{floquet-blue}\operatorname{bXX}$ & $\color{floquet-green}\operatorname{gXX}$ & $\color{floquet-green}\operatorname{gZZ}$ & $\color{floquet-red}\operatorname{rZZ}$ & $\color{floquet-red}\operatorname{rYY}$ \\
        & & & & & $\color{floquet-red}\operatorname{rZZ}$ & $\color{floquet-red}\operatorname{rYY}$ & $\color{floquet-blue}\operatorname{bYY}$ & $\color{floquet-blue}\operatorname{bXX}$ & $\color{floquet-green}\operatorname{gXX}$ \\\hline
        $\operatorname{Z_L}$ & $\color{floquet-blue}\operatorname{bXX}$ & $\color{floquet-green}\operatorname{gXX}$ & $\color{floquet-green}\operatorname{gZZ}$ & $\color{floquet-red}\operatorname{rZZ}$ & $\color{floquet-red}\operatorname{rZZ}$ & $\color{floquet-red}\operatorname{rYY}$ & $\color{floquet-blue}\operatorname{bYY}$ & $\color{floquet-blue}\operatorname{bXX}$ & $\color{floquet-green}\operatorname{gXX}$ \\\hline
    \end{tabular}
\end{table}

There are also other fold-transversal gates which can be implemented beyond the two shown above. For example, in \Cref{app:sqrt-x-type-gate} we show how it is possible to implement a fold-transversal $\sqrt{\operatorname{X}}$-type gate on a Floquet code, and discuss how this might lead to better performance via smaller detecting regions.

\subsection{Folding the $4.8.8$ Floquet code}
\label{ssec:folded-4-8-8}

We conclude this section by discussing how these fold-transversal operations can be applied to Floquet codes defined on other lattices, in particular the $4.8.8$ lattice shown in \Cref{fig:4-8-8}. For the $4.8.8$ lattice with closed boundaries, we can see that a $\operatorname{ZX}$-duality runs horizontally across the code. This duality will transform detecting regions in the same way as was previously discussed for the honeycomb code in \Cref{ssec:logical-hadamard-type-gate} and \Cref{ssec:logical-s-type-gate}. However, while this will change the Pauli bases of the logical observables, it will not change the \textit{homology} of the logical observables: a logical observable running across the horizontal (resp.\ vertical) boundary would still run across the horizontal (resp.\ vertical) boundary, thereby implementing different logical gates to what was seen for the honeycomb code. More precisely, the action of a fold-transversal Hadamard-type gate is a logical $\operatorname{SWAP}_{0, 1}(\operatorname{H}_0\otimes\operatorname{H}_1)$ operation\footnote{Note that this is the same operation that is produced by automorphism of the period-three schedule.}, and the action of a fold-transversal $\operatorname{S}$-type gate is $\operatorname{CZ}_{0,1}$.

\begin{figure}
    \centering
    \begin{subfigure}{0.3\linewidth}
        \includegraphics[width=\linewidth]{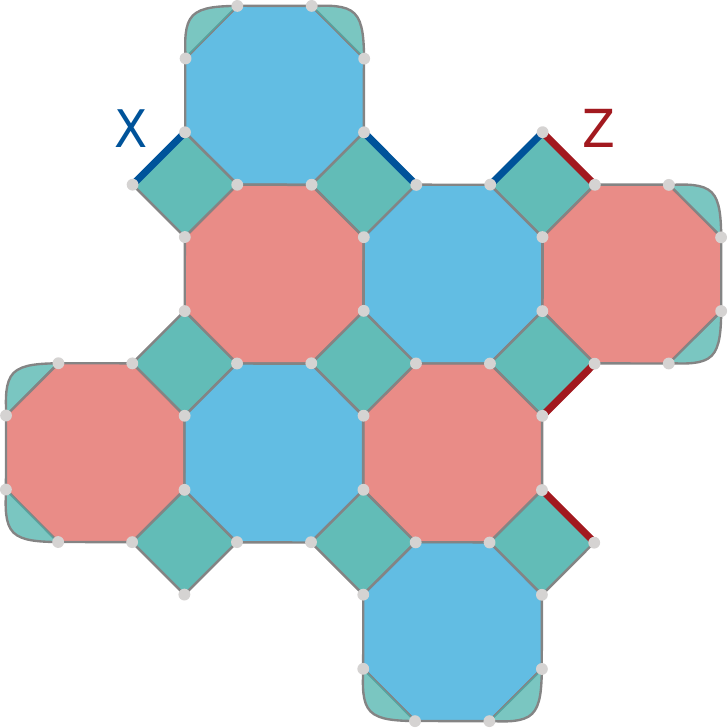}
        \caption{}
        \label{fig:4-8-8-planar}
    \end{subfigure}
    \hfill
    \begin{subfigure}{0.6\linewidth}
        \includegraphics[width=\linewidth]{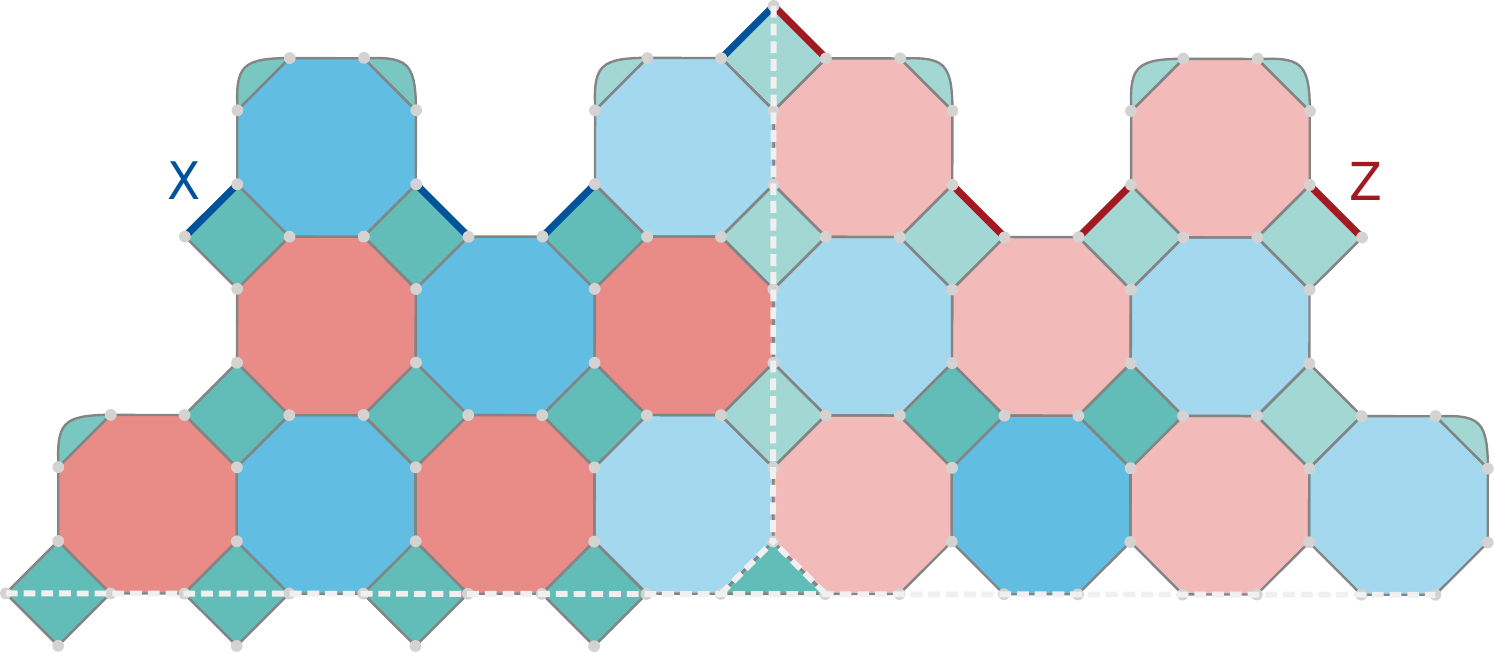}
        \caption{}
        \label{fig:4-8-8-cone}
    \end{subfigure}
    \caption{Implementing a fold-transversal gate on the planar $4.8.8$ code. (a) The planar $4.8.8$ code does not support a $\operatorname{ZX}$-duality, as the $\operatorname{Z}$ (resp.\ $\operatorname{X}$) stabilisers on the boundaries do not have a corresponding $\operatorname{X}$ (resp.\ $\operatorname{Z}$) stabiliser to map to. (b) A $\operatorname{ZX}$-duality can be formed by embedding the planar $4.8.8$ lattice in a larger $4.8.8$ lattice on the surface of a cone. We have followed the same convention as Moussa \cite{Moussa2016FoldTransversal} of presenting the cone with a cut so that it can be shown in 2-D, the cut runs along the bottom of the figure, with the green triangle being the tip of the cone. Plaquettes in the original code follow the same colour scheme as (a), new plaquettes are depicted in lighter shades of red, blue, and green. Logical operators are depicted in red and blue, running along the top boundary. The fold line for the $\operatorname{ZX}$-duality is highlighted in white and dashed.}
    \label{fig:4-8-8-zx-duality}
\end{figure}

We can also consider the planar $4.8.8$ lattice from Paetznick \etal, shown in \Cref{fig:4-8-8-planar} \cite{Paetznick2023PlanarFloquetPerformance}. Note that while this code uses a rewinding measurement schedule in order to preserve the logical qubit on a planar lattice, the instantaneous stabiliser group between $\color{floquet-blue}\operatorname{bZZ}$ and $\color{floquet-red}\operatorname{rXX}$ sub-rounds is the same as the examples we discuss earlier in this work. Therefore our main focus in this section is identifying a suitable mapping between plaquettes.

Note that the structure of the boundaries in the $4.8.8$ lattice are such that there is not a $\operatorname{ZX}$-duality like we have used for the honeycomb lattice. A similar challenge was identified by Moussa for implementing fold-transversal gates on the rotated surface code, which was resolved by using lattice surgery to embed the lattice into a larger lattice on the surface of a cone \cite{Moussa2016FoldTransversal}. The cone surface creates a symmetry in the boundaries which can then be used to perform the $\operatorname{ZX}$-duality. Here we use the same solution, with the cone presented in \Cref{fig:4-8-8-cone}. A disadvantage to this approach is that the process of using lattice surgery to embed this code on the surface of a cone requires $O(d)$ QEC rounds to be fault-tolerant, meaning that implementing a fold-transversal logical gate this way will take longer. Another challenge with this code is the higher number of physical qubits required for implementing the fold-transversal logical gates.

In \Cref{app:rotated-4-8-8} we present an alternative $4.8.8$ lattice, which we call the ``rotated'' $4.8.8$ code. This code has a direct $\operatorname{ZX}$-duality, allowing for faster implementation of fold-transversal logical gates. However, this benefit comes at the cost of requiring more physical qubits in order to preserve the code distance.

Chen \etal showed that a logical Hadamard gate on the rotated surface code can be implemented through a layer of transversal Hadamard gates and two layers of $\operatorname{SWAP}$ gates. It is also known that a logical Hadamard gate on the rotated surface code can be realised by applying a layer of transversal Hadamard gates and rotating the code by $90^\circ$ \cite{Blunt2024Compilation, Geher2024LogicalHadamard}. Both of these techniques can also be used to implement a logical Hadamard gate on the planar $4.8.8$ code.

Chen \etal also showed that while a rotated surface code does not support a $\operatorname{ZX}$-duality as a code, a logical $S$ gate can still be realised by implementing a fold-transversal operation on the mid-cycle state of the code \cite{Chen2024TransversalLogic}. This utilises the fact that the mid-cycle state of the rotated surface code is an unrotated surface code \cite{McEwen2023RelaxingHardware}. A similar technique could be applied here, by noting that after a $\color{floquet-green}\operatorname{gXX}$ measurement, $4.8.8$ code embeds an unrotated surface code \cite{Davydova2023CSSFloquet}. We will discuss how we can implement fold-transversal gates on Floquet codes via the embedded codes in \Cref{sec:embedded-codes}. We caution however that there are technical details regarding decoding at the boundaries of the rotated surface code when using the technique of Chen \etal. The solutions they propose might be extendable to the planar $4.8.8$ Floquet code, but we leave confirming this to future work.

Finally, we note that the fold-transversal gates on a planar Floquet code produce logical $\operatorname{H}$ and $\operatorname{S}$ gates, thus giving us the complete set of single-qubit Clifford gates. When combined with a logical $\operatorname{CNOT}$ operation across two planar Floquet codes through e.g.\ transversal operations \cite{Chen2024TransversalLogic} or lattice surgery \cite{Haah2022BoundaryFloquet}, we get the full Clifford gate set. The inclusion of noisy magic state preparation would then provide a universal logical gate set \cite{Bravyi2005UniversalGateSet}.

\section{Dehn twists on Floquet codes}
\label{sec:dehn-twist-floquet}

We will now discuss how fault-tolerant $\operatorname{CNOT}$ gates can be implemented on 2-D topological Floquet codes by distorting the lattice along a non-trivial loop. Our technique relies on the non-trivial loop having a pattern where alternating edges have the same colour, for instance $\color{floquet-red}\operatorname{r}\shortrightarrow\color{floquet-green}\operatorname{g}\shortrightarrow\color{floquet-blue}\operatorname{b}\shortrightarrow\color{floquet-green}\operatorname{g}$ is a valid pattern, as is  $\color{floquet-red}\operatorname{r}\shortrightarrow\color{floquet-green}\operatorname{g}\shortrightarrow\color{floquet-red}\operatorname{r}\shortrightarrow\color{floquet-green}\operatorname{g}\shortrightarrow\color{floquet-blue}\operatorname{b}\shortrightarrow\color{floquet-green}\operatorname{g}\shortrightarrow\color{floquet-blue}\operatorname{b}\shortrightarrow\color{floquet-green}\operatorname{g}$, but $\color{floquet-red}\operatorname{r}\shortrightarrow\color{floquet-green}\operatorname{g}\shortrightarrow\color{floquet-blue}\operatorname{b}\shortrightarrow\color{floquet-red}\operatorname{r}$ is not a valid pattern. In \Cref{ssec:edge-swapping} we describe a gadget which exchanges the non-green edges in this loop, thus creating a distortion in the lattice. When this distortion is repeatedly applied, a Dehn twist of the lattice is completed, thus producing a logical $\operatorname{CNOT}$ gate.

As with in \Cref{sec:fold-transversal-floquet}, our primary example will be the CSS Floquet code defined on a honeycomb lattice with periodic boundaries. On this code we are able to identify logical gates corresponding to $\operatorname{CNOT}_{0,1}$ and $\operatorname{CNOT}_{1,0}$. We will also show how this work can be extended to other Floquet codes: in \Cref{ssec:dehn-twist-gidney-schedule}, we show how Dehn twists can be implemented on the period-3 measurement schedule; and in \Cref{ssec:dehn-twist-4-8-8}, we use the edge exchange gadget on a more complicated example: the $4.8.8$ code with periodic boundaries.

\subsection{Edge-swapping gadget}
\label{ssec:edge-swapping}

The central component in our Dehn twist is what we call the edge-swapping gadget. This gadget works by swapping edges of different colours in a non-trivial loop.

As in \Cref{sec:fold-transversal-floquet}, let us assume we are between the $\color{floquet-blue}\operatorname{bZZ}$ and $\color{floquet-red}\operatorname{rXX}$ sub-rounds in the CSS Floquet code schedule. For convenience, we will restate the instantaneous stabiliser group at this point which consists of the following (see \Cref{tab:floquet-css-schedule}):
\begin{itemize}
    \item $\operatorname{X}$ checks on the red plaquettes;
    \item $\operatorname{Z}$ checks on the blue plaquettes;
    \item $\operatorname{X}$ checks on the green plaquettes;
    \item $\operatorname{Z}$ checks on the green plaquettes; and
    \item $\operatorname{Z}$ checks on the blue edges.
\end{itemize}

\begin{figure}
    \centering
    \begin{subfigure}{0.11\linewidth}
        \centering
        \includegraphics[width=\linewidth]{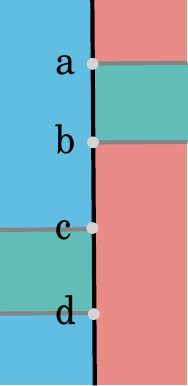}
        \caption{}
        \label{fig:edge-original}
    \end{subfigure}
    \begin{subfigure}{0.275\linewidth}
        \begin{adjustbox}{width=\linewidth}
            \begin{quantikz}
            \\
                \lstick{a} & \ctrl{2} & \qw & \ctrl{3} & \qw & \qw \\
                \lstick{b} & \qw & \ctrl{2} & \qw & \ctrl{1} & \qw \\
                \lstick{c} & \targ{} & \qw & \qw & \targ{} & \qw \\
                \lstick{d} & \qw & \targ{} & \targ{} & \qw & \qw \\
            \end{quantikz}
        \end{adjustbox}
        \caption{}
        \label{fig:edge-swapping-gadget}
    \end{subfigure}
    \begin{subfigure}{0.11\linewidth}
        \centering
        \includegraphics[width=\linewidth]{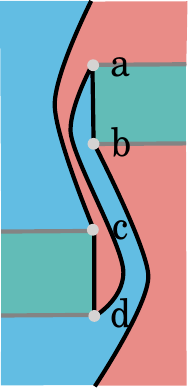}
        \caption{}
        \label{fig:edge-swapped}
    \end{subfigure}
    \caption{(a) A sequence of four qubits connected in a chain of edges coloured $\color{floquet-green}\operatorname{g}\shortrightarrow\color{floquet-red}\operatorname{r}\shortrightarrow\color{floquet-green}\operatorname{g}\shortrightarrow\color{floquet-blue}\operatorname{b}\shortrightarrow\color{floquet-green}\operatorname{g}$ from top to bottom. Parts of the plaquettes these qubits contribute to have not been shown. (b) The edge exchange gadget consists of applying four $\operatorname{CNOT}$ gates across pairs of qubits in this chain, with the two control qubits (a and b) connected by a red edge, and the two target qubits (c and d) connected by a blue edge. Note that the $\operatorname{CNOT}$ gates can be applied in two layers in parallel. (c) The transformation of the stabilisers due to this gadget. The sequence of edge colours is now $\color{floquet-green}\operatorname{g}\shortrightarrow\color{floquet-blue}\operatorname{b}\shortrightarrow\color{floquet-green}\operatorname{g}\shortrightarrow\color{floquet-red}\operatorname{r}\shortrightarrow\color{floquet-green}\operatorname{g}$. The resulting lattice is still trivalent and three-face-colourable.}
    \label{fig:edge-swapping}
\end{figure}

Let us pick a path of edges coloured $\color{floquet-green}\operatorname{g}\shortrightarrow\color{floquet-red}\operatorname{r}\shortrightarrow\color{floquet-green}\operatorname{g}\shortrightarrow\color{floquet-blue}\operatorname{b}\shortrightarrow\color{floquet-green}\operatorname{g}$. These edges are shown in \Cref{fig:edge-original}. Edges along this path contribute to six plaquettes, two of each colour.

The edge-swapping gadget works by performing four $\operatorname{CNOT}$ gates as shown in \Cref{fig:edge-swapping-gadget}. Based on the action of the $\operatorname{CNOT}$ gate as shown in \Cref{eq:cnot_heisenberg}, we can see that the action of this gadget is to move edges between plaquettes: the first and fourth $\operatorname{CNOT}$ gates remove the two qubits connected by a red edge from a $\operatorname{Z}$ check on a blue, and the second and third $\operatorname{CNOT}$ gates will add those two qubits to a $\operatorname{Z}$ check on a different blue plaquette. The same happens in reverse for the $\operatorname{X}$ checks on red plaquettes: the two qubits connected by a blue edge will be removed from one red plaquette and added to the other red plaquette. Note that checks on green plaquettes are unaffected by these $\operatorname{CNOT}$ gates, as are the $\operatorname{Z}$ checks on blue edges.

If we consider how these stabilisers have evolved on the lattice, it corresponds to the edges on the path being re-ordered to $\color{floquet-green}\operatorname{g}\shortrightarrow\color{floquet-blue}\operatorname{b}\shortrightarrow
\color{floquet-green}\operatorname{g}\shortrightarrow\color{floquet-red}\operatorname{r}\shortrightarrow\color{floquet-green}\operatorname{g}$, as shown in \Cref{fig:edge-swapped}. This produces a valid Floquet code, as the lattice is still trivalent and three-face-colourable. The structure of the instantaneous stabiliser group is the same as before, but now the plaquettes are defined on this modified lattice.

\subsection{Logical $\operatorname{CNOT}$ gates}
\label{ssec:dehn-twist-css-honeycomb}

We will now discuss how the edge-swapping gadget can be applied across a Floquet code to implement logical $\operatorname{CNOT}$ gates. As before, we shall use the CSS honeycomb Floquet code as our main example.

\begin{figure}
    \centering
    \begin{subfigure}{0.35\linewidth}
        \includegraphics[width=\linewidth]{honeycomb.pdf}
        \caption{}
        \label{fig:honeycomb-dehn-twist-original}
    \end{subfigure}
    \hspace{2cm}
    \begin{subfigure}{0.35\linewidth}
        \includegraphics[width=\linewidth]{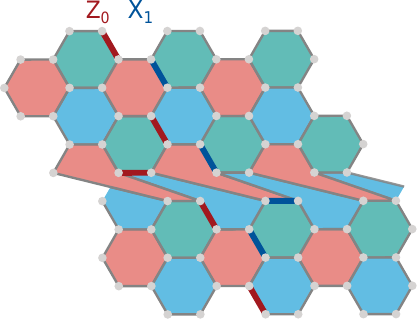}
        \caption{}
        \label{fig:honeycomb-dehn-twist-step-1}
    \end{subfigure}
    \begin{subfigure}{0.35\linewidth}
        \includegraphics[width=\linewidth]{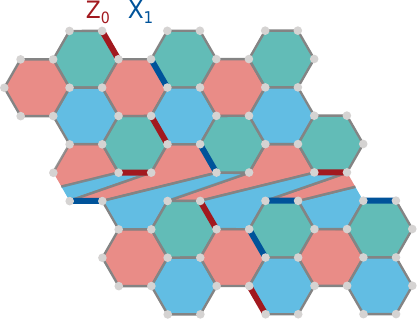}
        \caption{}
        \label{fig:honeycomb-dehn-twist-step-2}
    \end{subfigure}
    \hspace{2cm}
    \begin{subfigure}{0.35\linewidth}
        \includegraphics[width=\linewidth]{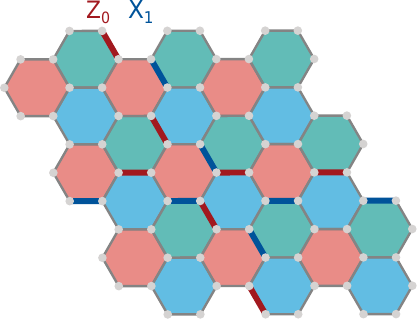}
        \caption{}
        \label{fig:honeycomb-dehn-twist-step-3}
    \end{subfigure}
    \caption{Implementing a horizontal Dehn twist on the honeycomb code. (a) The original honeycomb lattice along with logicals as in \Cref{fig:example-lattices}.  The support of $\operatorname{X}_0$ and $\operatorname{Z}_1$ remains the same in the successive figures, thus omitted. (b) The honeycomb lattice with the edge-swapping gadget applied three times in parallel along the horizontal loop. Updated logicals are also highlighted. (c) The lattice from (b) with the edge-swapping gadget applied three more times. (d) The lattice from (c) with the edge-swapping gadget applied three more times. The code is now in the original honeycomb lattice. However, $\operatorname{Z}_0$ and $\operatorname{X}_1$ has been modified to now include the components of $\operatorname{Z}_1$ and $\operatorname{X}_0$, respectively, thus implementing a logical $\operatorname{CNOT}_{1,0}$ gate.
    }
    \label{fig:non-trivial-loop-honeycomb}
\end{figure}

To implement a logical $\operatorname{CNOT}$ gate using this approach, we need a non-trivial loop which alternates between green edges and red/blue edges. In \Cref{fig:honeycomb-dehn-twist-original}, we can see that the horizontal red and blue edges form such a loop.

In \Cref{fig:honeycomb-dehn-twist-step-1} we show what happens to the CSS honeycomb Floquet code after applying the edge-swapping gadget $d$ times along this non-trivial loop. We can see that the result is still a valid Floquet code which looks like the honeycomb lattice, but now there is a one-step horizontal distortion. The vertical logical observables have now been twisted horizontally in order to still form operators which commute with all the stabilisers on the lattice. This is one step of the Dehn twist.

We can see that this process is iterable. In \Cref{fig:honeycomb-dehn-twist-step-2} and \Cref{fig:honeycomb-dehn-twist-step-3}, we show what happens by repeatedly applying the edge-swapping gadget to distort the lattice. After repeating this distortion $d$ times---$d^2$ applications of the edge-swapping gadget in total---we return to the original lattice, but now the vertical observables now run around the horizontal boundary as well as the vertical boundary\footnote{Here we use $d$ to mean the distance under an entangling measurements noise model \cite{Gidney2021FloquetBenchmark}. This distance is the same under circuit-level noise when using an in-place measurement circuit, but is lower than the distance possible when implementing the measurement circuits via auxiliary qubits such as in a Shor-style syndrome extraction circuit \cite{Shor1996FaultTolerance, Gidney2021FloquetBenchmark, Gidney2022BenchmarkingPlanarHoneycombCode}.}. From a logical information perspective this is akin to the horizontal logical observables being included in the vertical logical observables. This is equivalent to a logical $\operatorname{CNOT}$ gate.

In order to prevent the spread of undetectable logical errors, we must apply one QEC round between each step of the Dehn twist. Therefore to fault-tolerantly implement the full Dehn twist, $d$ QEC rounds are required\footnote{As Guernut and Vuillot note \cite{Guernut2024ToricCliffordGates}, it is possible to implement multiple steps of the Dehn twist between each QEC round at a cost of reducing the code distance. Specifically, performing $c$ steps of the Dehn twist between each QEC round reduces the distance of the code by a factor of $O(c)$.}. For the example shown in \Cref{fig:non-trivial-loop-honeycomb}, this twist can be implemented with nine applications of the edge-swapping gadget across three QEC rounds.

\begin{figure}
    \centering
    \begin{subfigure}{0.45\linewidth}
        \includegraphics[width=\linewidth]{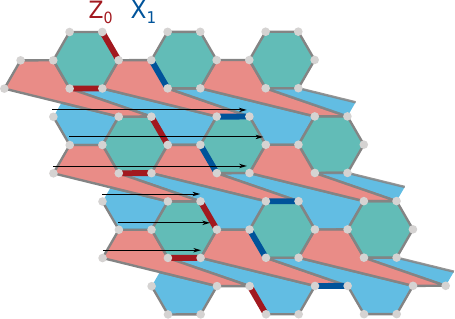}
        \caption{}
        \label{fig:honeycomb-instantaneous-dehn-twist-step-1}
    \end{subfigure}
    \hfill
    \begin{subfigure}{0.4\linewidth}
        \includegraphics[width=\linewidth]{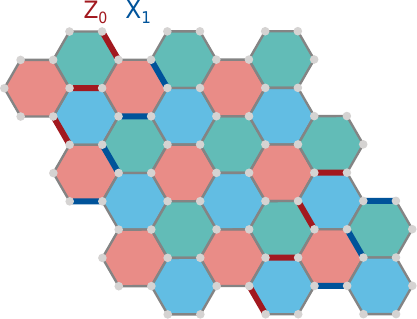}
        \caption{}
        \label{fig:honeycomb-instantaneous-dehn-twist-step-2}
    \end{subfigure}
    \caption{An instantaneous horizontal Dehn twist applied to the honeycomb Floquet code. (a) Distorted lattice after a single step of the Dehn twist applied in parallel along three horizontal loops in \Cref{fig:honeycomb-dehn-twist-original}. (b) The same twisted lattice but with the qubits shifted horizontally in order to return the code to the original honeycomb lattice. The original vertical loop now moves in a diagonal direction, thus acquiring the horizontal component. The whole process performs a logical $\operatorname{CNOT}_{1,0}$.}
    \label{fig:instantaneous-floquet-dehn-twist}
\end{figure}

Alternatively, the Dehn twist can be implemented using fewer QEC rounds by identifying multiple non-trivial loops that can be distorted in parallel \cite{Breuckmann2017HyperbolicSurfaceCodes, Zhu2020DehnTwist, Guernut2024ToricCliffordGates}. In \Cref{fig:honeycomb-instantaneous-dehn-twist-step-1}, we show how the Dehn twist can be implemented instantaneously on the CSS Floquet honeycomb code by applying distortions along parallel loops. Note that this approach does not produce the original honeycomb lattice, but this can be restored by shifting the qubits in a horizontal cycle. This shifted operation is shown in \Cref{fig:honeycomb-instantaneous-dehn-twist-step-2}. The number of QEC rounds required to remain fault-tolerant is $\left\lceil\frac{d}{p}\right\rceil$, where $d$ is the code distance and $p\leq d$ is the number of parallel loops. For the example honeycomb Floquet code discussed in this section, the Dehn twist can be implemented in three QEC rounds if using a single Dehn twist, or one QEC round if using three parallel loops.

\begin{figure}
    \centering
    \begin{subfigure}{0.4\linewidth}
        \includegraphics[width=\linewidth]{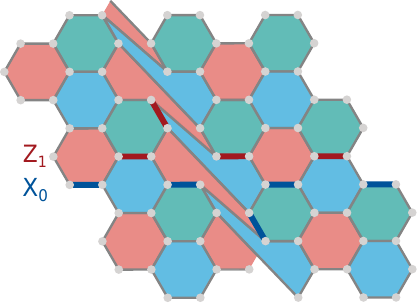}
        \caption{}
    \end{subfigure}
    \hfill
    \begin{subfigure}{0.4\linewidth}
        \includegraphics[width=\linewidth]{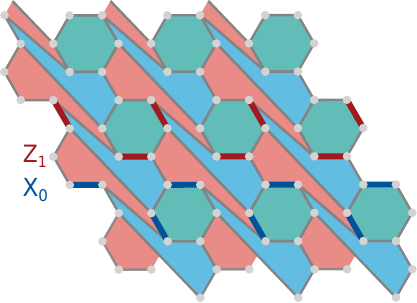}
        \caption{}
    \end{subfigure}
    \caption{First steps in realising logical $\operatorname{CNOT}_{0,1}$ when applying (a) linear time and (b) instantaneous vertical Dehn twists to the honeycomb Floquet code.}
    \label{fig:vertical-dehn-twist}
\end{figure}

So far we have only considered Dehn twists along a horizontal non-trivial loop in the honeycomb code. However, the honeycomb Floquet code also includes vertical non-trivial loop which follows the correct pattern. We won't show the full transformation, but the first steps of both the linear time and instantaneous vertical Dehn twists to the honeycomb Floquet code can be seen in \Cref{fig:vertical-dehn-twist}.

Finally, we note that it is possible to implement Dehn twists on the honeycomb code by operating on the $3.6.3.6$ code that is embedded at each sub-round. We discuss this embedding further in \Cref{sec:embedded-codes}, and provide explicit constructions for Dehn twists on this embedded code in \Cref{app:toric_hex_triangle}.

\subsection{Extensions to other measurement schedules}
\label{ssec:dehn-twist-gidney-schedule}

As with the fold-transversal gates, we will now briefly touch on how this technique can be applied to other measurement schedules. As a reminder, the instantaneous stabiliser group for the period-three schedule from Gidney \etal consists of the following:
\begin{itemize}
    \item $\operatorname{X}$ checks on all red plaquettes;
    \item $\operatorname{Y}$ checks on all green plaquettes;
    \item $\operatorname{Z}$ checks on all blue plaquettes; and
    \item $\operatorname{Z}$ checks on all blue edges.
\end{itemize}

We will focus on the green plaquettes, which are the only difference compared to \Cref{ssec:dehn-twist-css-honeycomb}. Crucially, we note in \Cref{ssec:edge-swapping} that the green plaquettes are unaffected by the edge-swapping gadget. This is still true for this instantaneous stabiliser group: any $\operatorname{Y}$ checks on the green plaquettes will not be modified after applying the gadget. The rest of the instantaneous stabiliser group is modified the same as before, as are the logical operators. Note that a slight complication is that the period-three measurement schedule on the honeycomb code induces a logical $\operatorname{SWAP}_{0, 1}(\operatorname{H_0}\otimes \operatorname{H_1})$ operation every QEC round, thus the specific logical operators will change between steps of the Dehn twist, but this does not affect the overall logical $\operatorname{CNOT}$ gate.

\subsection{Dehn twists on the periodic $4.8.8$ Floquet code}
\label{ssec:dehn-twist-4-8-8}

\begin{figure}
    \centering
    \begin{subfigure}{0.25\linewidth}
        \includegraphics[width=\linewidth]{4-8-8.pdf}
        \caption{}
        \label{fig:4-8-8-dehn-twist-original}
    \end{subfigure}
    \hfill
    \begin{subfigure}{0.25\linewidth}
        \includegraphics[width=\linewidth]{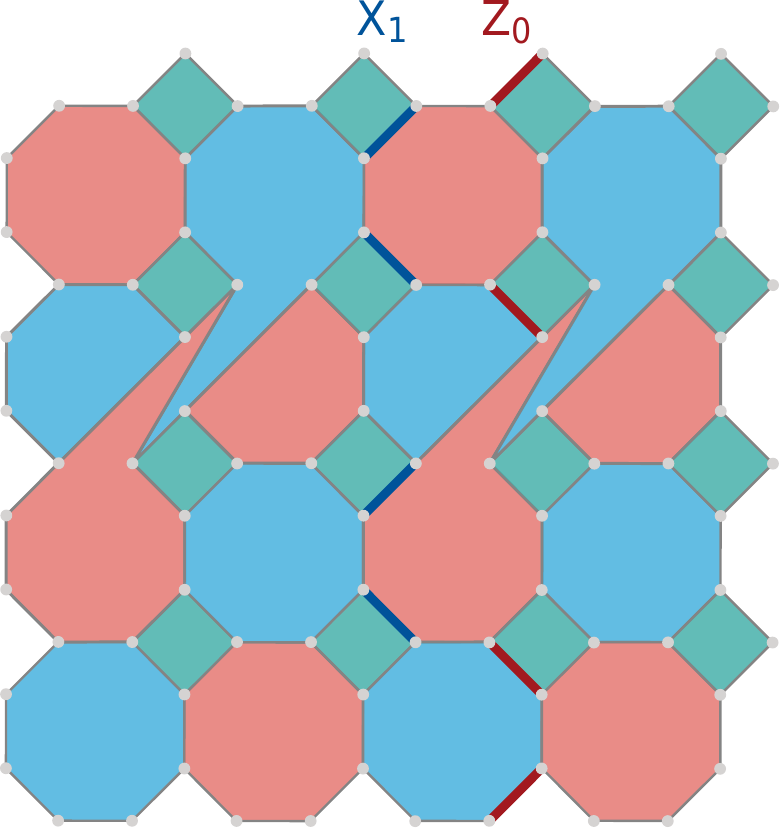}
        \caption{}
        \label{fig:4-8-8-dehn-twist-step-1}
    \end{subfigure}
    \hfill
    \begin{subfigure}{0.25\linewidth}
        \includegraphics[width=\linewidth]{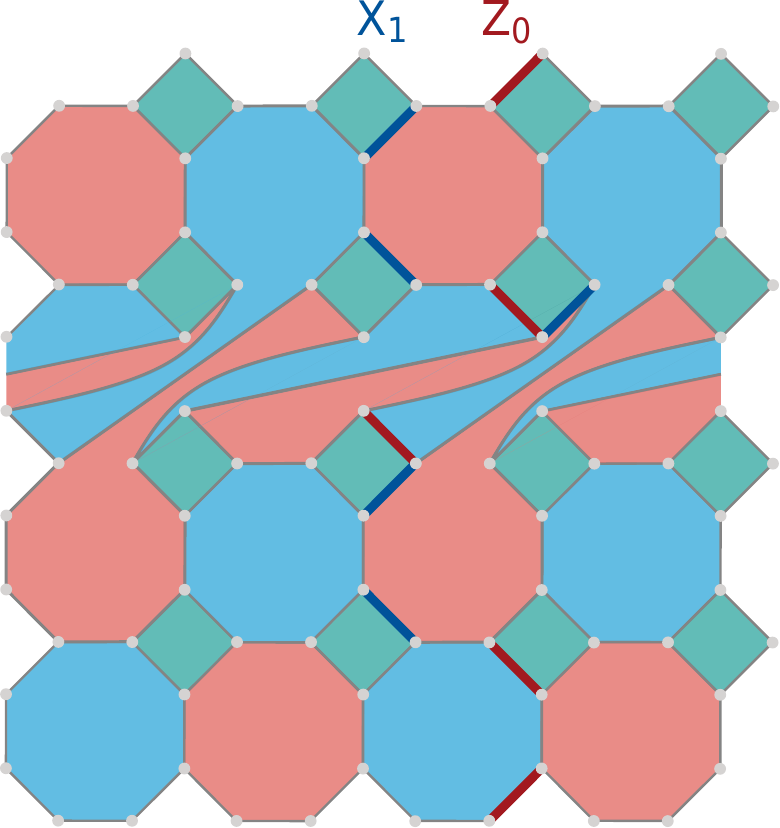}
        \caption{}
        \label{fig:4-8-8-dehn-twist-step-2}
    \end{subfigure}

    \begin{subfigure}{0.25\linewidth}
        \includegraphics[width=\linewidth]{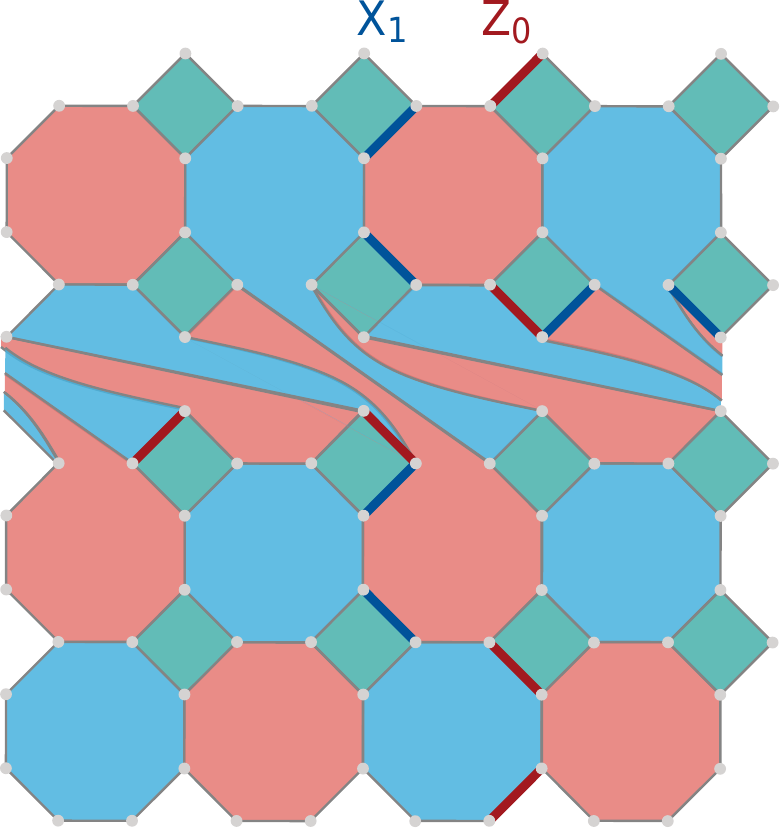}
        \caption{}
        \label{fig:4-8-8-dehn-twist-step-3}
    \end{subfigure}
    \hfill
    \begin{subfigure}{0.25\linewidth}
        \includegraphics[width=\linewidth]{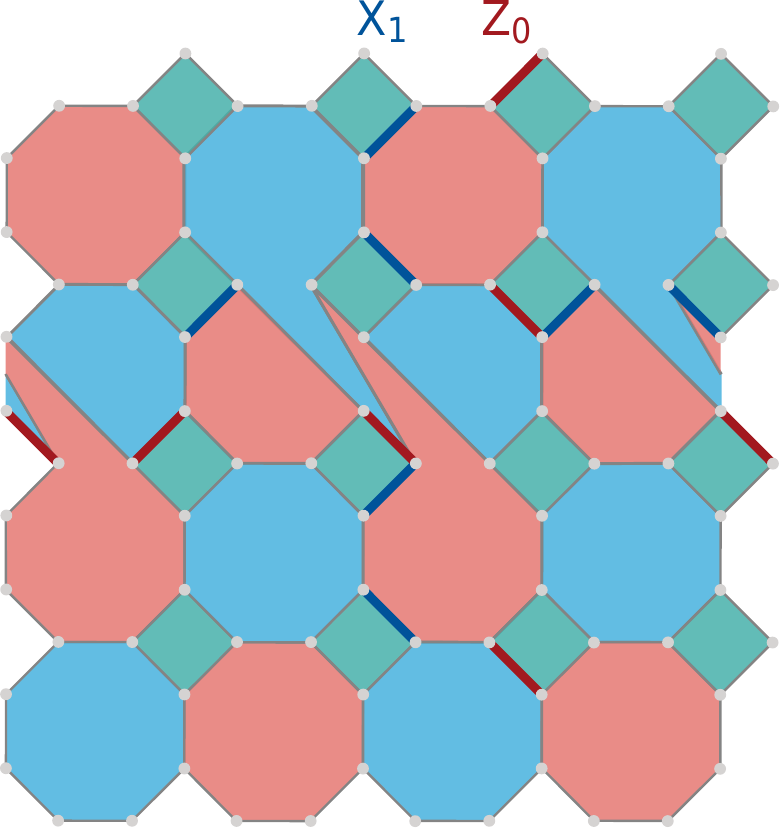}
        \caption{}
        \label{fig:4-8-8-dehn-twist-step-4}
    \end{subfigure}
    \hfill
    \begin{subfigure}{0.25\linewidth}
        \includegraphics[width=\linewidth]{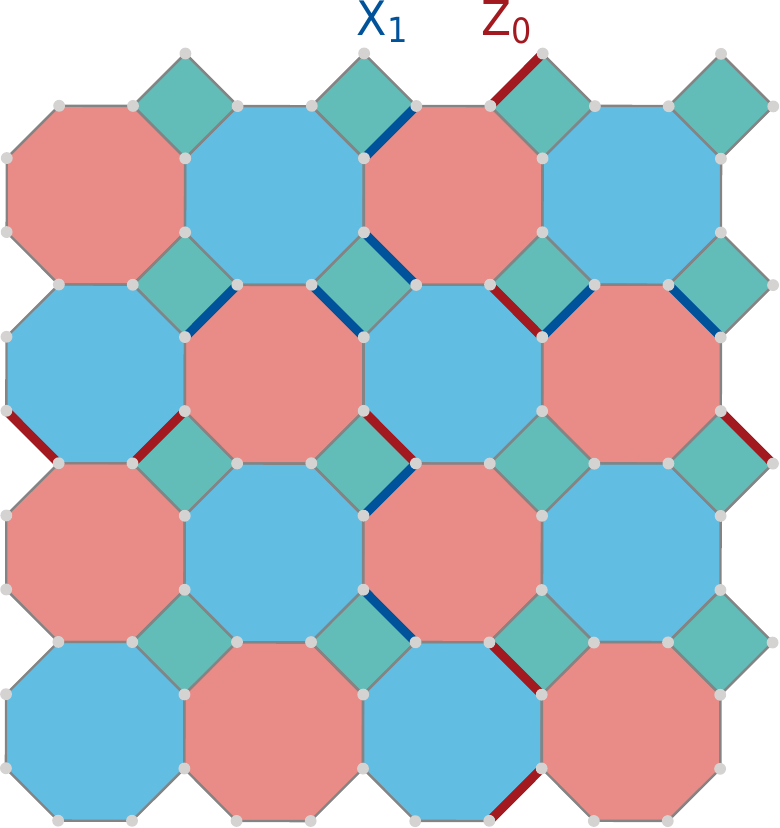}
        \caption{}
        \label{fig:4-8-8-dehn-twist-step-5}
    \end{subfigure}
    \caption{Repeatedly applying the edge-swapping gadget to the $4.8.8$ Floquet code to produce a Dehn twist. The original code as in \Cref{fig:4-8-8} is presented in (a), intermediate steps are presented in (b)-(e), and the final lattice is presented in (f). The distortion of the lattice in (d) has been reversed for ease of presentation, similarly to \Cref{fig:honeycomb-dehn-twist-step-2}. (b) and (f) are produced by applying the edge-swapping gadget twice in parallel, while (c)-(e) are produced by applying the edge-swapping gadget four times in parallel. A single QEC round is required after each step. The support of $\operatorname{X}_0$ and $\operatorname{Z}_1$ remains unchanged throughout the procedure, while $\operatorname{X}_1$ and $\operatorname{Z}_0$ get modified devicing the logical $\operatorname{CNOT}_{1,0}$.}
    \label{fig:4-8-8-dehn-twist}
\end{figure}

We finish the section by discussing how to implement a Dehn twist on a more complicated example: a Floquet code on a $4.8.8$ lattice with periodic boundaries\footnote{Note that unlike in \Cref{ssec:folded-4-8-8}, we are not considering the planar $4.8.8$ code in this instance. This is because Dehn twists require closed loops.}. As shown in \Cref{fig:4-8-8-dehn-twist-original}, the $4.8.8$ code does have a non-trivial loop along the red and blue edges, but it now follows the more complicated pattern of $\color{floquet-red}\operatorname{r}\shortrightarrow\color{floquet-green}\operatorname{g}\shortrightarrow\color{floquet-red}\operatorname{r}\shortrightarrow\color{floquet-green}\operatorname{g}\shortrightarrow\color{floquet-blue}\operatorname{b}\shortrightarrow\color{floquet-green}\operatorname{g}\shortrightarrow\color{floquet-blue}\operatorname{b}\shortrightarrow\color{floquet-green}\operatorname{g}$. This means that we cannot apply the edge-swapping gadget along the full length of this non-trivial loop in one step, but we can still apply the edge-swapping gadget along parts of the non-trivial loop.

The result of applying this edge-swapping gadget twice is shown in \Cref{fig:4-8-8-dehn-twist-step-1}. As before, this process is iterable, so we can apply the edge-swapping gadget along parts of the loop again. The result of repeatedly applying the edge-swapping gadget is shown in \Cref{fig:4-8-8-dehn-twist-step-5}, with intermediate steps presented from \Cref{fig:4-8-8-dehn-twist-step-2} to \Cref{fig:4-8-8-dehn-twist-step-4}. The resulting lattice is a $4.8.8$ lattice with the vertical observables twisted around the horizontal boundaries. A vertical Dehn twist is similarly possible.

We need to apply a round of QEC after each intermediate step shown in \Cref{fig:4-8-8-dehn-twist}. The overall Dehn twist for a distance $d$ $4.8.8$ code will use $d^2$ applications of the edge-swapping gadget across $d+1$ QEC rounds. For the example in \Cref{fig:4-8-8-dehn-twist} this corresponds to sixteen applications of the edge-swapping gadget across five QEC rounds. The number of QEC rounds can be reduced through the use of parallel loops to $\left\lceil\frac{d}{p}\right\rceil+1$, where $d$ is the code distance and $p\leq d/2$ is the number of parallel loops used.

Note that unlike with the honeycomb code, applying the edge-swapping gadget to the $4.8.8$ code does result in a decrease in the vertical code distance. This is because the intermediate steps distort the plaquettes horizontally \textit{and} vertically. For each parallel loop, the distance is reduced by a constant amount. The most significant distance reduction is when using $d/2$ parallel loops to implement a Dehn twist instantaneously, where the distance is halved. This suggests some care must be taken when constructing Dehn twists for Floquet codes on general colour code lattices. We leave identifying good non-trivial loops for implementing Dehn twists on more complex Floquet codes for future work.

As noted in \Cref{ssec:dehn-twist-static}, it is possible to implement Dehn twists on the unrotated toric code on a square lattice. It is also known that the CSS Floquet $4.8.8$ code embeds a toric code after the $\color{floquet-green}\operatorname{gXX}$ sub-round \cite{Davydova2023CSSFloquet}, therefore, one could implement a Dehn twist on the $4.8.8$ code by operating on the embedded toric code. We will discuss this in further detail in \Cref{sec:embedded-codes}.

Finally, we note that a rotated version of the $4.8.8$ lattice supports a simpler non-trivial loop for implementing Dehn twists, at the cost of requiring more physical qubits. We discuss this code in further detail in \Cref{app:rotated-4-8-8}.

\section{Benchmarking}
\label{sec:benchmarking}

\subsection{Circuit and noise model}

\begin{figure}
    \centering
    \begin{quantikz}
        \lstick{P} & \gate[2]{\operatorname{QEC}(1, 0)}\slice{} & \gate[2]{\operatorname{QEC}(d, \epsilon)}\slice{} & \gate[2]{\operatorname{U}(\epsilon)}\slice{} & \gate[2]{\operatorname{QEC}(d, \epsilon)}\slice{}  & \gate[2]{\operatorname{QEC}(1, 0)} & \meter{\operatorname{P}'} \\
        \lstick{Q} & & & & & & \meter{\operatorname{Q}'}
    \end{quantikz}
    \caption{The protocol followed when implementing logical gates. $\operatorname{QEC}(d, \epsilon)$ indicates $d$ QEC rounds with physical error probability $\epsilon$, with $\epsilon=0$ indicating a noiseless QEC round. $\operatorname{U}(\epsilon)$ implements a logical gate $\operatorname{U}$ with physical error probability $\epsilon$. Implementation of $\operatorname{U}(\epsilon)$ requires either one QEC round, in the case of fold-transversal gates and instantaneous Dehn twists, or $d$ QEC rounds in the case of linear-time Dehn twists. For the memory simulations, $U(\epsilon)$ is simply an identity operation for the relevant number of QEC rounds. Only the first and last slices of the circuit are noise-free, other steps in the protocol feature circuit-level noise. $\operatorname{P}$ and $ \operatorname{Q}$ are $\operatorname{X}$ or $\operatorname{Z}$ bases, with $\operatorname{P}'$ and $\operatorname{Q}'$ chosen such that the logical information is preserved.}
    \label{fig:protocol}
\end{figure}
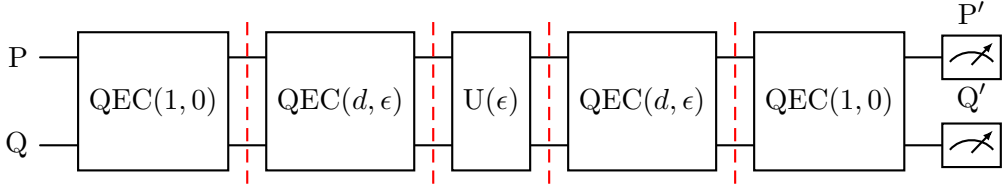

Our protocol for benchmarking the performance of logical gates is presented in \Cref{fig:protocol}. This is based on the protocol by Guernut and Vuillot \cite{Guernut2024ToricCliffordGates} but with a few differences. In order to test the performance of these techniques across a variety of Pauli bases, we use perfect logical state preparation and measurement, with one perfect QEC round after logical state preparation and one perfect QEC round before logical state measurement. We use this to noiselessly prepare both logical qubits in Pauli bases $\operatorname{P}, \operatorname{Q} \in \{\operatorname{X}, \operatorname{Z}\}$ independently. To imitate the effect of performing a logical operation on a noisy state in the middle of a computation, $d$ noisy QEC rounds are inserted before the logical operation. The logical operation $\operatorname{U}$ is then performed in a noisy way, along with sufficient noisy QEC rounds to preserve the code distance. This is then followed by $d$ noisy QEC rounds to imitate the rest of the logical computation. Finally, the logical qubits are perfectly measured out in Pauli bases $\operatorname{P}', \operatorname{Q}'$ such that

\begin{equation}
\operatorname{P}'\otimes \operatorname{Q}' = \operatorname{U}(\operatorname{P}\otimes \operatorname{Q})\operatorname{U}^\dagger.
\end{equation}

Perfect logical state preparation and measurement is required as some logical operations, such as the fold-transversal $\operatorname{S}$-type gate, map logical qubits in the Pauli $\operatorname{X}$ basis to the Pauli $\operatorname{Y}$ basis. It is unknown how to fault-tolerantly measure a logical qubit in the $\operatorname{Y}$ basis on a Floquet code\footnote{An alternative circuit in which logical operations for $\operatorname{U}$ and $\operatorname{U}^\dagger$ are applied in sequence would remove this problem, as $\operatorname{U}\operatorname{U}^\dagger = \operatorname{I}$ and therefore logical $\operatorname{X}$ (resp.\ $\operatorname{Z}$) operators would be the same at the start and end of the circuit. However, this would be akin to benchmarking two applications of the logical gate rather than one.}, though a construction using $d/2$ QEC rounds is known to exist for the rotated surface code \cite{Gidney2024YBasis}. Similarly, Dehn twists can map logical operations to bases which are non-trivial to measure simultaneously in a fault-tolerant way: from \Cref{eq:cnot_heisenberg} we can note that

\begin{equation}
    \operatorname{CNOT}_{0, 1}(\operatorname{X}_0\otimes\operatorname{Z}_1)\operatorname{CNOT}_{0, 1}^\dagger = (\operatorname{X}_0\otimes\operatorname{X}_1)(\operatorname{Z}_0\otimes\operatorname{Z}_1).
\end{equation}

\noindent It is unclear how to fault-tolerantly measure both observables fault-tolerantly on a honeycomb Floquet code, though it is likely techniques which exist in static codes can be adapted to Floquet codes \cite{lodyga2015}.

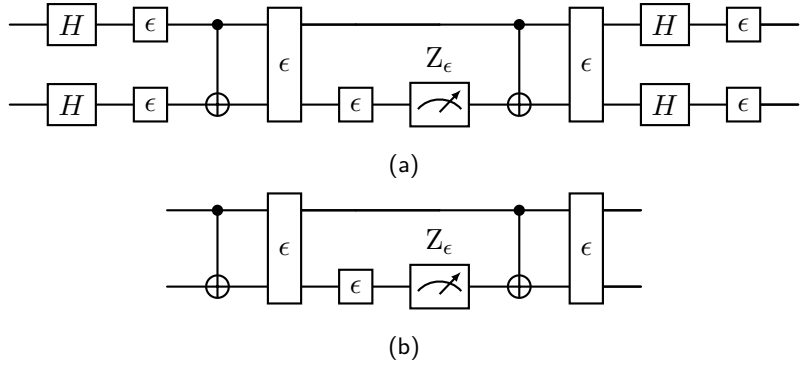
\begin{figure}
    \centering
    \begin{subfigure}{\linewidth}
        \centering
        \begin{quantikz}
            & \gate{H} & \gate{\epsilon} & \ctrl{1} & \gate[2]{\epsilon} & \qw & \qw & \ctrl{1} & \gate[2]{\epsilon} & \gate{H} & \gate{\epsilon} & \qw \\
            & \gate{H} & \gate{\epsilon} & \targ{} & & \gate{\epsilon} & \meter{\operatorname{Z}_\epsilon} & \targ{}&  & \gate{H} & \gate{\epsilon} & \qw
        \end{quantikz}
        \caption{}
    \end{subfigure}

    \begin{subfigure}{\linewidth}
        \centering
        \begin{quantikz}
            & \ctrl{1} & \gate[2]{\epsilon} & \qw & \qw & \ctrl{1} & \gate[2]{\epsilon} & \qw \\
            & \targ{} & & \gate{\epsilon}& \meter{\operatorname{Z}_\epsilon} & \targ{} & & \qw
        \end{quantikz}
        \caption{}
    \end{subfigure}
    \caption{Auxiliary-free check circuits for (a) $\operatorname{XX}$ and (b) $\operatorname{ZZ}$ bases, respectively. The $\epsilon$ gates correspond to depolarising noise channels with strength $\epsilon$. $\operatorname{Z}_\epsilon$ corresponds to a measurement in the $Z$ basis, where the measurement outcome is flipped with probability $\epsilon$.}
    \label{fig:check-circuits}
\end{figure}

Each QEC round consists of one implementation of the period-six schedule. We implement the two-qubit Pauli checks using auxiliary-free measurement circuits, shown in \Cref{fig:check-circuits}. These check circuits offer a lower distance than circuits with additional qubits for measurements, but have been shown to offer better performance in other properties such as qubit footprint \cite{Benito2025HexVersusHeavyHex}. The noisy QEC rounds feature circuit-level noise, with single-qubit and two-qubit gates being followed by single-qubit and two-qubit depolarising channels with strength $\epsilon$, and measurement outcomes also flipping with probability $\epsilon$. The physical gates used to implement the logical operations are subject to the same noise channels.

We implemented these circuits using Stim \cite{Gidney2021Stim}. Example Stim circuits which we constructed during this work can be found in \cite{Moylett2025StimCircuits}. Stim utilises the fact that the observables and detectors are both deterministic in the absence of errors to identify which errors anti-commute and therefore flip the observables and detectors. Stim is then able to sample measurement outcomes for both the observables and detectors by sampling from the error channels. The objective of a decoder is to then correctly identify the samples where errors have caused an observable to flip, based on the detector outcomes.

By decoding many measurement samples from Stim we are able to approximate a logical error rate for a given physical error rate $\epsilon$, code distance $d$, and initial bases $\operatorname{P}, \operatorname{Q}$. In this section we will report the maximum logical error rate observed for each value of $\epsilon$ and $d$; in \Cref{app:full-numerics} we present full numerical results for all initial bases\footnote{Note that an arbitrary logical state might have a higher logical error rate. Analysis of the logical error rate for arbitrary logical states has previously assumed that logical errors for each observable are independent \cite{Gidney2022BenchmarkingPlanarHoneycombCode, Geher2024LogicalHadamard, McLauchlan2024FabDefects}. This is not the case in our work however, due to the existence of correlated logical errors: a logical $\operatorname{X}$ error after the fold-transversal $\operatorname{S}$-type gate, for example, would cause a decoder to fail to correct a logical qubit prepared in the $\operatorname{X}$ basis or in the $\operatorname{Z}$ basis. This becomes even more complex when assessing the performance of two logical qubits on a torus: a physical $\operatorname{X}$ error, for example, can contribute to logical $\operatorname{X}$ errors on \textit{both} logical qubits. Analysis of an arbitrary logical state would therefore require estimating conditional logical error probabilities.}.

From the logical error rate for given values of $\epsilon$ and $d$ we can approximate two properties of interest. The first is the threshold, the physical error rate below which quantum error correction becomes feasible \cite{Aharonov2008Threshold, Knill1998Threshold, Kitaev2003Threshold}. While the threshold is important to understanding when fault-tolerant quantum computing is possible, it doesn't necessarily indicate practical relevance, as error rates close to threshold will require significant resources \cite{Gidney2021FloquetBenchmark}. Therefore we also simulate these protocols at sub-threshold error rates, to see how well they are able to exponentially suppress errors.

\subsection{Quantum memory}

To understand the baseline performance of these logical operations, we begin by simulating quantum memory experiments. These correspond to the setups as described above, but with our logical operation being the identity gate.

\begin{figure}
    \centering
    \begin{subfigure}{0.49\linewidth}
        \includegraphics[width=\linewidth]{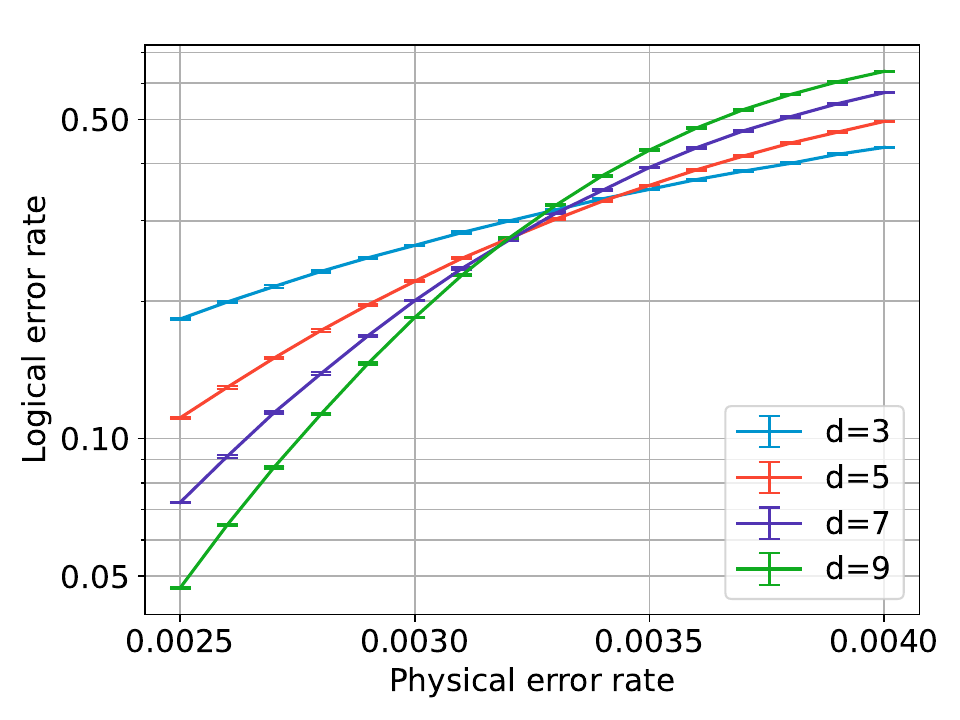}
        \caption{}
    \end{subfigure}
    \begin{subfigure}{0.49\linewidth}
        \includegraphics[width=\linewidth]{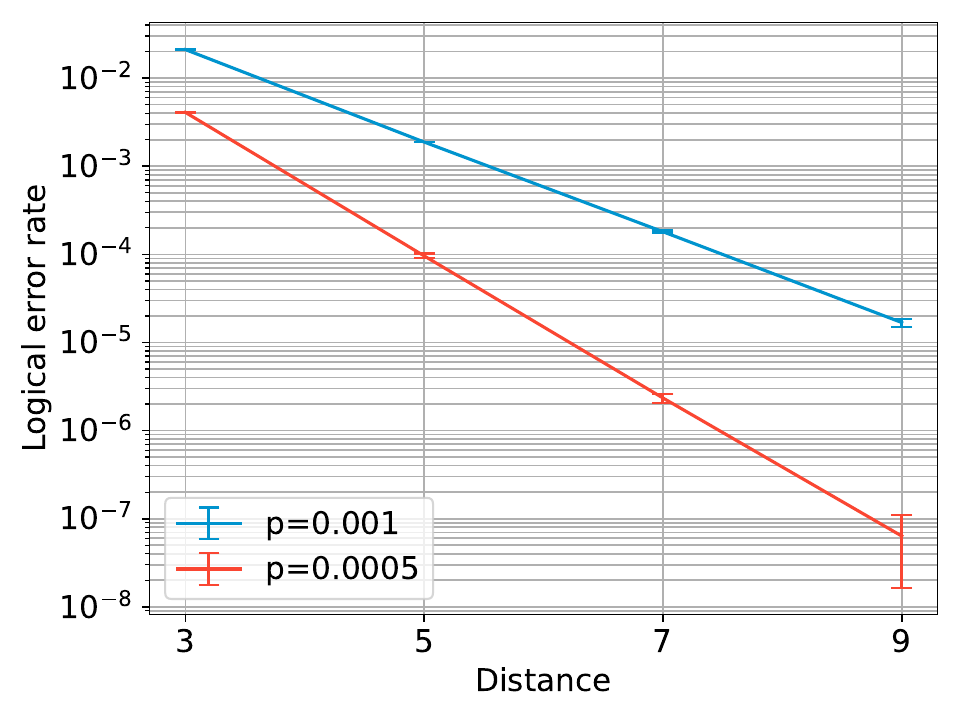}
        \caption{}
        \label{fig:subth-memory-2d-plus-1-pymatching}
    \end{subfigure}
    \caption{Benchmarking a $(2d+1)$-round quantum memory on the CSS honeycomb Floquet code when decoded with PyMatching \cite{Higgott2025SparseBlossom}. (a) We estimate a threshold of around $0.32\%$. (b) Sub-threshold analysis for physical error rates of $0.1\%$ and $0.05\%$ show exponential error suppression. Error bars indicate a $95\%$ confidence interval. This quantum-memory benchmark sets a reference when assessing the fold-transversal Hadamard-type gate (\Cref{fig:hadamard}), the instantaneous horizontal Dehn twist (\Cref{fig:instantaneous-horizontal-dehn-twist}), and the instantaneous vertical Dehn twist (\Cref{fig:instantaneous-vertical-dehn-twist}).}
    \label{fig:memory-2d-plus-1-pymatching}
\end{figure}

In \Cref{fig:memory-2d-plus-1-pymatching}, we look at the threshold and sub-threshold performance of a quantum memory on the CSS Floquet honeycomb code, which preserves the two logical qubits for $2d+1$ QEC rounds. This number of rounds is comparable to the fold-transversal Hadamard-type circuits and the instantaneous Dehn twist circuits, which we benchmark in \Cref{ssec:hadamard-experiment} and \Cref{ssec:instantaneous-dehn-twist-experiment}, respectively. The experiments are decoded using PyMatching, a decoder based on solving the minimum-weight perfect matching problem on a graph, and which has been particularly optimised for quantum error correction \cite{Higgott2025SparseBlossom}. We estimate a threshold of approximately $0.32\%$, comparable to the threshold for the honeycomb code with a period-three schedule, as studied by Benito \etal \cite{Benito2025HexVersusHeavyHex}. For physical error rates of $0.1\%$ and $0.05\%$, we estimate logical error rates of approximately $2 \times 10^{-4}$ and $2 \times 10^{-6}$ for a distance-7 CSS honeycomb Floquet code, respectively. Such a code uses $3d\times2d = 294$ physical qubits. For a distance-9 code, which uses 486 physical qubits, we estimate logical error rates of approximately $2 \times 10^{-5}$ and $6 \times 10^{-8}$, respectively.

\begin{figure}
    \centering
    \begin{subfigure}{0.49\linewidth}
        \includegraphics[width=\linewidth]{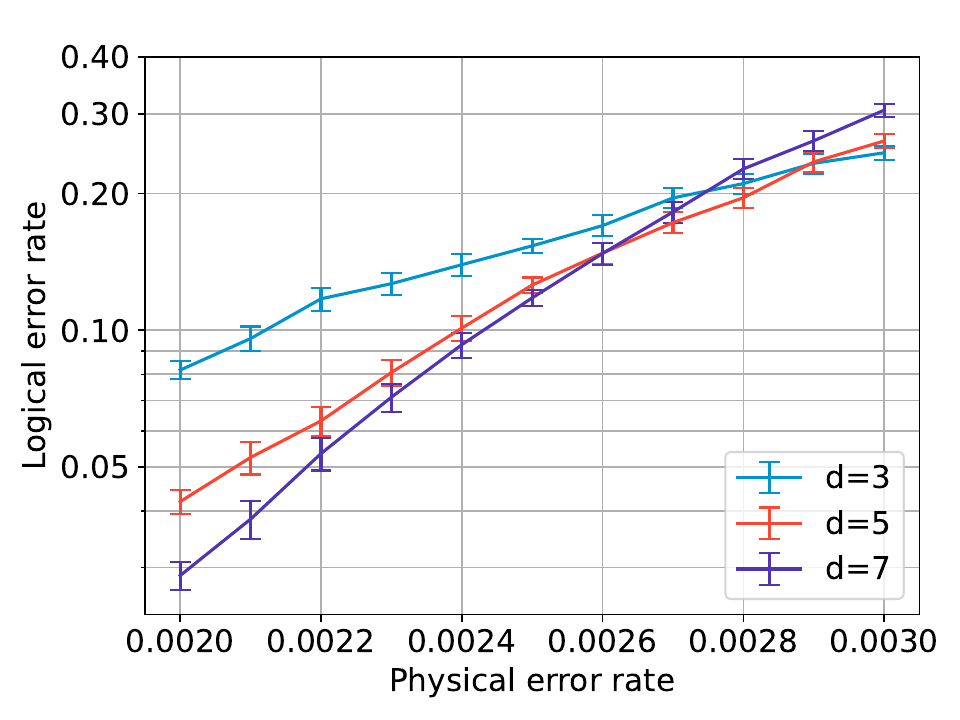}
        \caption{}
    \end{subfigure}
    \begin{subfigure}{0.49\linewidth}
        \includegraphics[width=\linewidth]{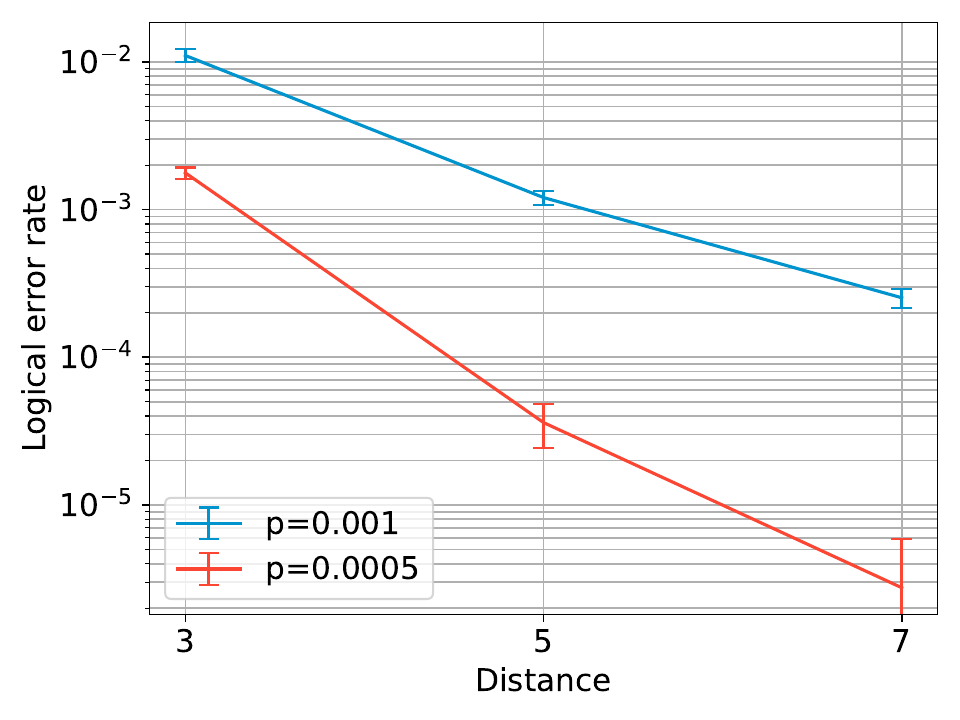}\caption{}
        \label{fig:sub-threshold-memory-2d-plus-1-bp-lsd-subth-max-ler}
    \end{subfigure}
    \caption{Benchmarking a $(2d+1)$-round quantum memory on the CSS honeycomb Floquet code when decoded with BP+LSD-0 \cite{Hillmann2025BPLSD}. (a) We estimate a threshold of around $0.25$-$0.3\%$. (b) Sub-threshold analysis for physical error rates of $0.1\%$ and $0.05\%$. We attribute the flattening of the error suppression at higher distances to sub-optimal performance of zeroth-order BP+LSD-0, as reported in previous works \cite{Higgott2023DecodingHypergraphProductCodes, MagdalenadelaFuente2025XYZRubyCode}. Error bars indicate a $95\%$ confidence interval. This quantum-memory benchmark sets a reference when assessing the fold-transversal $\operatorname{S}$-type gate (\Cref{fig:s-gate}).}
    \label{fig:memory-2d-plus-1-bp-lsd}
\end{figure}

The logical $S$-type gate we benchmark in \Cref{ssec:s-gate-experiment} introduces non-graph-like errors, which cannot be decoding using matching decoders such as PyMatching. Therefore when benchmarking the logical $S$-type gate, we use Belief Propagation and zeroth-order Localised Statistics Decoding (``BP+LSD-0'') \cite{Hillmann2025BPLSD}, a decoder for general QEC codes which can decode non-graph-like errors. As a baseline to compare against, we also benchmark a $(2d+1)$-QEC round quantum memory, decoded using BP+LSD-0. The threshold and sub-threshold performance are shown in \Cref{fig:memory-2d-plus-1-bp-lsd}. Note that because of the slower runtime of BP+LSD-0 compared to PyMatching, smaller-scale experiments had to be performed. We estimate a threshold somewhere between $0.26\%$ and $0.3\%$. We see in \Cref{fig:sub-threshold-memory-2d-plus-1-bp-lsd-subth-max-ler} that at physical error rates of $0.1\%$ and $0.05\%$, the logical error rates for a memory experiment with distance-7 CSS honeycomb Floquet code (294 physical qubits) decoded with BP+LSD-0 are approximately $3 \times 10^{-4}$ and $3 \times 10^{-6}$, respectively. This is slightly higher than the sub-threshold performance seen when decoded with PyMatching. We can also see that towards the higher code distances there is a tail-off in terms of error suppression. This is because of a sub-optimality in BP+LSD-0 for larger system sizes \cite{Higgott2023DecodingHypergraphProductCodes, MagdalenadelaFuente2025XYZRubyCode}.

\begin{figure}
    \centering
    \begin{subfigure}{0.49\linewidth}
        \includegraphics[width=\linewidth]{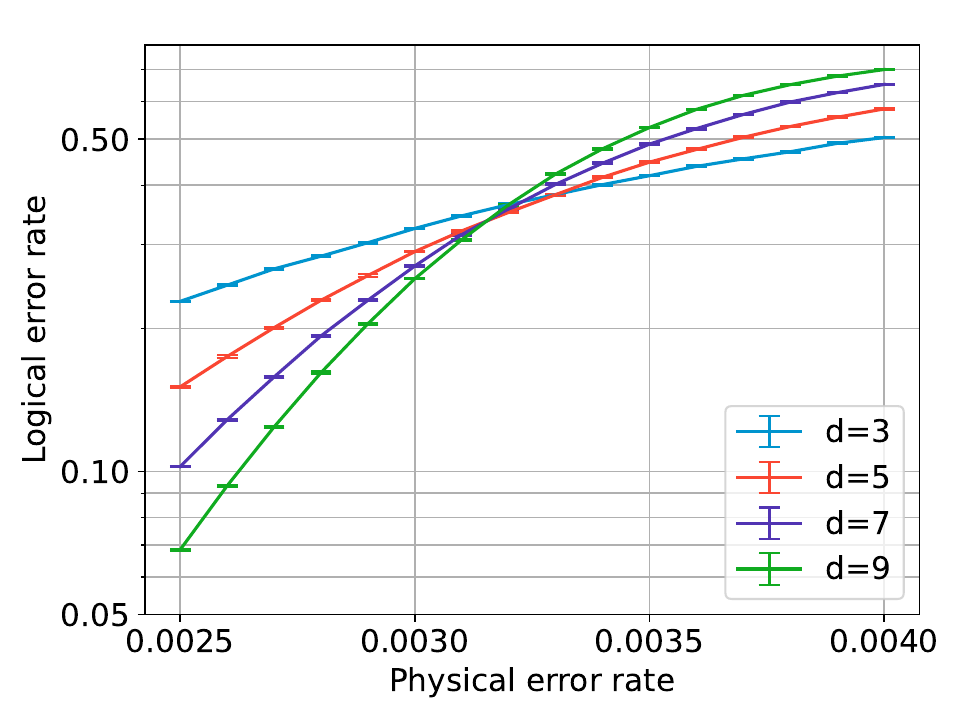}
        \caption{}
    \end{subfigure}
    \begin{subfigure}{0.49\linewidth}
        \includegraphics[width=\linewidth]{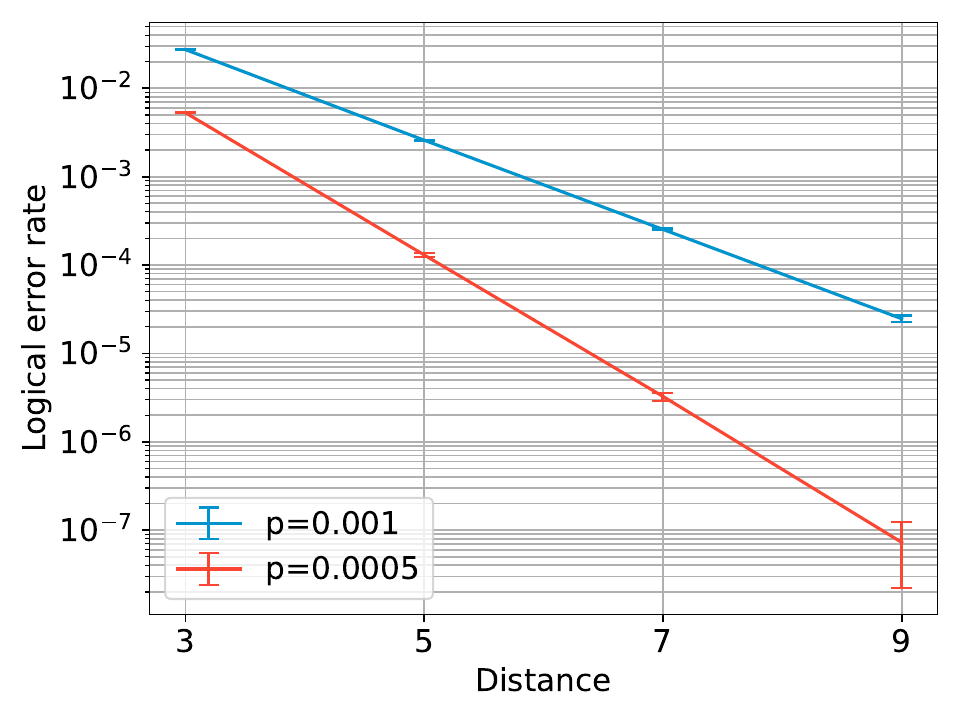}
        \caption{}
        \label{fig:subth-memory-3d}
    \end{subfigure}
    \caption{Benchmarking a $3d$-round quantum memory on the CSS honeycomb Floquet code using PyMatching. (a) We estimate a threshold of around $0.32\%$. (b) Sub-threshold analysis for physical error rates of $0.1\%$ and $0.05\%$ shows exponential error suppression. Error bars indicate a $95\%$ confidence interval. This quantum-memory benchmark sets a reference when assessing the linear-time horizontal Dehn twist (\Cref{fig:linear-horizontal-dehn-twist}), and the linear-time vertical Dehn twist (\Cref{fig:linear-vertical-dehn-twist}).}
    \label{fig:memory-3d}
\end{figure}

Finally, we benchmark a $3d$-QEC round quantum memory. This is a comparable number of rounds to the linear-time Dehn twists, which we simulate in \Cref{ssec:linear-dehn-twist-experiment}. The threshold and sub-threshold performance are presented in \Cref{fig:memory-3d}. These simulations were decoded using PyMatching \cite{Higgott2025SparseBlossom}. We estimate a threshold of approximately $0.32\%$. For physical error rates of $0.1\%$ and $0.05\%$, we estimate logical error rates are approximately $3 \times 10^{-4}$ and $3 \times 10^{-6}$ for a distance-7 CSS honeycomb Floquet code. For a distance-9 code, these logical error rates improve to approximately $3 \times 10^{-5}$ and $7 \times 10^{-8}$, respectively. These logical error probabilities are slightly higher than the $(2d+1)$-round quantum memory experiment due to the increased number of QEC rounds.

\subsection{Logical Hadamard-type gate}
\label{ssec:hadamard-experiment}

To benchmark the fold-transversal Hadamard type gate, we simulated a logical $\operatorname{H_0}\otimes \operatorname{H_1}$ gate on the CSS Floquet honeycomb code. The total number of noisy QEC rounds for this circuit is $2d+1$: $d$ rounds to prepare an initial noisy logical state, one round for the logical $\operatorname{H_0}\otimes \operatorname{H_1}$ gate, and $d$ rounds to mimic any further processing of the logical information. The fold-transversal Hadamard-type gate preserves graph-like errors, meaning that PyMatching can be used as a decoder.

\begin{figure}
    \centering
    \begin{subfigure}{0.49\linewidth}
        \includegraphics[width=\linewidth]{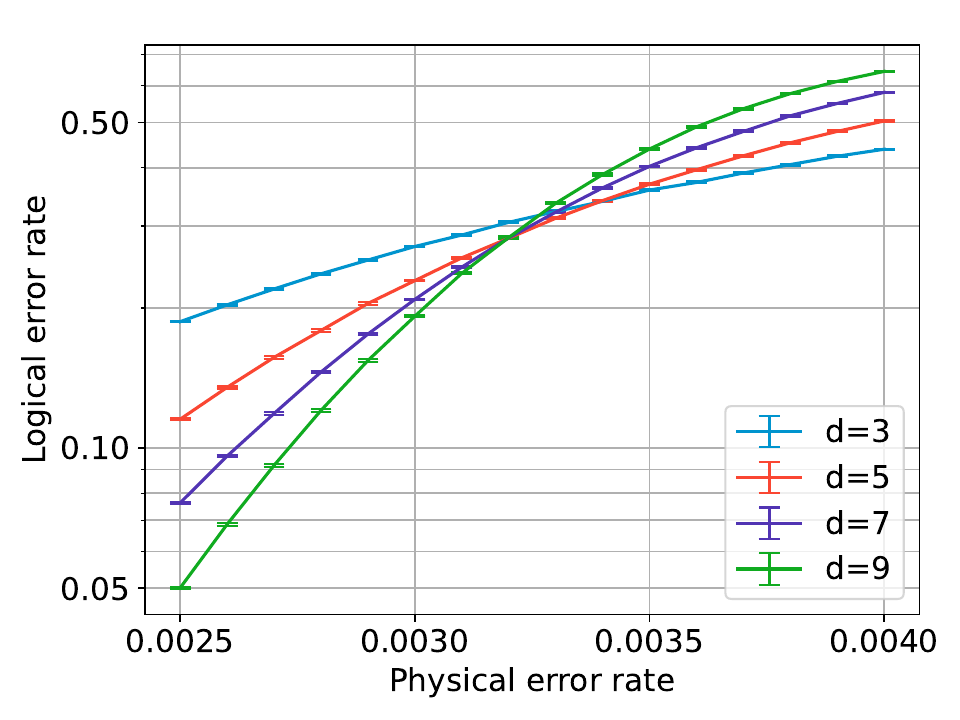}
        \caption{}
    \end{subfigure}
    \begin{subfigure}{0.49\linewidth}
        \includegraphics[width=\linewidth]{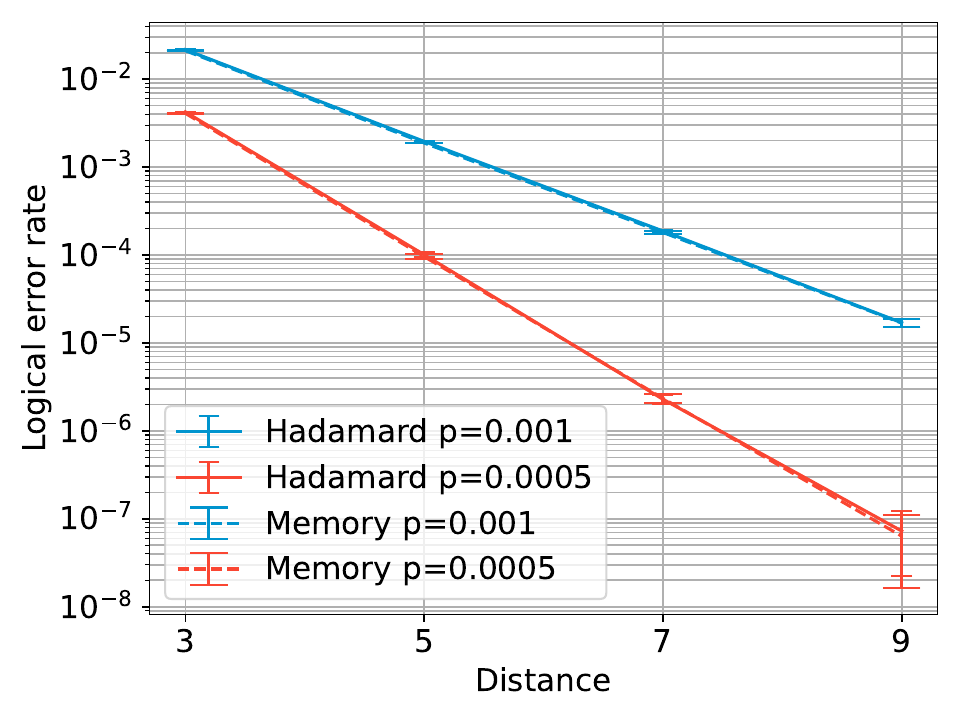}
        \caption{}
    \end{subfigure}
    \caption{Benchmarking a logical $\operatorname{H_0}\otimes \operatorname{H_1}$ gate on the CSS honeycomb Floquet code. (a)  We estimate a threshold of around $0.32\%$. (b) Sub-threshold analysis for physical error rates of $0.1\%$ and $0.05\%$ shows exponential error suppression. Dashed lines show subthreshold performance for a $(2d+1)$-round quantum memory experiment decoded with PyMatching (replotted from \Cref{fig:subth-memory-2d-plus-1-pymatching}). Error bars indicate a $95\%$ confidence interval.}
    \label{fig:hadamard}
\end{figure}

The threshold and sub-threshold performance for the logical $\operatorname{H_0}\otimes \operatorname{H_1}$ gate can be seen in \Cref{fig:hadamard}. We estimate a threshold of approximately $0.32\%$, similarly to the memory experiment. For a distance-7 code, we estimate logical error rates of approximately $2 \times 10^{-4}$ and $2 \times 10^{-6}$ for physical error rates of $0.1\%$ and $0.05\%$ when implementing a fold-transversal Hadamard-type logical gate, respectively. When we increase to a distance-9 code, we estimate logical error rates of approximately $2 \times 10^{-5}$ and $7 \times 10^{-8}$ for the same physical error rates. Of the logical gates we benchmark in this paper, the Hadamard is the one whose performance best matches the performance of a quantum memory experiment. This is understandable as the action of the fold-transversal Hadamard gate does not introduce correlated errors which can impact the logical state: if a logical qubit is prepared in the $\operatorname{X}$ (resp.\ $\operatorname{Z}$) basis it will be sensitive to only $\operatorname{Z}$ (resp.\ $\operatorname{X}$) errors before the logical Hadamard gate and only $\operatorname{X}$ (resp.\ $\operatorname{Z}$) errors after the logical Hadamard gate.

\subsection{Logical $\operatorname{S}$-type gate}
\label{ssec:s-gate-experiment}

We simulated a logical $\operatorname{S_0}\otimes \operatorname{S_1}$ gate on the CSS Floquet honeycomb code in order to assess the performance of the fold-transversal $\operatorname{S}$-type gate. As with the fold-transversal Hadamard-type gate, the total number of noisy QEC rounds for this circuit is $2d+1$, with $d$ rounds to prepare the initial noisy state, one round to implement the fold-transversal $S$-type gate, and $d$ rounds to mimic any further processing.

\begin{figure}
    \centering
    \begin{subfigure}{0.49\linewidth}
        \includegraphics[width=\linewidth]{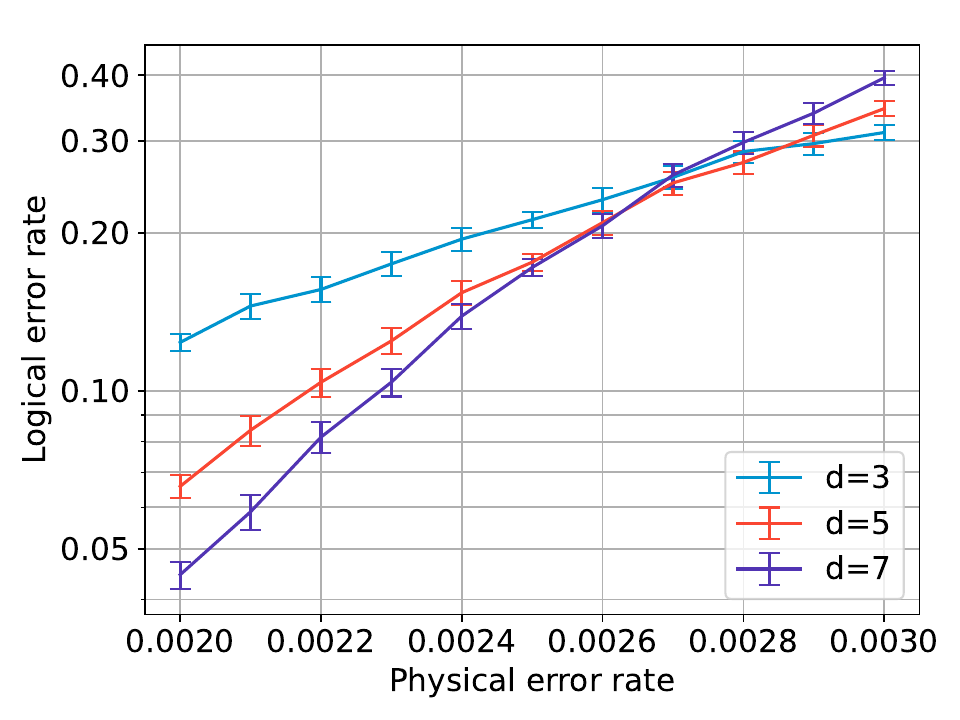}
        \caption{}
    \end{subfigure}
    \begin{subfigure}{0.49\linewidth}
        \includegraphics[width=\linewidth]{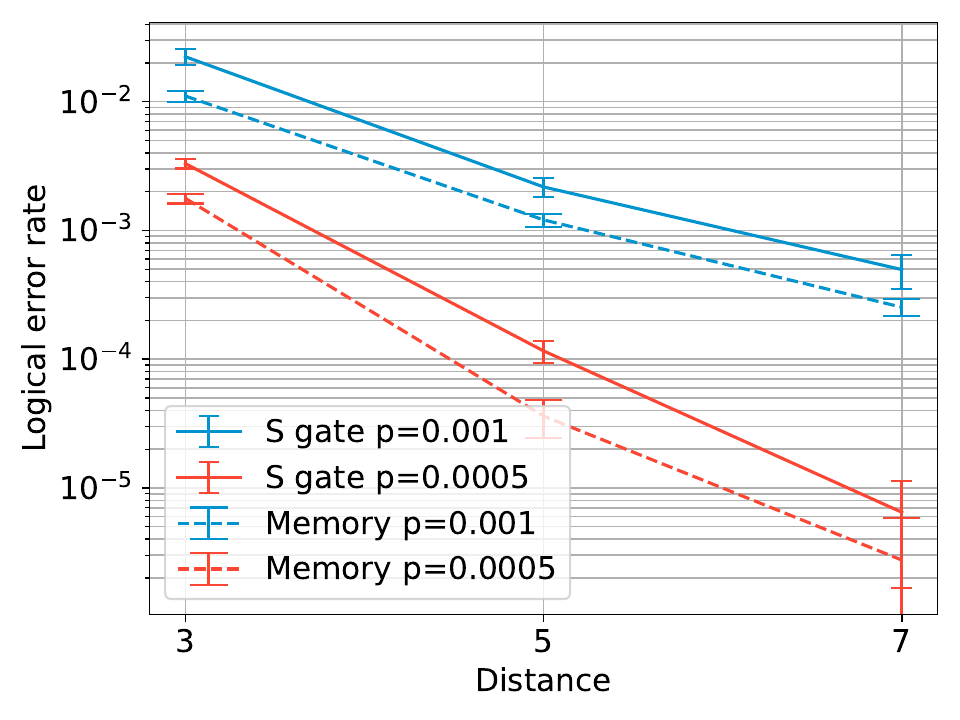}
        \caption{}
    \end{subfigure}
    \caption{Benchmarking a logical $\operatorname{S_0}\otimes \operatorname{S_1}$ gate on the CSS honeycomb Floquet code. (a)  We estimate a threshold of around $0.25$-$0.3\%$. (b) Sub-threshold analysis for physical error rates of $0.1\%$ and $0.05\%$. Dashed lines show subthreshold performance for a $(2d+1)$-round quantum memory experiment decoded with BP+LSD-0 (replotted from \Cref{fig:sub-threshold-memory-2d-plus-1-bp-lsd-subth-max-ler}). We attribute the flattening of the error suppression at higher distances to sub-optimality of zeroth-order BP+LSD-0, this is also seen in the memory simulations and in previous works \cite{Higgott2023DecodingHypergraphProductCodes, MagdalenadelaFuente2025XYZRubyCode}. Error bars indicate a $95\%$ confidence interval.}
    \label{fig:s-gate}
\end{figure}

The results of simulating the logical $\operatorname{S}$-type gate are presented in \Cref{fig:s-gate}. We estimate a threshold around $0.25$-$0.3\%$. For sub-threshold performance, we estimate that a fold-transversal $\operatorname{S}$-type gate on a distance-7 CSS honeycomb Floquet code achieves logical error rates of approximately $5 \times 10^{-4}$ and $7 \times 10^{-6}$ for physical error rates of $0.1\%$ and $0.05\%$, respectively. This is approximately twice the logical error rate seen for a quantum memory experiment, which can be explained from the fact that a logical qubit prepared in the $\operatorname{X}$ basis is mapped to the $\operatorname{Y}$ basis after the logical $\operatorname{S}$ gate is applied, which is sensitive to both logical $\operatorname{X}$ and logical $\operatorname{Z}$ errors. In comparison, a quantum memory experiment with a logical qubit prepared in the $\operatorname{X}$ (resp.\ $\operatorname{Z}$) basis will only be sensitive to logical $\operatorname{Z}$ (resp.\ $\operatorname{X}$) errors. Therefore the number of possible error strings after the logical gate is approximately double compared to the quantum memory experiment.

As with the memory simulations in \Cref{fig:memory-2d-plus-1-bp-lsd}, we can see a tail-off in the error suppression at higher distances. This is because of sub-optimal performance of zeroth-order BP+LSD-0, rather than a limitation of the logical $\operatorname{S}$-type gate \cite{Higgott2023DecodingHypergraphProductCodes, MagdalenadelaFuente2025XYZRubyCode}. In recent months there has been a significant interest in developing matching-based decoders for (fold-)transversal logical gates on static codes, with a particular interest in the rotated surface code \cite{Chen2024TransversalLogic, Wan2025CNOTDecoder, Cain2025FastCNOTDecoder, Serraperalta2025FastCNOTDecoder, Turner2025FastCNOTDecoder}. It is likely that these decoders can also be extended to dynamic codes, however we leave this task for future work.

\subsection{Linear-time Dehn twist simulations}
\label{ssec:linear-dehn-twist-experiment}

We simulated a logical $\operatorname{CNOT}$ gate implemented across $d$ QEC rounds on the CSS Floquet honeycomb code via linear-time horizontal and vertical Dehn twists. The total number of noisy QEC rounds for this circuit is $3d$: $d$ rounds to prepare an initial noisy logical state, $d$ rounds for fault-tolerantly implementing the Dehn twist, and $d$ rounds to mimic any further processing of the logical information. Dehn twists preserve graph-like errors, meaning that PyMatching can be used as a decoder \cite{Guernut2024ToricCliffordGates}.

\begin{figure}
    \centering
    \begin{subfigure}{0.49\linewidth}
        \includegraphics[width=\linewidth]{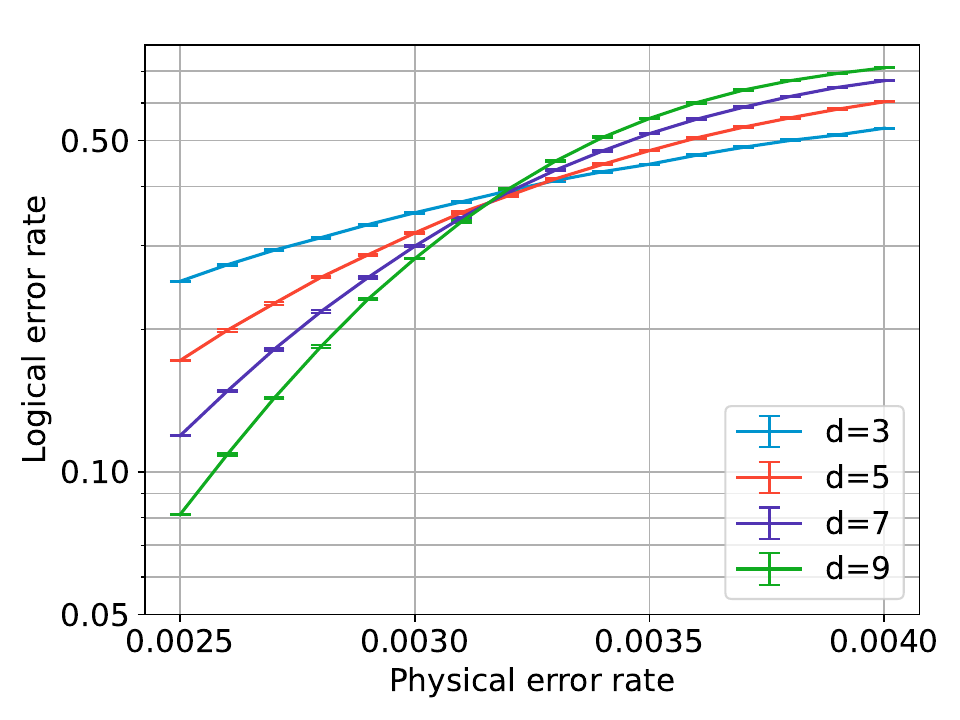}
        \caption{}
    \end{subfigure}
    \begin{subfigure}{0.49\linewidth}
        \includegraphics[width=\linewidth]{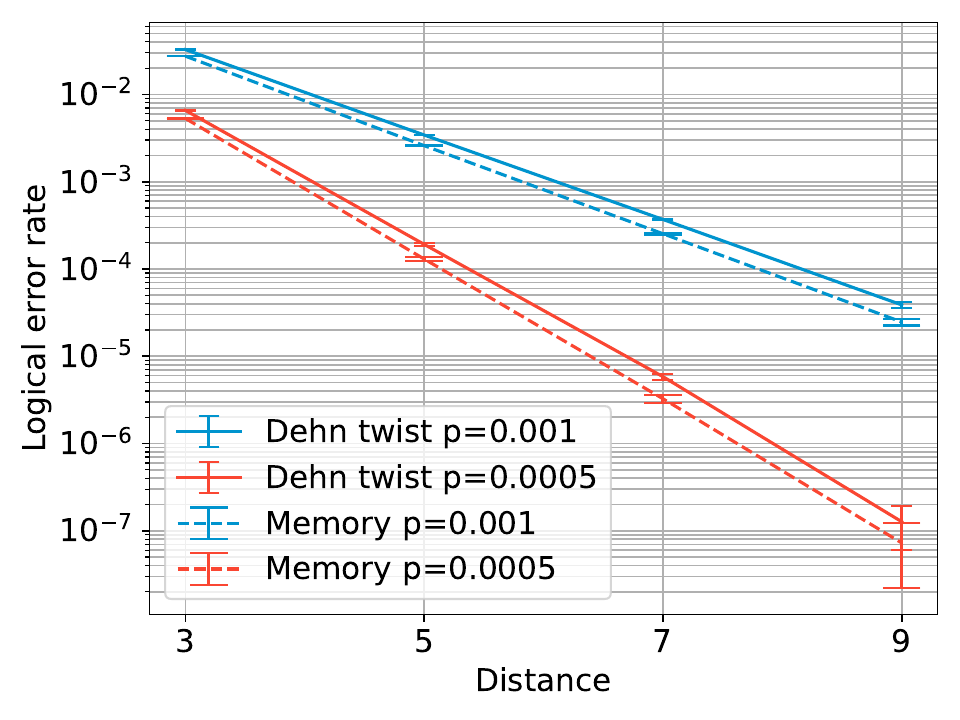}
        \caption{}
    \end{subfigure}
    \caption{Benchmarking a linear-time horizontal Dehn twist on the CSS honeycomb Floquet code. (a)  We estimate a threshold of around $0.32\%$. (b) Sub-threshold analysis for physical error rates of $0.1\%$ and $0.05\%$ show exponential error suppression. Dashed lines show subthreshold performance for a $3d$-round quantum memory experiment decoded with PyMatching (replotted from \Cref{fig:subth-memory-3d}). Error bars indicate a $95\%$ confidence interval.}
    \label{fig:linear-horizontal-dehn-twist}
\end{figure}

\begin{figure}
    \centering
    \begin{subfigure}{0.49\linewidth}
        \includegraphics[width=\linewidth]{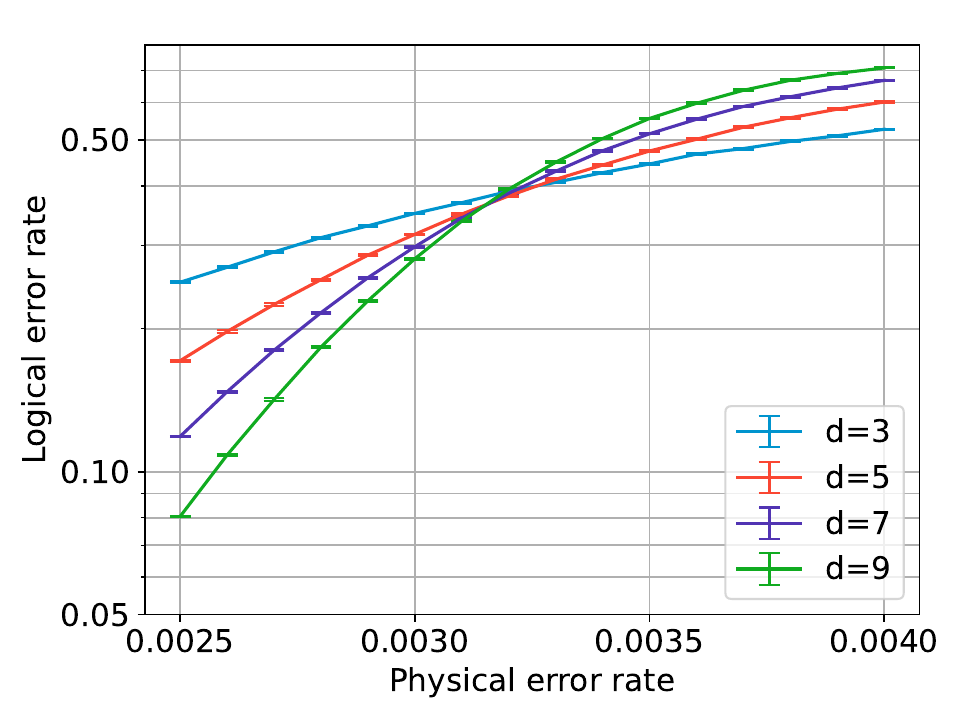}
        \caption{}
    \end{subfigure}
    \begin{subfigure}{0.49\linewidth}
        \includegraphics[width=\linewidth]{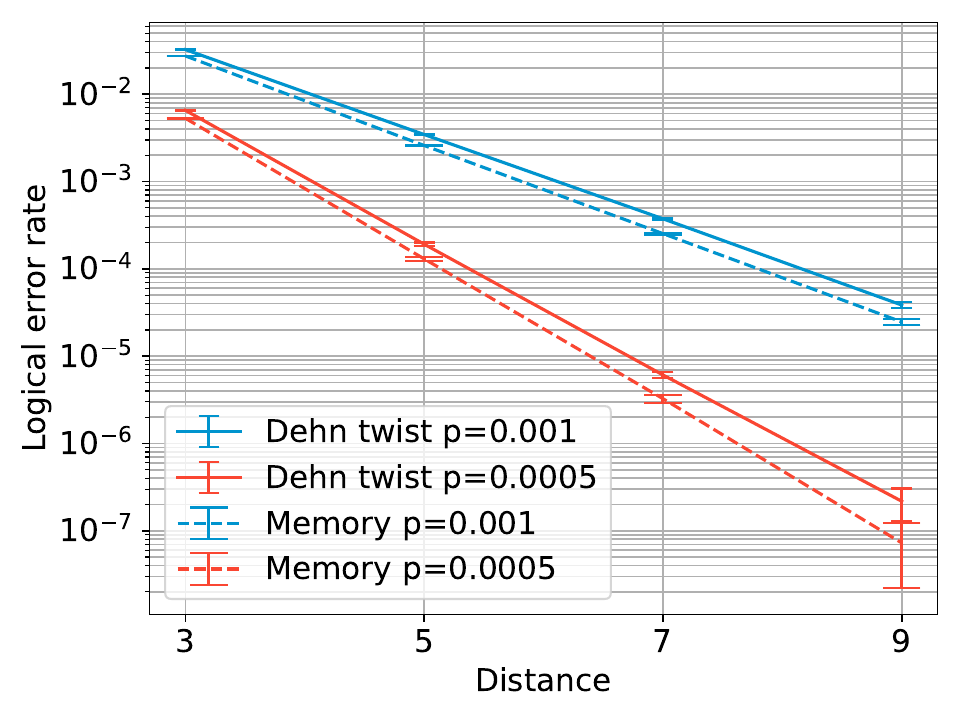}
        \caption{}
    \end{subfigure}
    \caption{Benchmarking a linear-time vertical Dehn twist on the CSS honeycomb Floquet code. (a)  We estimate a threshold of around $0.32\%$. (b) Sub-threshold analysis for physical error rates of $0.1\%$ and $0.05\%$ show exponential error suppression. Dashed lines show subthreshold performance for a $3d$-round quantum memory experiment decoded with PyMatching (replotted from \Cref{fig:subth-memory-3d}). Error bars indicate a $95\%$ confidence interval.}
    \label{fig:linear-vertical-dehn-twist}
\end{figure}

The performance for the horizontal and vertical Dehn twists is shown in Figures \ref{fig:linear-horizontal-dehn-twist} and \ref{fig:linear-vertical-dehn-twist}, respectively. In both cases we see similar results: thresholds of approximately $0.32\%$; logical error rates of approximately $4 \times 10^{-4}$ and $6 \times 10^{-6}$ for physical error rates of $0.1\%$ and $0.05\%$ on a distance-7 code; and logical error rates of approximately $4 \times 10^{-5}$ and $1 \times 10^{-7}$ for a distance-9 code. Note that similarly to the $\operatorname{S}$ gate, Dehn twists can introduce correlated logical errors: if logical qubit 0 is prepared in the $\operatorname{X}$ basis, then $\operatorname{CNOT}_{0, 1}$ will map the logical observable to $\operatorname{X}_0\otimes\operatorname{X}_1$, which is sensitive in both directions.

\subsection{Instantaneous Dehn twist simulations}
\label{ssec:instantaneous-dehn-twist-experiment}

Finally, we simulated logical $\operatorname{CNOT}$ gates implemented across a single QEC round on the CSS Floquet honeycomb code via instantaneous horizontal and vertical Dehn twists. The total number of noisy QEC rounds for this circuit is $2d + 1$: $d$ rounds to prepare an initial noisy logical state, one round for the instantaneous Dehn twist, and $d$ rounds to mimic any further processing of the logical information. As with the linear-time Dehn twists, instantaneous Dehn twists preserve graph-like errors, meaning that PyMatching can be used as a decoder \cite{Guernut2024ToricCliffordGates}.

One step that we have excluded from these simulations is the shifting operation to restore the code to the original honeycomb lattice. This is because the only qubit-exchange operation within Stim is the two-qubit $\operatorname{SWAP}$ gate \cite{Gidney2021Stim}, whereas the shuttling required for this circuit is a cyclic shift. It is always possible to decompose a permutation into a sequence of $\operatorname{SWAP}$ operations \cite{Clark1984ElementsAbstractAlgebra}. However, doing so without additional QEC rounds can halve the code distance, as previously discussed in \Cref{ssec:dehn-twist-static}. It is unclear how many additional QEC rounds are required in order to preserve fault tolerance. It would also not necessarily be a realistic depiction of how this circuit would be implemented on actual hardware: atomic qubits for instance can be physically moved to implement more general qubit permutations \cite{Bergou2021AtomicQubits}, thus removing the need for $\operatorname{SWAP}$ gates between qubits. We considered emulating such an operation, but it would then be unclear how to capture the noise from said qubit shuttling, as Stim only features single-qubit and two-qubit error models. For simplicity we therefore opted to omit this step from the simulation. This is consistent with simulations in the literature of instantaneous Dehn twist on static codes \cite{Guernut2024ToricCliffordGates}.

\begin{figure}
    \centering
    \begin{subfigure}{0.49\linewidth}
        \includegraphics[width=\linewidth]{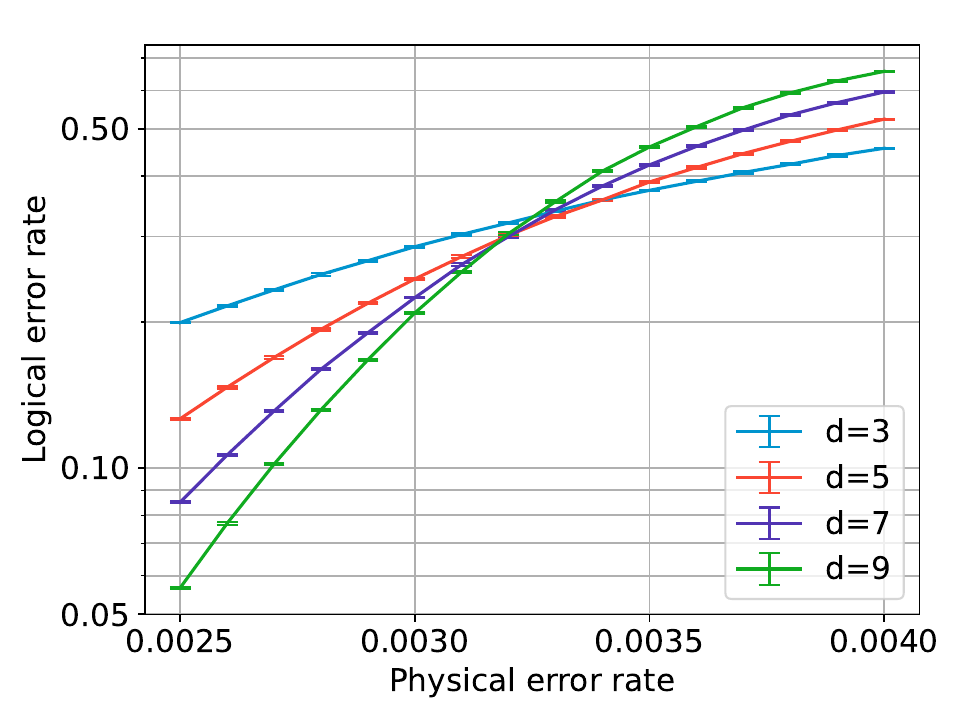}
        \caption{}
    \end{subfigure}
    \begin{subfigure}{0.49\linewidth}
        \includegraphics[width=\linewidth]{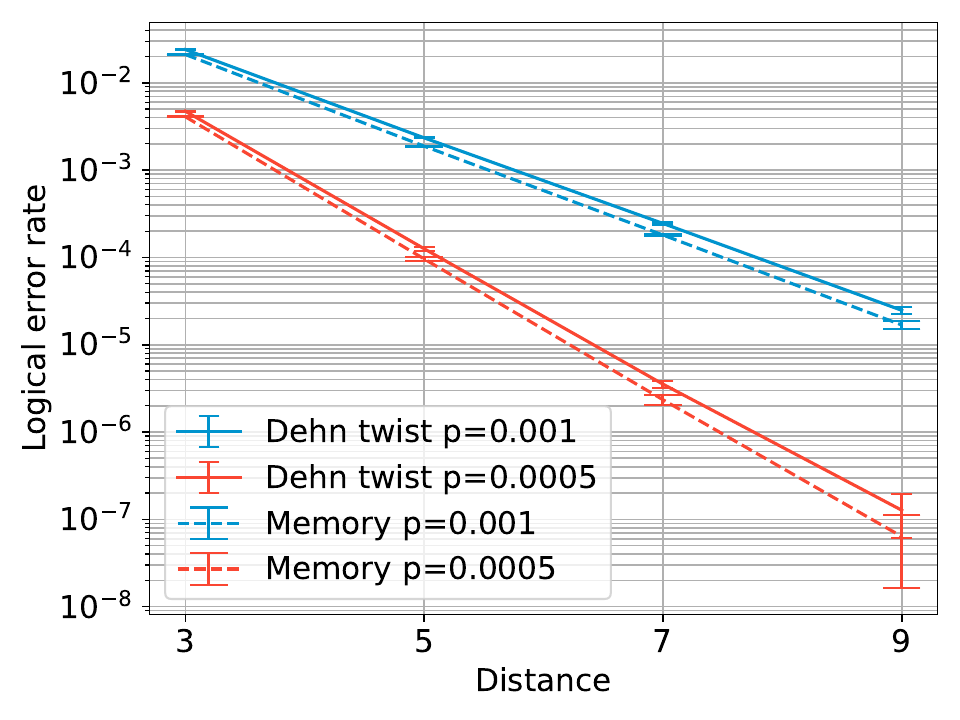}
        \caption{}
    \end{subfigure}
    \caption{Benchmarking an instantaneous horizontal Dehn twist on the CSS honeycomb Floquet code. (a) We estimate a threshold of around $0.32\%$. (b) Sub-threshold analysis for physical error rates of $0.1\%$ and $0.05\%$. Dashed lines show subthreshold performance for a $(2d+1)$-round quantum memory experiment decoded with PyMatching (replotted from \Cref{fig:subth-memory-2d-plus-1-pymatching}). Error bars indicate a $95\%$ confidence interval.}
    \label{fig:instantaneous-horizontal-dehn-twist}
\end{figure}

\begin{figure}
    \centering
    \begin{subfigure}{0.49\linewidth}
        \includegraphics[width=\linewidth]{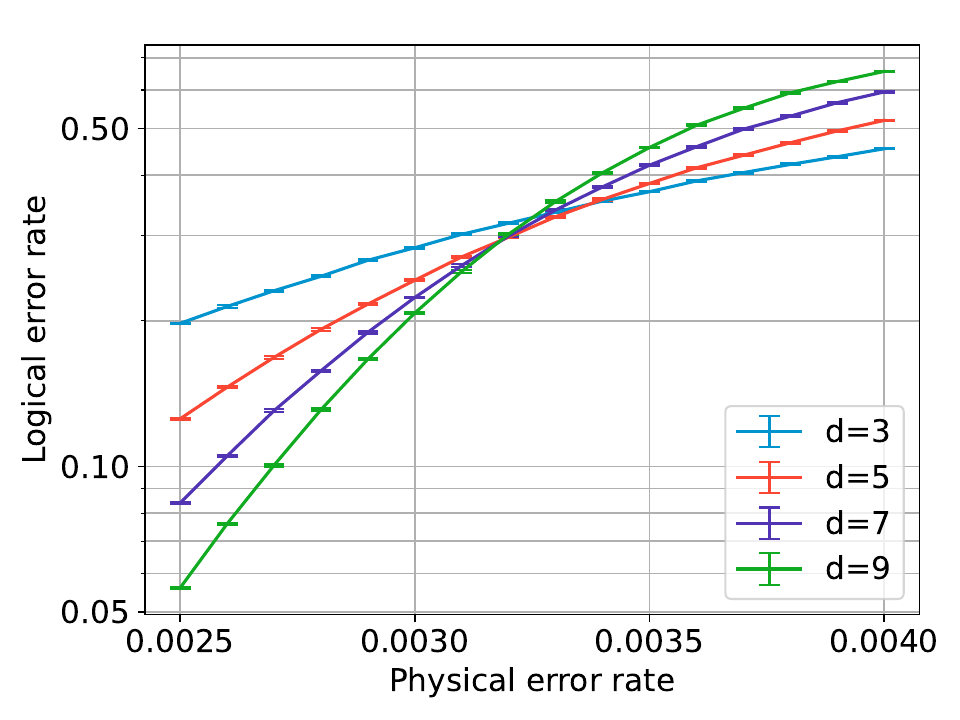}
        \caption{}
    \end{subfigure}
    \begin{subfigure}{0.49\linewidth}
        \includegraphics[width=\linewidth]{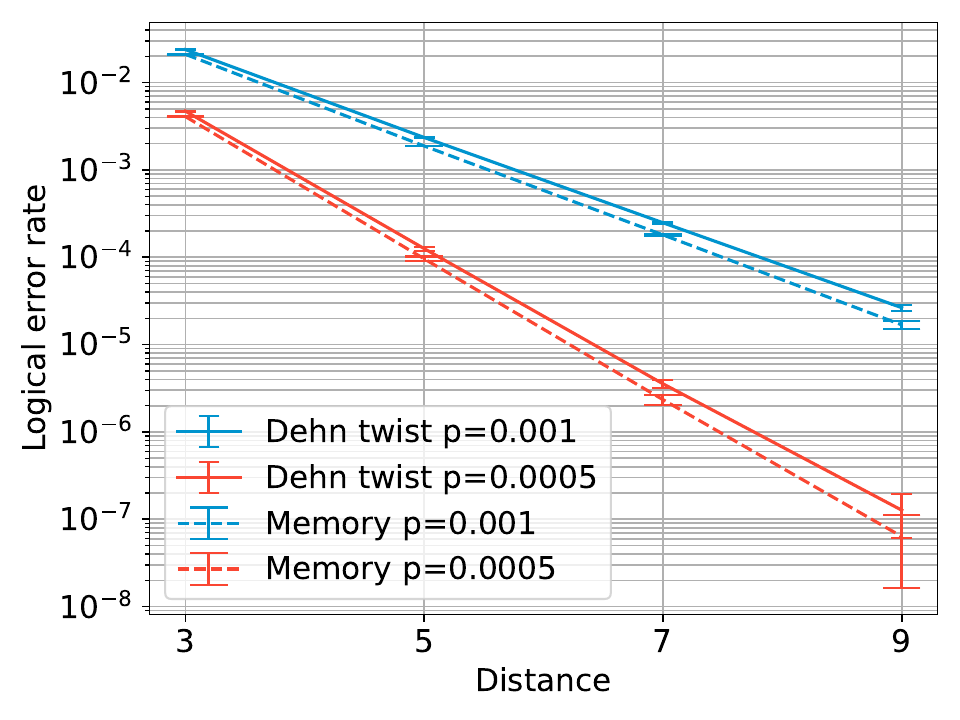}
        \caption{}
    \end{subfigure}
    \caption{Benchmarking an instantaneous vertical Dehn twist on the CSS honeycomb Floquet code. (a)  We estimate a threshold of around $0.32\%$. (b) Sub-threshold analysis for physical error rates of $0.1\%$ and $0.05\%$ show exponential error suppression. Dashed lines show subthreshold performance for a $(2d+1)$-round quantum memory experiment decoded with PyMatching (replotted from \Cref{fig:subth-memory-2d-plus-1-pymatching}). Error bars indicate a $95\%$ confidence interval.}
    \label{fig:instantaneous-vertical-dehn-twist}
\end{figure}

The performance for the horizontal and vertical Dehn twists is displayed in Figures \ref{fig:instantaneous-horizontal-dehn-twist} and \ref{fig:instantaneous-vertical-dehn-twist}, respectively. Once again we see similar performance for both the horizontal and vertical twists. In both cases we identify a threshold of approximately $0.32\%$. For sub-threshold performance, we identify logical error rates of approximately $3 \times 10^{-4}$ and $4 \times 10^{-6}$ for physical error rates of $0.1\%$ and $0.05\%$ on a distance-7 code in both cases. For a distance-9 code, we identify logical error rates of approximately $3 \times 10^{-5}$ and $1 \times 10^{-7}$ for a distance-9 code.

\section{Logical gates via embedded codes}
\label{sec:embedded-codes}

After each measurement sub-round, the Floquet code embeds an instantaneous stabiliser code \cite{Davydova2023CSSFloquet}. In this section, we discuss how to perform logical gates via these embedded codes using techniques discussed in \Cref{ssec:fold-transversal-static} and \Cref{ssec:dehn-twist-static}.

\begin{figure}
    \centering
    \begin{subfigure}{0.3\linewidth}
        \includegraphics[width=\linewidth]{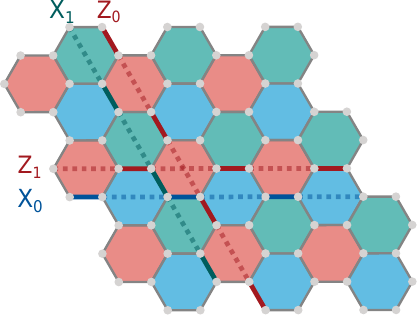}
        \caption{}
        \label{fig:honeycomb-embedding}
    \end{subfigure}
    \hfill
    \begin{subfigure}{0.3\linewidth}
        \includegraphics[width=\linewidth]{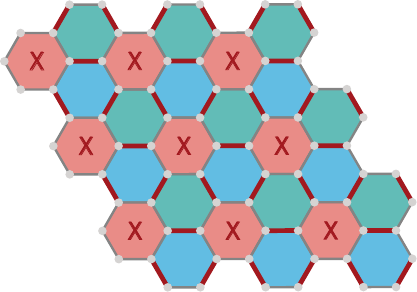}
        \caption{}
    \end{subfigure}
    \hfill
    \begin{subfigure}{0.3\linewidth}
        \includegraphics[width=\linewidth]{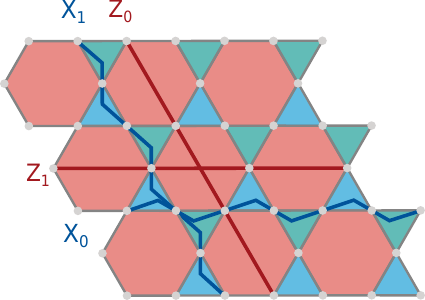}
        \caption{}
        \label{fig:honeycomb-embedded}
    \end{subfigure}
    \caption{Embedding in the honeycomb Floquet code. Honeycomb Floquet code, along with a choice of logicals, is shown in (a). Figure (b) depicts the Floquet code during $\color{floquet-red}\operatorname{rXX}$ sub-round. Each parity measurement, indicated as red edges, can be regarded as a $[[2,1,1]]$ repetition code.  Note that $\operatorname{X}_0$ can be combined with $\color{floquet-red}\operatorname{rXX}$ measurements to obtain an $\operatorname{X}_0$ passing through green plaquettes similar to $\operatorname{X}_1$, and consistent with \Cref{tab:floquet-css-schedule}. Figure (c) highlights the instantaneous toric code that embeds the Floquet code after the $\color{floquet-red}\operatorname{rXX}$ sub-round. The $[[2,1,1]]$ code serves as an effective single qubit displayed as vertices. The $\operatorname{X}$ stabilsers are located on red plaquettes, and $\operatorname{Z}$ stabilisers are located on blue and green plaquettes.}
    \label{fig:honeycomb-embedding-embedded}
\end{figure}

First, we consider the CSS honeycomb Floquet code. After each sub-round, the code embeds a toric code tessellated with hexagons and triangles. Each parity measurement between two qubits, which can be thought of as a $[[2,1,1]]$ repetition code, yields an effective single qubit that lies at the vertices of the instantaneous code. For example, after the $\color{floquet-red}\operatorname{rXX}$ sub-round, the instantaneous toric code is demonstrated in \Cref{fig:honeycomb-embedded} with $\operatorname{X}$ checks on red plaquettes and $\operatorname{Z}$ checks on green and blue plaquettes. The configuration of the embedded code changes after each measurement sub-round with a period of 6. The embedded code conserves logical information and possesses a threshold. Note that there is no ZX-duality in the instantaneous code due to $\operatorname{X}$ and $\operatorname{Z}$ checks having different weights. Hence, fold-transversal gates are not possible via the embedded code. Nevertheless, logical gates through Dehn twists are still possible.  The $\operatorname{CNOT}$ operations during the Dehn twist in \Cref{fig:honeycomb-embedded} can be perceived as transversal $\operatorname{CNOT}$ gates between two $[[2,1,1]]$ repetition codes, i.e., between qubits involved in the two-body parity measurements. Maintaining the 3-colourability of the code during the application of Dehn twists is essential for retrieving the Floquet code from its embedding. Both instantaneous and linear Dehn twists, as described in \Cref{app:toric_hex_triangle} for the toric code in \Cref{fig:honeycomb-embedded}, preserve this 3-colourability property. The linear Dehn twist shown in \Cref{fig:toric_hex_trianle_dehn_linear} maps $\operatorname{X}$ checks to $\operatorname{X}$ checks, $\operatorname{Z}$ checks to $\operatorname{Z}$ checks, and red, green, and blue plaquettes to their respective colours. The instantaneous Dehn twist illustrated in \Cref{fig:toric_hex_triangle_dehn_instantaneous} maps $\operatorname{X}$ checks on red plaquettes to the same check type on red plaquettes. Also, it maps $\operatorname{Z}$ checks on green plaquettes to $\operatorname{Z}$ checks on blue plaquettes, and vice versa. 

\begin{figure}
    \centering
    \begin{subfigure}{0.3\linewidth}
        \includegraphics[width=\linewidth]{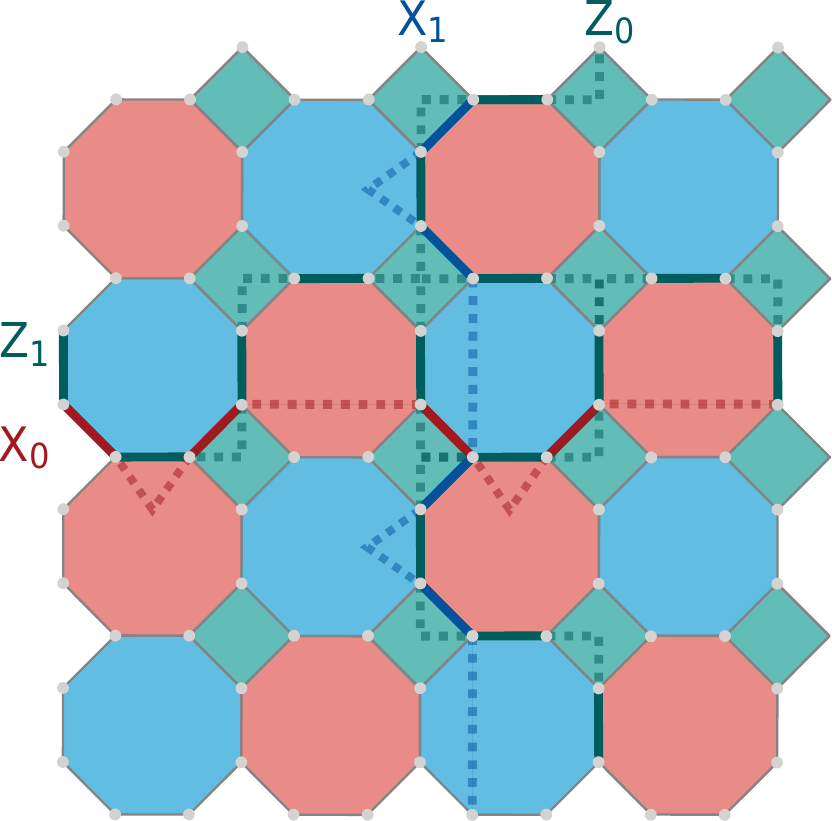}
        \caption{}
        \label{fig:unrotated-4-8-8-embedding}
    \end{subfigure}
    \hfill
    \begin{subfigure}{0.3\linewidth}
        \includegraphics[width=\linewidth]{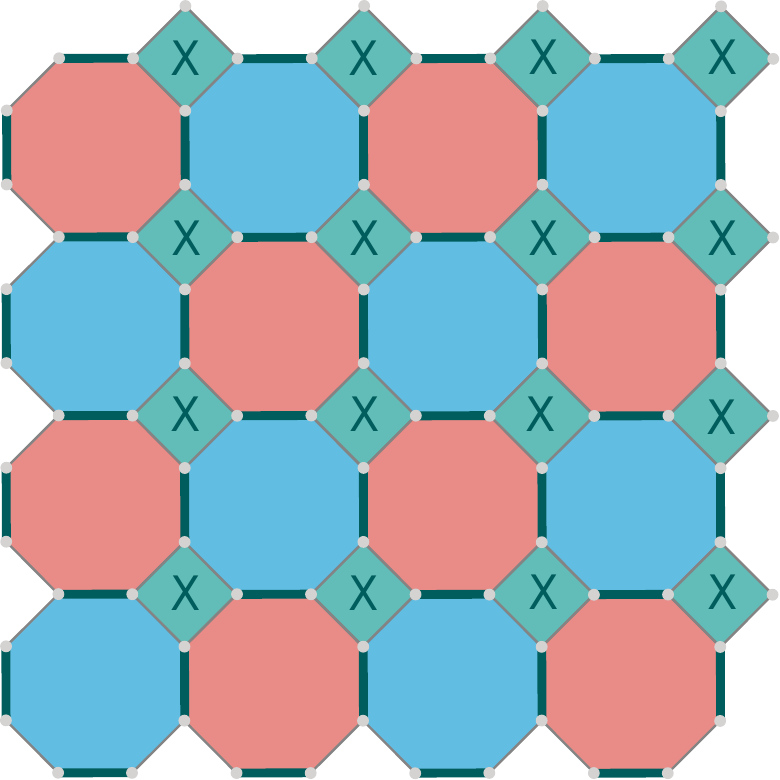}
        \caption{}
    \end{subfigure}
    \hfill
    \begin{subfigure}{0.3\linewidth}
        \includegraphics[width=\linewidth]{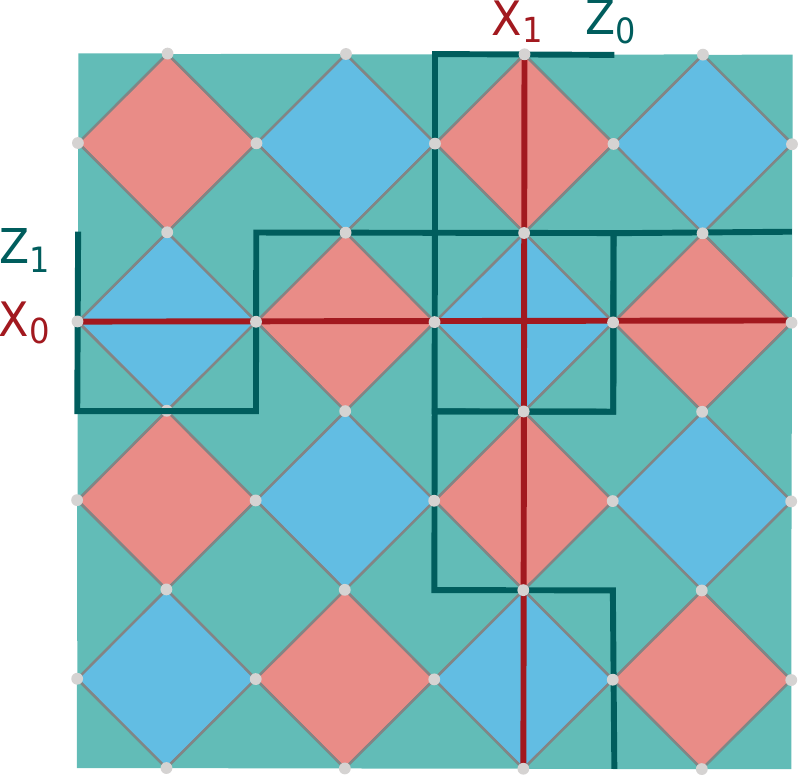}
        \caption{}
        \label{fig:unrotated-4-8-8-embedded}
    \end{subfigure}
    \caption{Embedding an unrotated toric code in the $4.8.8$ Floquet code. (a) Logical operators running along horizontal and vertical loops are pictured. (b) The Floquet code during $\color{floquet-green}\operatorname{gXX}$ sub-round. (c) The instantaneous unrotated toric code that embeds the Floquet code. Effective qubits are displayed as vertices. The $\operatorname{X}$ stabilsers are located on green plaquettes, and $\operatorname{Z}$ stabilisers are located on blue and red plaquettes. Note that red stabilisers can be combined with $\operatorname{Z}_0$ and $\operatorname{Z}_1$ to obtains the operators as shown in \Cref{fig:duality-toric-code}.}
    \label{fig:unrotated-4-8-8-embedding-embedded}
\end{figure}

Next, we consider the $4.8.8$ Floquet code. An instance of embedded code after the $\color{floquet-green}{\operatorname{gXX}}$ is presented in \Cref{fig:unrotated-4-8-8-embedded}. The instantaneous toric code comprises $\operatorname{X}$ checks on green plaquettes, and $\operatorname{Z}$ checks on red and blue plaquettes. Logical gates can be realised on the $4.8.8$ Floquet code via the embedded toric code. The embedded code possesses a ZX-duality, with the fold running from the bottom-left to top-right as shown in \Cref{fig:fold-surface-code}. Thus, fold-transversal Hadamard and $\operatorname{S}$ gates can be performed using the tools in \Cref{ssec:fold-transversal-static}. Note that a toric code in \Cref{fig:duality-toric-code} rotated by $45^\circ$ embeds the $4.8.8$ code after $\color{floquet-red}{\operatorname{rXX}}$ sub-round. The rotated code also demonstrates ZX-duality, utilizing an alternative selection of logical operators and fold relative to that depicted in \Cref{fig:fold-surface-code}. The Hadamard gates during the fold in the embedded code (\Cref{fig:unrotated-4-8-8-embedded}) do not have a corresponding transversal operation in the 4.8.8 Floquet code (\Cref{fig:unrotated-4-8-8-embedding}) since a transversal Hadamard gate on a repetition code is not available. Techniques such as gate teleportation \cite{Gottesman1999Demonstrating} or code switching \cite{Kubica2015, Butt2024} can be used at the expense of performance. Logical $\operatorname{CNOT}$ can also be implemented using linear or instantaneous Dehn twists discussed in \cref{ssec:dehn-twist-static}, both of which maintain the 3-colourable property. Similar to the honeycomb Floquet code's embedding, the linear Dehn twists map each check type on a coloured plaquette to the corresponding type and colour. While the instantaneous Dehn twist maps $\operatorname{X}$ checks to $\operatorname{X}$ checks on green plaquettes. $\operatorname{Z}$ checks on red plaquettes gets mapped to $\operatorname{Z}$ checks on blue plaquettes, and vice versa. 

As mentioned, stabilisers change during Dehn twists and fold-transversal gate implementations. Therefore, they should be appropriately compared when performing quantum error correction (QEC) rounds. Furthermore, the distance of the code during Dehn twists can be affected for reasons given in \Cref{ssec:dehn-twist-static}.

\section{Conclusion}
\label{sec:conclusion}

In this work we have explored possible techniques for implementing fault-tolerant logical gates on Floquet codes. In particular we have shown how two techniques originally designed for static codes can be extended to dynamic codes. Finally we have shown through Monte Carlo simulation that these novel techniques exhibit a threshold of $0.25$-$0.35\%$ on the CSS honeycomb Floquet code, comparable with a quantum memory implemented on the same code for the same number of QEC rounds. We also benchmarked the code under sub-threshold noise and showed that exponential error suppression can be achieved. The sub-threshold performance achieved when implementing these logical gates was slightly worse than the logical error rates achieved for quantum memory circuits, but the increase in logical error rate was shown to be negligible at higher code distances.

An important consideration with the approaches considered here is the additional connectivity requirements of the underlying quantum hardware. Both fold-transversal operations and Dehn twists require long-range connections. While long-range connections are non-trivial for a solid-state architecture, these techniques might work well on an atomic architecture where long-range operations can be implemented via qubit shuttling \cite{Bergou2021AtomicQubits}. Long-range operations can also be implemented on a distributed quantum computing architecture \cite{Sutcliffe2025DistributedQEC} through techniques such as gate teleportation \cite{Gottesman1999Demonstrating}.

One particular challenge when performing the simulations required for this work was decoding measurement results for the logical $\operatorname{S}$-type gate, which introduces non-graph-like errors. Recent works have investigated how to extend matching decoders to this problem on static codes, with a particular focus on the rotated surface code \cite{Chen2024TransversalLogic, Wan2025CNOTDecoder, Cain2025FastCNOTDecoder, Serraperalta2025FastCNOTDecoder, Turner2025FastCNOTDecoder, Guernut2024ToricCliffordGates}. We believe it is possible to extend these techniques to Floquet codes, but leave this for future work.

In this work we have only presented a limited set of gate operations. As we discuss in \Cref{app:rotated-4-8-8}, the planar rotated $4.8.8$ code has fold-transversal logical Hadamard and $\operatorname{S}$ gates, which are sufficient to implement the single-qubit Clifford group. When accompanied by a logical $\operatorname{CNOT}$ gate implemented through, for example, transversal operations, we can generate the full $k$-qubit Clifford group across $k$ instances of the planar $4.8.8$ code. But for other codes such as the honeycomb code on a torus, the operations presented here are not sufficient to generate the full Clifford group. In the static toric code, it is possible to implement single-qubit Hadamard and $\operatorname{S}$ gates, thus producing the full two-qubit Clifford group, by teleporting the logical qubit of interest onto an auxiliary code with the relevant logical gate \cite{Guernut2024ToricCliffordGates}. As Floquet codes embed static codes during each sub-round, this approach would also work here, but we leave the details for future work.

A more complicated question is what logical operations are possible on higher-rate Floquet codes, such as (semi)-hyperbolic Floquet codes \cite{Higgott2024HyperbolicFloquet, Fahimniya2025Hyperbolic, Sutcliffe2025DistributedQEC}. We believe these techniques can also be applied to (semi-)hyperbolic Floquet codes, but further work is required to ensure the stabilisers and logical strings have the necessary structure and preserve fault tolerance. Even if these techniques are applicable to these high-rate Floquet codes, generating a full set of logical Clifford gates on these codes would likely require more than the techniques detailed in this work. Possible extensions include code automorphisms \cite{Breuckmann2024FoldTransversal, Sayginel2025AutQEC, Malcolm2025SHYPS} and code surgery \cite{Horsman2012LatticeSurgery, Breuckmann2017HyperbolicSurfaceCodes, Baspin2025FastSurgeryQuantumLDPC}.

\section*{Code availability}

Example Stim circuits generated for the simulations presented in this paper are available in \cite{Moylett2025StimCircuits}.

\section*{Acknowledgements}

This work was supported by Innovate UK grant 10074653. We thank Evan Sutcliffe, Coral Westoby, Ed Wood, and Claire Le Gall, for providing us with helpful comments during the preparation of this manuscript. We thank our colleagues at Nu Quantum for insightful discussions, and thank Carmen Palacios-Berraquero and Claire Le Gall for providing an environment where this work was possible.

\appendix

\section{Fold-transversal $\sqrt{\operatorname{X}}$-type gates}
\label{app:sqrt-x-type-gate}

Here we will explain how to implement a logical $\sqrt{\operatorname{X}}$-type gate on a Floquet code. This operation is closely related to the $\operatorname{S}$-type gate discussed in \Cref{ssec:logical-s-type-gate}. We will focus on the period-three Floquet code schedule for this example, but we can note that the same technique also works on the period-six CSS Floquet code schedule.

Consider the following series of operations, which is equivalent up to single-qubit Clifford gates to the fold-transversal $\operatorname{S}$-type gate presented in \Cref{eq:s-type}:

\begin{equation}
    \bigotimes_{\substack{i=1\\i<\tau(i)}}^n\operatorname{XCX}_{i, \tau(i)}\bigotimes_{\substack{i=1\\i=\tau(i)}}^n\sqrt{\operatorname{X}_i}=
    \bigotimes_{i=1}^n\operatorname{H}_i\bigotimes_{\substack{i=1\\i<\tau(i)}}^n\operatorname{CZ}_{i, \tau(i)}\bigotimes_{\substack{i=1\\i=\tau(i)}}^n\operatorname{S}_i\bigotimes_{i=1}^n\operatorname{H}_i,
    \label{eq:sqrt-x-type}
\end{equation}

\noindent where $\operatorname{XCX}$ is a controlled Pauli gate with both the control and target qubits in the $\operatorname{X}$ basis.

We can see from \Cref{tab:floquet-sqrtx-gate-gidney-schedule} that after applying this gate sequence the instantaneous stabiliser group modifies to become the following:
\begin{itemize}
    \item $\operatorname{X}$ checks on the red plaquettes;
    \item $\operatorname{Z}$ checks on the blue plaquettes combined with $\operatorname{X}$ checks on the red plaquettes;
    \item $\operatorname{Y}$ checks on the green plaquettes combined with $\operatorname{X}$ checks on the green plaquettes; and
    \item $\operatorname{Z}$ checks on the blue edges combined with $\operatorname{X}$ checks on the red edges.
\end{itemize}

As before, the checks on red and blue plaquettes will be measured using the standard measurement schedule. Where this approach might offer an advantage however is that measuring the $\operatorname{X}$ check on the green plaquettes is now simpler, as this component will be measured in the proceeding $\color{floquet-red}\operatorname{rXX}$ sub-round. This might be beneficial as it would lead to smaller detecting regions.

The action of this operation on the logical operators of the honeycomb Floquet code is to leave $\operatorname{X}$ operators invariant, and map $\operatorname{Z}$ operators to $\operatorname{Y}$ operators, thus implementing a logical $\sqrt{\operatorname{X_0}}\otimes\sqrt{\operatorname{X_1}}$ gate. This operation is presented in \Cref{tab:floquet-sqrtx-gate-gidney-schedule}.

\begin{table}[]
    \centering
    \caption{Evolution of the stabilisers, detectors, and observables when implementing a fold-transversal $\sqrt{\operatorname{X}}$-type gate via $\operatorname{ZX}$-duality $\tau$ using the period-three schedule. The $\operatorname{Z}$ stabilisers on the blue plaquettes become a product with the $\operatorname{X}$ stabilisers on the red plaquettes. After one full QEC round, the $\operatorname{Z}$ stabilisers on the blue plaquettes return to their original form. The $\operatorname{Y}$ stabilisers on the green plaquettes become a product with $\operatorname{X}$ stabilisers on other green plaquettes. This can be immediately measured out during the $\color{floquet-red} \operatorname{rXX}$ sub-round.}
    \label{tab:floquet-sqrtx-gate-gidney-schedule}
    \begin{tabular}{|c|c|c|c|c|c|c|c|c|c|}
        \hline
        \multirow{2}{2.5em}{Pauli string} & \multicolumn{9}{|c|}{Sub-round}\\\cline{2-10}
        & $\color{floquet-blue}\operatorname{bZZ}$ & $\color{floquet-red}\operatorname{rXX}$ & $\color{floquet-green}\operatorname{gYY}$ & $\color{floquet-blue}\operatorname{bZZ}$ & $\sqrt{\operatorname{X}}$/$\operatorname{XCX}$ & $\color{floquet-red}\operatorname{rXX}$ & $\color{floquet-green}\operatorname{gYY}$ & $\color{floquet-blue}\operatorname{bZZ}$ & $\color{floquet-red}\operatorname{rXX}$ \\\hline
        \includegraphics[width=1.5em]{green-plaquette.pdf} & $\color{floquet-green}\operatorname{Y}_i$ & $\color{floquet-green}\operatorname{Y}_i$ & $\color{floquet-green}\operatorname{Y}_i$ & $\color{floquet-green}\operatorname{Y}_i$ & $\color{floquet-green}\operatorname{Y}_i\operatorname{X}_{\tau(i)}$ & $\color{floquet-green}\operatorname{Y}_i\operatorname{I}_{\tau(i)}$ & $\color{floquet-green}\operatorname{Y}_i$ & $\color{floquet-green}\operatorname{Y}_i$ & $\color{floquet-green}\operatorname{Y}_i$ \\\hline
        \includegraphics[width=1.5em]{blue-plaquette.pdf} & $\color{floquet-blue}\operatorname{Z}_i$ & $\color{floquet-blue}\operatorname{Z}_i$ & $\color{floquet-blue}\operatorname{Z}_i$ & $\color{floquet-blue}\operatorname{Z}_i$ & $\color{floquet-blue}\operatorname{Z}_i\color{floquet-red}\operatorname{X}_{\tau(i)}$ & $\color{floquet-blue}\operatorname{Z}_i\color{floquet-red}\operatorname{X}_{\tau(i)}$ & $\color{floquet-blue}\operatorname{Z}_i\color{floquet-red}\operatorname{Z}_{\tau(i)}$ & $\color{floquet-blue}\operatorname{Z}_i\color{floquet-red}\operatorname{I}_{\tau(i)}$ & $\color{floquet-blue}\operatorname{Z}_i$ \\\hline
        \includegraphics[width=1.5em]{red-plaquette.pdf} & $\color{floquet-red}\operatorname{X}_i$ &  $\color{floquet-red}\operatorname{X}_i$ & $\color{floquet-red}\operatorname{X}_i$ & $\color{floquet-red}\operatorname{X}_i$ & $\color{floquet-red}\operatorname{X}_i$ & $\color{floquet-red}\operatorname{X}_i$ & $\color{floquet-red}\operatorname{X}_i$ & $\color{floquet-red}\operatorname{X}_i$ & $\color{floquet-red}\operatorname{X}_i$ \\\hline
        \includegraphics[width=1.5em]{green-plaquette.pdf} & $\color{floquet-green}\operatorname{Z}_i$ & $\color{floquet-green}\operatorname{Y}_i$ & $\color{floquet-green}\operatorname{Y}_i$ & $\color{floquet-green}\operatorname{X}_i$ & $\color{floquet-green}\operatorname{X}_i$ & $\color{floquet-green}\operatorname{I}_i$ & & & \\\hline
        \includegraphics[width=1.5em]{blue-plaquette.pdf} & & $\color{floquet-blue}\operatorname{X}_i$ & $\color{floquet-blue}\operatorname{Z}_i$ & $\color{floquet-blue}\operatorname{Z}_i$ & $\color{floquet-blue}\operatorname{Z}_i\color{floquet-red}\operatorname{X}_{\tau(i)}$ & $\color{floquet-blue}\operatorname{Y}_i\color{floquet-red}\operatorname{X}_{\tau(i)}$ & $\color{floquet-blue}\operatorname{I}_i\color{floquet-red}\operatorname{Z}_{\tau(i)}$ & $\color{floquet-red}\operatorname{I}_{\tau(i)}$ & \\\hline
        \includegraphics[width=1.5em]{red-plaquette.pdf} & &  & $\color{floquet-red}\operatorname{Y}_i$ & $\color{floquet-red}\operatorname{X}_i$ & $\color{floquet-red}\operatorname{X}_i$ & $\color{floquet-red}\operatorname{X}_i$ & $\color{floquet-red}\operatorname{Z}_i$ & $\color{floquet-red}\operatorname{I}_i$ & \\\hline
        \includegraphics[width=1.5em]{green-plaquette.pdf} & & & & $\color{floquet-green}\operatorname{Z}_i$ & $\color{floquet-green}\operatorname{Z}_i\operatorname{X}_{\tau(i)}$ & $\color{floquet-green}\operatorname{Y}_i\operatorname{I}_{\tau(i)}$ & $\color{floquet-green}\operatorname{Y}_i$ & $\color{floquet-green}\operatorname{X}_i$ & $\color{floquet-green}\operatorname{I}_i$ \\\hline
        $\operatorname{X_L}$ & $\color{floquet-red}\operatorname{rZZ}$ & $\color{floquet-red}\operatorname{rYY}$ & $\color{floquet-blue}\operatorname{bYY}$ & $\color{floquet-blue}\operatorname{bXX}$ & $\color{floquet-blue}\operatorname{bXX}$ & $\color{floquet-green}\operatorname{gXX}$ & $\color{floquet-green}\operatorname{gZZ}$ & $\color{floquet-red}\operatorname{rZZ}$ & $\color{floquet-red}\operatorname{rYY}$ \\\hline
        \multirow{2}{1.5em}{$\operatorname{Z_L}$} & \multirow{2}{2.5em}{$\color{floquet-blue}\operatorname{bXX}$} & \multirow{2}{2.5em}{$\color{floquet-green}\operatorname{gXX}$} & \multirow{2}{2.5em}{$\color{floquet-green}\operatorname{gZZ}$} & \multirow{2}{2.5em}{$\color{floquet-red}\operatorname{rZZ}$} & $\color{floquet-red}\operatorname{rZZ}$ & $\color{floquet-red}\operatorname{rYY}$ & $\color{floquet-blue}\operatorname{bYY}$ & $\color{floquet-blue}\operatorname{bXX}$ & $\color{floquet-green}\operatorname{gXX}$ \\
        & & & & & $\color{floquet-blue}\operatorname{bXX}$ & $\color{floquet-green}\operatorname{gXX}$ & $\color{floquet-green}\operatorname{gZZ}$ & $\color{floquet-red}\operatorname{rZZ}$ & $\color{floquet-red}\operatorname{rYY}$ \\\hline
    \end{tabular}
\end{table}

Finally, we note that whilst in this appendix we have focused on the period-3 schedule, we can see that this gate sequence can also be applied to the period-6 CSS Floquet code schedule. The resulting evolution of the instantaneous stabiliser group and the logical observables can be seen in \Cref{tab:floquet-sqrtx-gate-css-schedule}.

\begin{table}[]
    \centering
    \caption{Evolution of the detecting regions and observables when implementing a logical $\sqrt{\operatorname{X}}$-type gate via $\operatorname{ZX}$-duality $\tau$ on a CSS Floquet code schedule. Note that $Z$ detecting regions now have a $\operatorname{X}$ component included.}
    \label{tab:floquet-sqrtx-gate-css-schedule}
    \begin{tabular}{|c|c|c|c|c|c|c|c|c|c|}
        \hline
        \multirow{2}{2.9em}{Pauli string} & \multicolumn{9}{|c|}{Sub-round}\\\cline{2-10}
        & $\color{floquet-blue}\operatorname{bXX}$ & $\color{floquet-red}\operatorname{rZZ}$ & $\color{floquet-green}\operatorname{gXX}$ & $\color{floquet-blue}\operatorname{bZZ}$ & $\sqrt{\operatorname{X}}$/$\operatorname{XCX}$ & $\color{floquet-red}\operatorname{rXX}$ & $\color{floquet-green}\operatorname{gZZ}$ & $\color{floquet-blue}\operatorname{bXX}$ & $\color{floquet-red}\operatorname{rZZ}$ \\\hline
        \includegraphics[width=1.5em]{green-plaquette.pdf} & $\color{floquet-green}\operatorname{X}_i$ & $\color{floquet-green}\operatorname{X}_i$ & $\color{floquet-green}\operatorname{X}_i$ & $\color{floquet-green}\operatorname{X}_i$ & $\color{floquet-green}\operatorname{X}_i$ & $\color{floquet-green}\operatorname{I}_i$ &  &  &  \\\hline
        \includegraphics[width=1.5em]{blue-plaquette.pdf} & & $\color{floquet-blue}\operatorname{Z}_i$ & $\color{floquet-blue}\operatorname{Z}_i$ & $\color{floquet-blue}\operatorname{Z}_i$ & $\color{floquet-blue}\operatorname{Z}_i\color{floquet-red}\operatorname{X}_{\tau(i)}$ & $\color{floquet-blue}\operatorname{Z}_i\color{floquet-red}\operatorname{X}_{\tau(i)}$ & $\color{floquet-blue}\operatorname{I}_i\color{floquet-red}\operatorname{X}_{\tau(i)}$ & $\color{floquet-red}\operatorname{I}_{\tau(i)}$ & \\\hline
        \includegraphics[width=1.5em]{red-plaquette.pdf} & &  & $\color{floquet-red}\operatorname{X}_i$ & $\color{floquet-red}\operatorname{X}_i$ & $\color{floquet-red}\operatorname{X}_i$ & $\color{floquet-red}\operatorname{X}_i$ & $\color{floquet-red}\operatorname{X}_i$ & $\color{floquet-red}\operatorname{I}_i$ & \\\hline
        \includegraphics[width=1.5em]{green-plaquette.pdf} & & & & $\color{floquet-green}\operatorname{Z}_i$ & $\color{floquet-green}\operatorname{Z}_i\operatorname{X}_{\tau(i)}$ & $\color{floquet-green}\operatorname{Z}_i\operatorname{I}_{\tau(i)}$ & $\color{floquet-green}\operatorname{Z}_i$ & $\color{floquet-green}\operatorname{Z}_i$ & $\color{floquet-green}\operatorname{I}_i$ \\\hline
        $\operatorname{X_L}$ & $\color{floquet-red}\operatorname{rXX}$ & $\color{floquet-red}\operatorname{rXX}$ & $\color{floquet-blue}\operatorname{bXX}$ & $\color{floquet-blue}\operatorname{bXX}$ & $\color{floquet-blue}\operatorname{bXX}$ & $\color{floquet-green}\operatorname{gXX}$ & $\color{floquet-green}\operatorname{gXX}$ & $\color{floquet-red}\operatorname{rXX}$ & $\color{floquet-red}\operatorname{rXX}$ \\\hline
        \multirow{2}{1.5em}{$\operatorname{Z_L}$} & \multirow{2}{2.5em}{$\color{floquet-blue}\operatorname{bZZ}$} & \multirow{2}{2.5em}{$\color{floquet-green}\operatorname{gZZ}$} & \multirow{2}{2.5em}{$\color{floquet-green}\operatorname{gZZ}$} & \multirow{2}{2.5em}{$\color{floquet-red}\operatorname{rZZ}$} & $\color{floquet-red}\operatorname{rZZ}$ & $\color{floquet-red}\operatorname{rZZ}$ & $\color{floquet-blue}\operatorname{bZZ}$ & $\color{floquet-blue}\operatorname{bZZ}$ & $\color{floquet-green}\operatorname{gZZ}$ \\
        & & & & & $\color{floquet-blue}\operatorname{bXX}$ & $\color{floquet-green}\operatorname{gXX}$ & $\color{floquet-green}\operatorname{gXX}$ & $\color{floquet-red}\operatorname{rXX}$ & $\color{floquet-red}\operatorname{rXX}$ \\\hline
    \end{tabular}
\end{table}

\section{Logical gates on the rotated $4.8.8$ code}
\label{app:rotated-4-8-8}

In this appendix we present an alternative Floquet code on the $4.8.8$ lattice. This code, which we will refer to as the ``rotated $4.8.8$'' code, provides a layout which is more amenable to the logical operations considered in the rest of the paper, at the cost of requiring additional qubits to maintain the same code distance as the original $4.8.8$ Floquet code. This is similar to the relationship between the rotated and unrotated surface codes.

\begin{figure}
    \centering
    \begin{subfigure}{0.35\linewidth}
        \includegraphics[width=\linewidth]{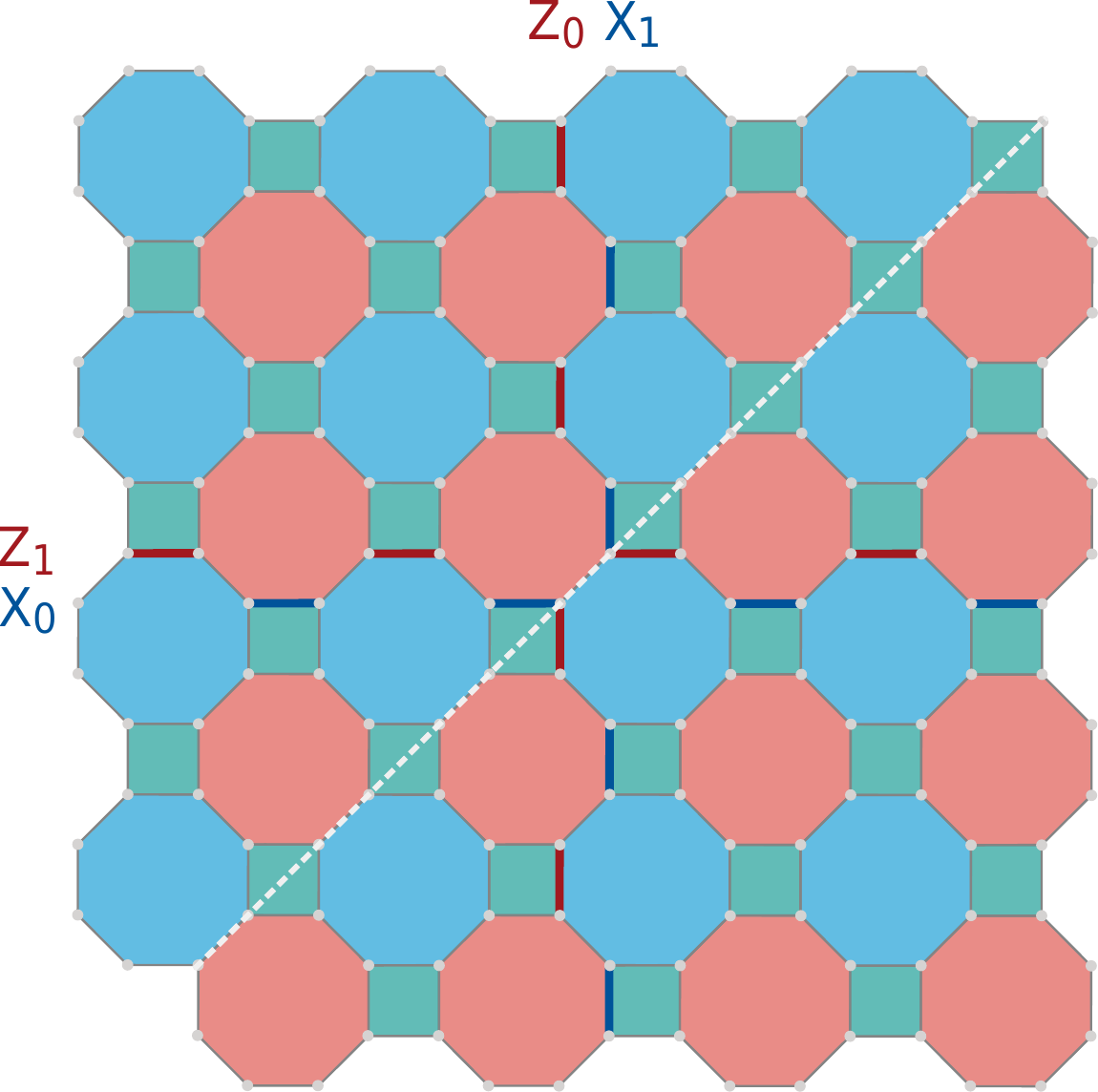}
        \caption{}
    \end{subfigure}
    \hspace{2cm}
    \begin{subfigure}{0.35\linewidth}
        \includegraphics[width=\linewidth]{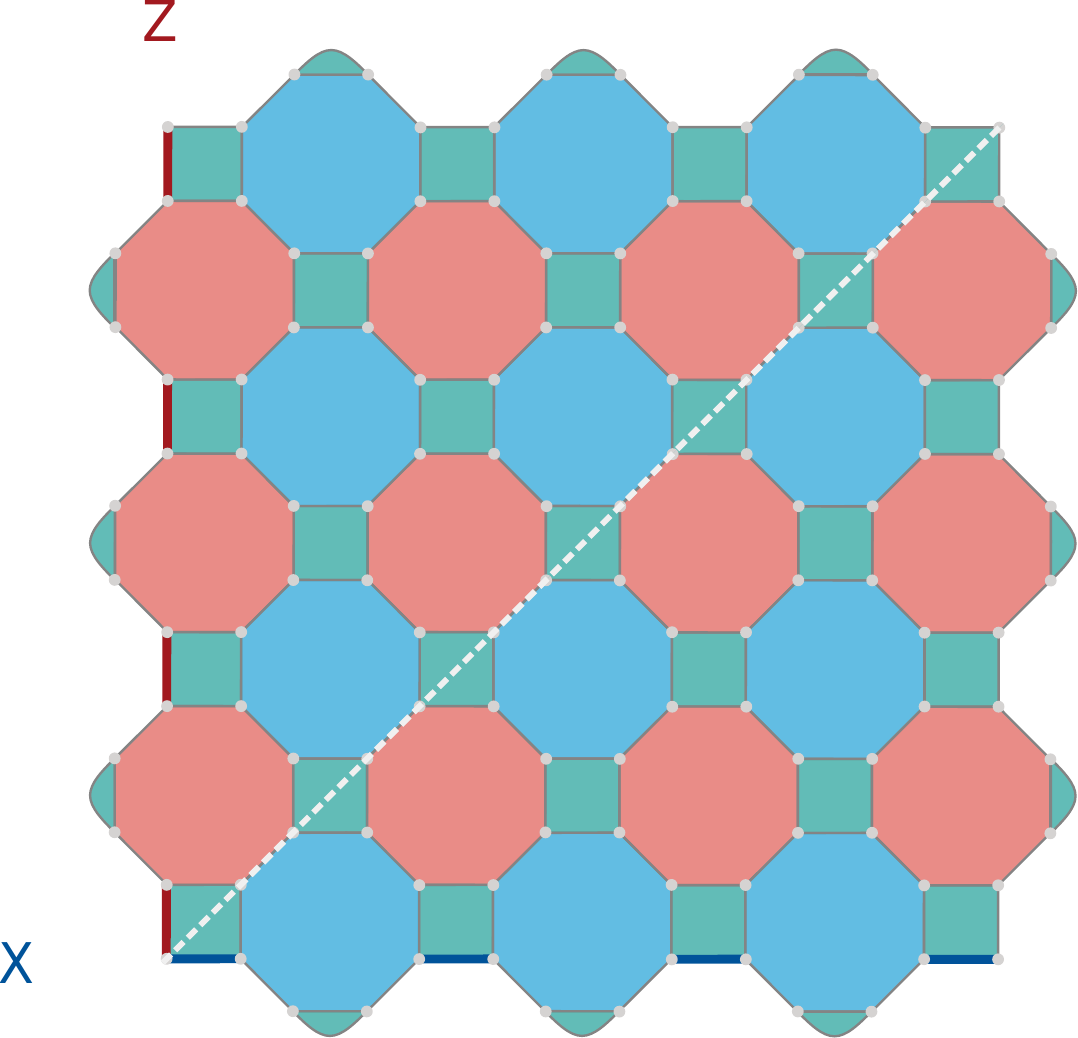}
        \caption{}
    \end{subfigure}
    \caption{The rotated $4.8.8$ Floquet code with both (a) periodic and (b) planar boundaries. A $ZX$-duality is shown as a white dashed line for both instances.}
    \label{fig:rotated-4-8-8}
\end{figure}

The rotated $4.8.8$ lattice is presented in \Cref{fig:rotated-4-8-8} for both periodic and planar boundaries. The bulk of this lattice is a $45^\circ$ rotation of the standard $4.8.8$ Floquet code lattice. The boundaries are modified but follow the same pattern of alternating between red/blue edges and green edges. The rotated lattice means that logical operators are shorter, hence more physical qubits are required to maintain the same code distance.

\begin{figure}
    \centering
    \begin{subfigure}{0.4\linewidth}
        \includegraphics[width=\linewidth]{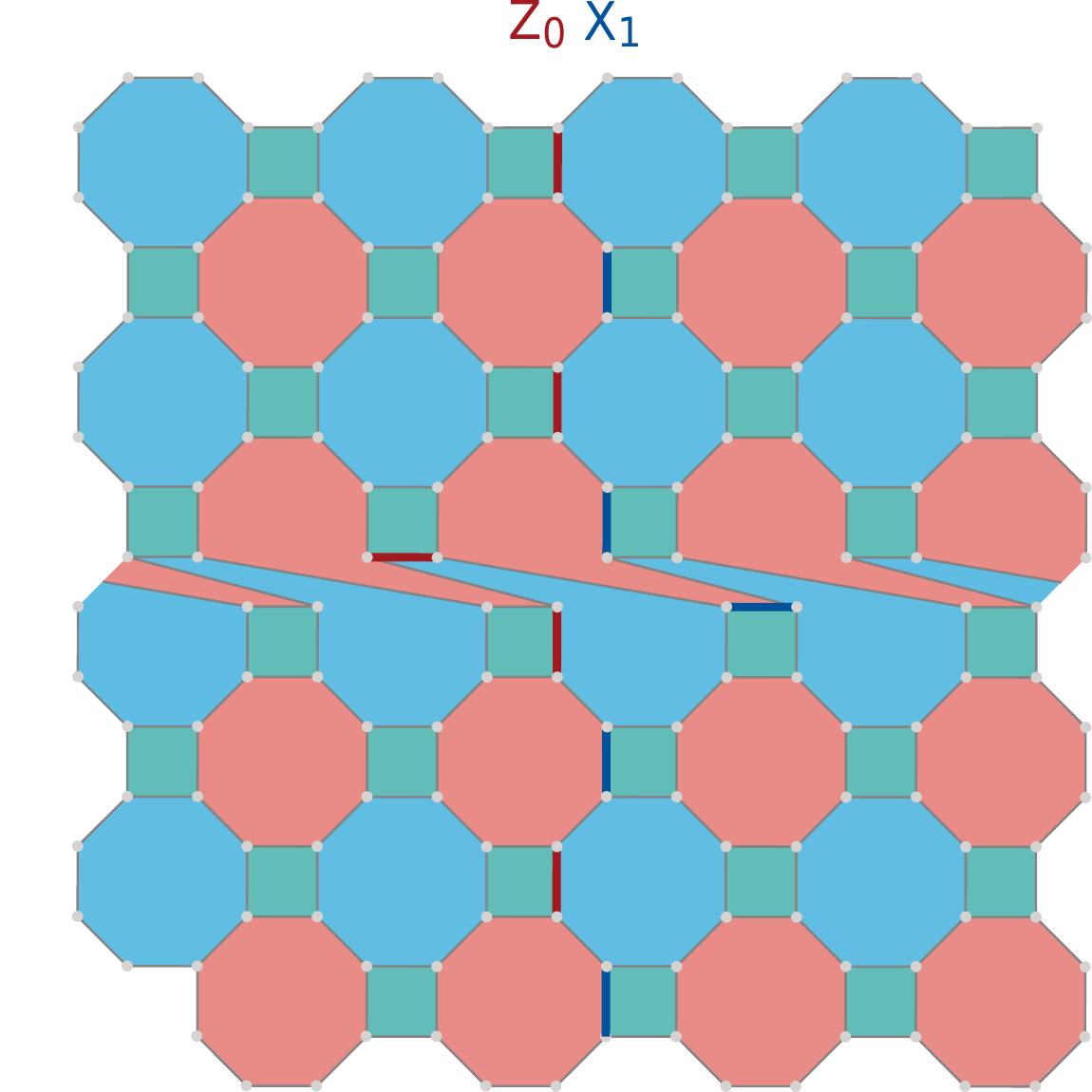}
        \caption{}
    \end{subfigure}
    \hfill
    \begin{subfigure}{0.4\linewidth}
        \includegraphics[width=\linewidth]{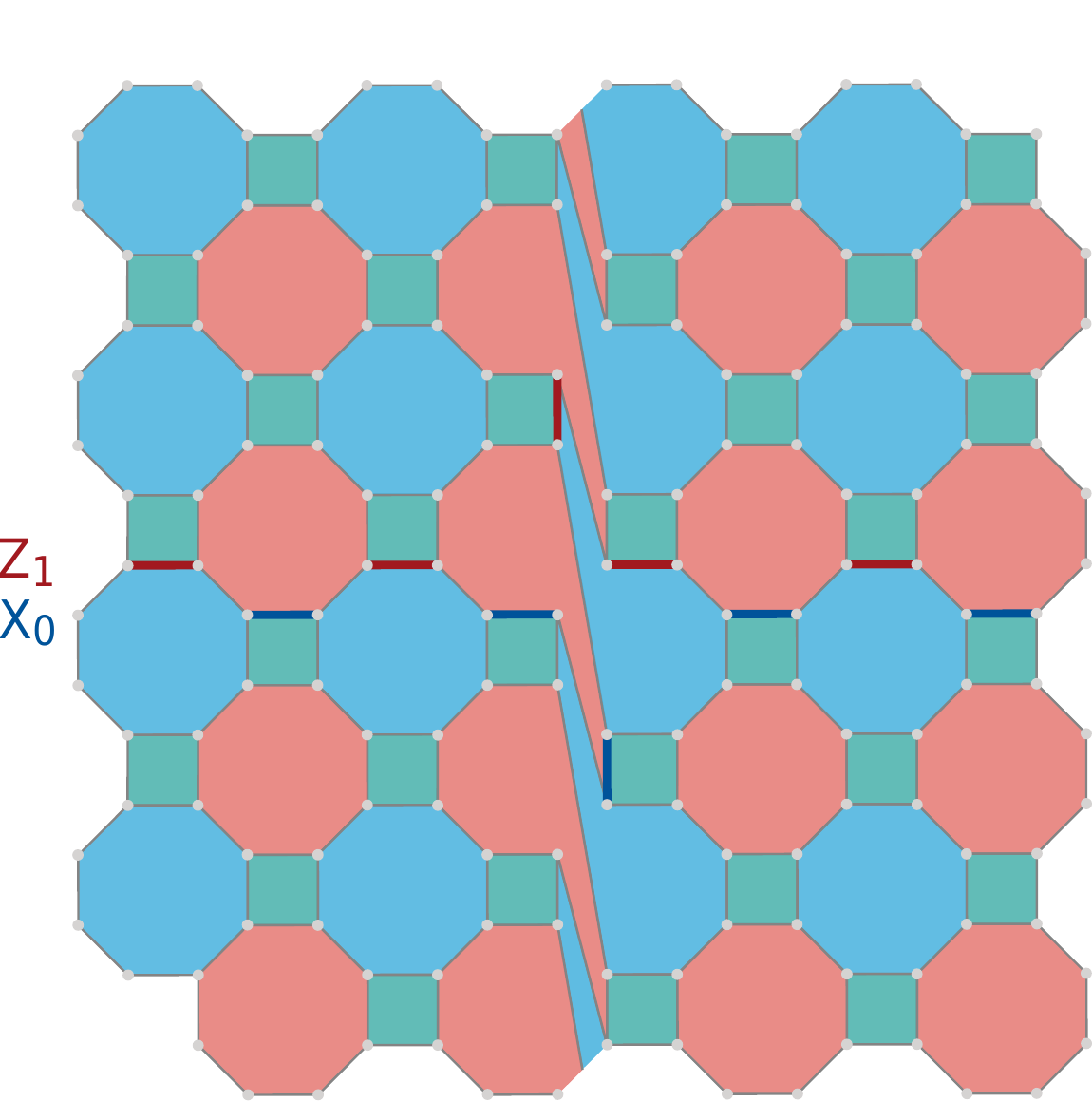}
        \caption{}
    \end{subfigure}
    \caption{First steps of (a) horizontal and (b) vertical Dehn twists on the rotated $4.8.8$ Floquet code, using the edge-swapping gadget described in \Cref{sec:dehn-twist-floquet}. Only those logical operators that change are highlighted.}
    \label{fig:rotated-4-8-8-dehn-twist}
\end{figure}

The benefit of this code compared to the standard $4.8.8$ code is that the structure is more amenable to logical gates. In \Cref{fig:rotated-4-8-8} we present the $\operatorname{ZX}$-duality for both the periodic and planar versions of this code, allowing for fold-transversal logical gates. And in \Cref{fig:rotated-4-8-8-dehn-twist}, we present the first steps of horizontal and vertical Dehn twists.

\begin{figure}
    \centering
    \begin{subfigure}{0.3\linewidth}
        \includegraphics[width=\linewidth]{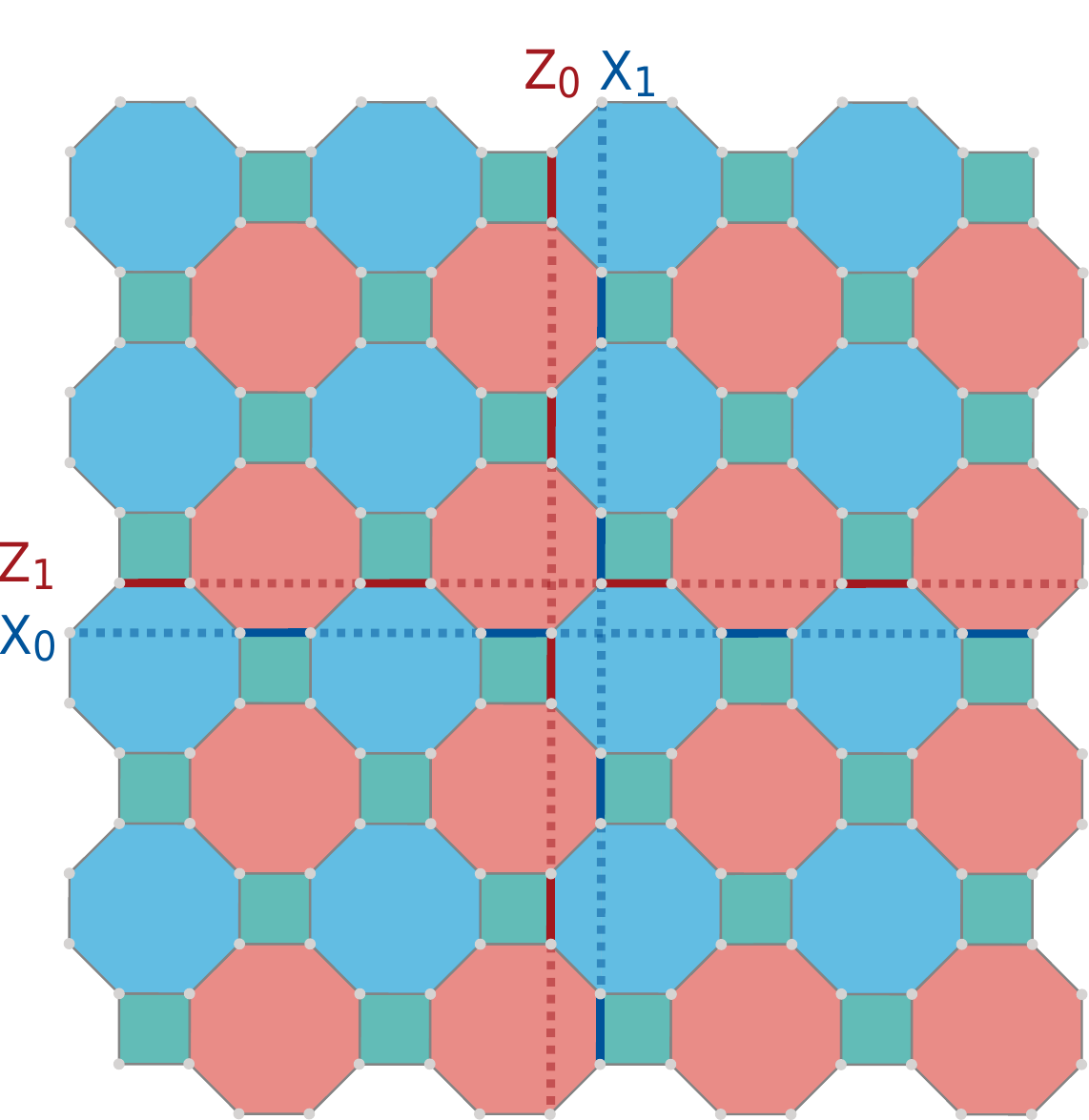}
        \caption{}
    \end{subfigure}
    \hfill
    \begin{subfigure}{0.3\linewidth}
        \includegraphics[width=\linewidth]{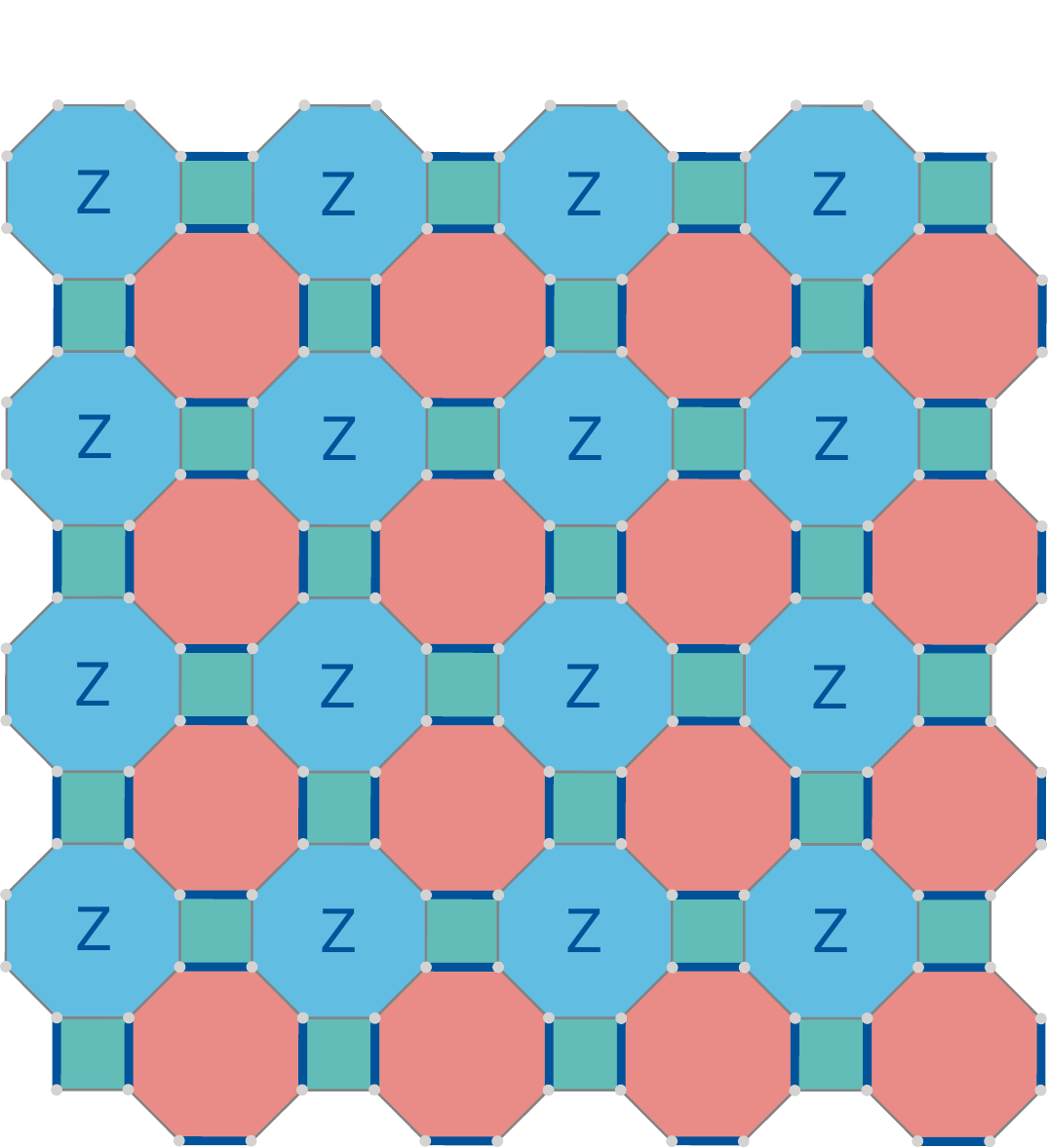}
        \caption{}
    \end{subfigure}
    \hfill
    \begin{subfigure}{0.3\linewidth}
        \includegraphics[width=\linewidth]{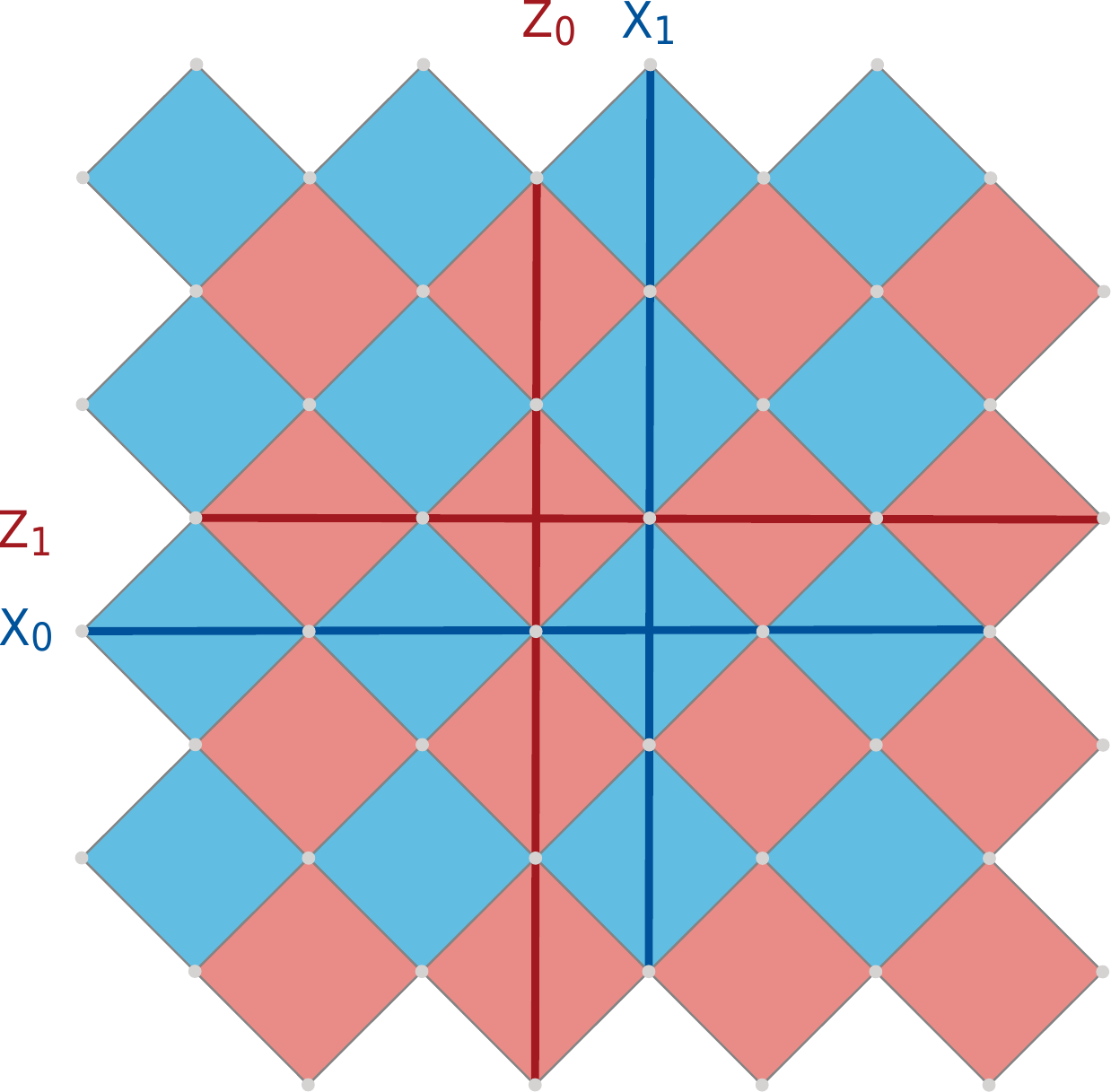}
        \caption{}
    \end{subfigure}
    \caption{Embedding an unrotated toric code in the rotated $4.8.8$ Floquet code. (a) A selection of logicals on the Floquet code. (b) Between the $\color{floquet-blue}\operatorname{bZZ}$ and $\color{floquet-red}\operatorname{rXX}$ sub-rounds, the $X$ stabilisers on the green plaquettes can be combined with the $\operatorname{Z}$ stabilisers on the blue edges (highlighted in blue) to form $[[4,1,2]]$ codes \cite{Davydova2023CSSFloquet}. (c) Treating these $[[4,1,2]]$ codes as a qubit produces a square lattice, which through periodicity of the boundaries is equivalent to \Cref{fig:duality-toric-code}  - the unrotated toric code as discussed in \Cref{sec:background}.}
    \label{fig:rotated-4-8-8-embedded}
\end{figure}

Finally we note that it is also possible to implement logical gates on this code via the embedded code. Between the $\color{floquet-blue}\operatorname{bZZ}$ and $\color{floquet-red}\operatorname{rXX}$ sub-rounds, the $\operatorname{X}$ stabilisers on the green plaquettes\footnote{For the Gidney \etal schedule where $\operatorname{Y}$ stabilisers are on the green plaquettes, the $\operatorname{Z}$ stabilisers on the blue edges can be multiplied in to produce $\operatorname{X}$ stabilisers on the green plaquettes.} can be combined with the $\operatorname{Z}$ stabilisers on the blue edges to create $[[4,1,2]]$ codes \cite{Davydova2023CSSFloquet}. The result when applied across the full $4.8.8$ code is an unrotated toric code, as we show in \Cref{fig:rotated-4-8-8-embedded}. We can then use the logical gates originally presented in \Cref{sec:background} to implement logical gates on the unrotated toric code, which in turn gives us logical gates on the rotated $4.8.8$ code.

\section{Logical gates on the $3.6.3.6$ toric code }
\begin{figure}
    \centering
    \begin{subfigure}{0.32\linewidth}
        \includegraphics[width=\linewidth]{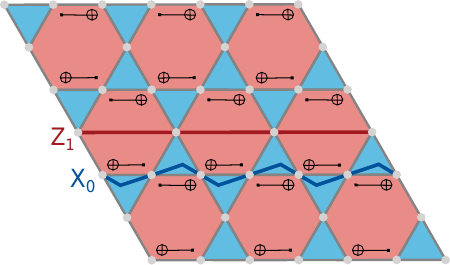}
        \caption{}
    \end{subfigure}
    \hfill
    \begin{subfigure}{0.32\linewidth}
        \includegraphics[width=\linewidth]{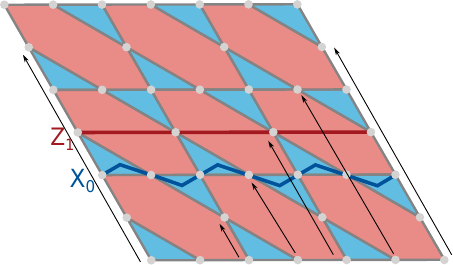}
        \caption{}
    \end{subfigure}
    \hfill
    \begin{subfigure}{0.32\linewidth}
        \includegraphics[width=\linewidth]{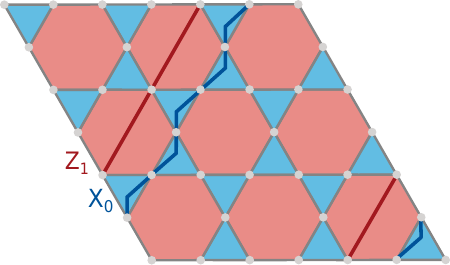}
        \caption{}
        \label{fig: torus-3-3-6-6-instantaneous-dehn}
    \end{subfigure}
    \caption{Instantaneous vertical Dehn twist on a $3.6.3.6$ toric code implementing $\operatorname{CNOT_{0,1}}$. (a) Logical gates are identified, along with the $\operatorname{CNOT}$ gates to deform the lattice. (b) Distorted lattice along with a representative of long-range cyclic shift operators to restore the lattice. (c) Recovered lattice indicating logical $\operatorname{CNOT_{0,1}}$ gate.}
    \label{fig:toric_hex_triangle_dehn_instantaneous}
\end{figure}

\begin{figure}
     \centering
     \begin{subfigure}[b]{0.3\textwidth}
         \centering
         \includegraphics[width=\textwidth]{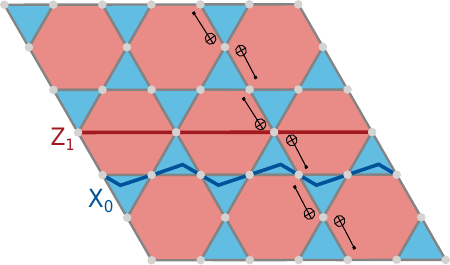}
         \caption{}
     \end{subfigure}
     \hfill
     \begin{subfigure}[b]{0.3\textwidth}
         \centering
         \includegraphics[width=\textwidth]{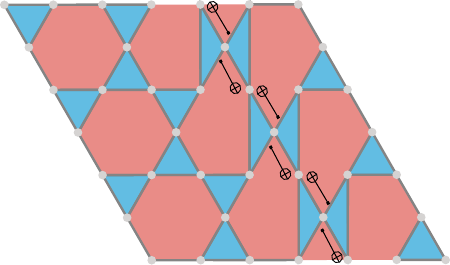}
         \caption{}
     \end{subfigure}
     \hfill
     \begin{subfigure}[b]{0.3\textwidth}
         \centering
         \includegraphics[width=\textwidth]{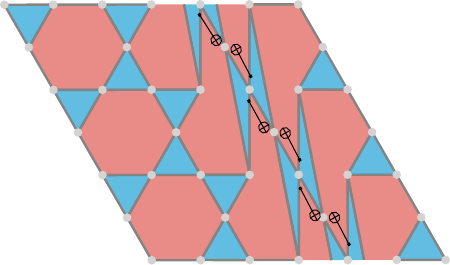}
         \caption{}
     \end{subfigure}
    \vspace{1em}
     \begin{subfigure}[b]{0.3\textwidth}
         \centering
         \includegraphics[width=\textwidth]{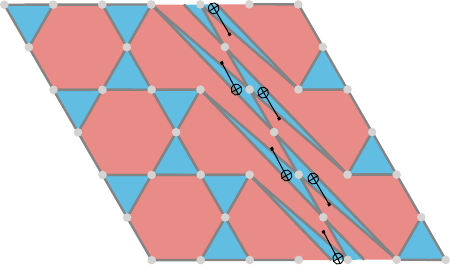}
         \caption{}
     \end{subfigure}
     \hfill
     \begin{subfigure}[b]{0.3\textwidth}
         \centering
         \includegraphics[width=\textwidth]{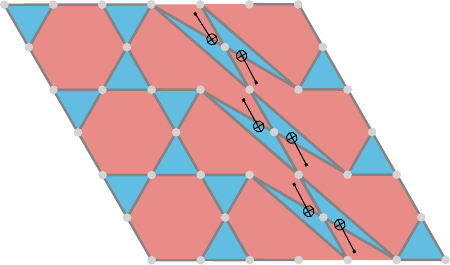}
         \caption{}
     \end{subfigure}
     \hfill
     \begin{subfigure}[b]{0.3\textwidth}
         \centering
         \includegraphics[width=\textwidth]{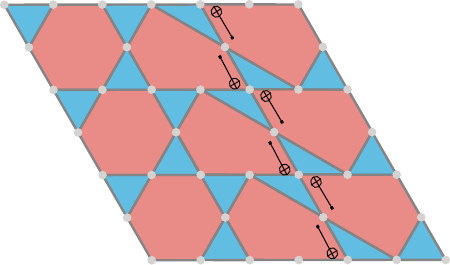}
         \caption{}
     \end{subfigure}
     \vspace{1em}
     \begin{subfigure}[b]{0.3\textwidth}
         \centering
         \includegraphics[width=\textwidth]{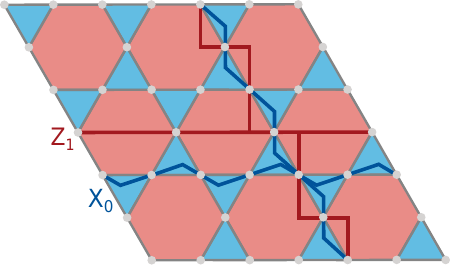}
         \caption{}
         \label{fig: torus-3-3-6-6-linear-dehn}
     \end{subfigure}
        \caption{Linear vertical Dehn twist implementing logical $\operatorname{CNOT}_{0,1}$. (a) Logical gates are identified, along with the $\operatorname{CNOT}$ gates to deform the lattice. (b)--(f) Lattice distortions during the Dehn twist. (g) Recovered lattice along with implementation of logical $\operatorname{CNOT}_{0,1}$. Note that the logical operators $\operatorname{X}_0$ and $\operatorname{Z}_1$ fully acquire the vertical components in (d). But we perform further deformations to restore the lattice to the original.}
        \label{fig:toric_hex_trianle_dehn_linear}
\end{figure}

In this section, we emphasise Dehn twists on a toric code that embeds the honeycomb Floquet code. The underlying lattice consists of hexagons and triangles. Each vertex is surrounded by two triangles and two hexagons, hence the name $3.6.3.6$ lattice. Similar to the unrotated toric code in \Cref{sec:background}, qubits lie at the vertices, $\operatorname{X}$ and $\operatorname{Z}$ stabilisers lie on red and blue plaquettes, respectively. The instantaneous and linear Dehn twist implementations to perform logical $\operatorname{CNOT}_{0,1}$ are demonstrated in \Cref{fig:toric_hex_triangle_dehn_instantaneous} and \Cref{fig: torus-3-3-6-6-linear-dehn}, respectively.

\label{app:toric_hex_triangle}

\section{Full numerical results}
\label{app:full-numerics}

\begin{figure}
    \centering
    \begin{subfigure}{0.49\linewidth}
        \includegraphics[width=\linewidth]{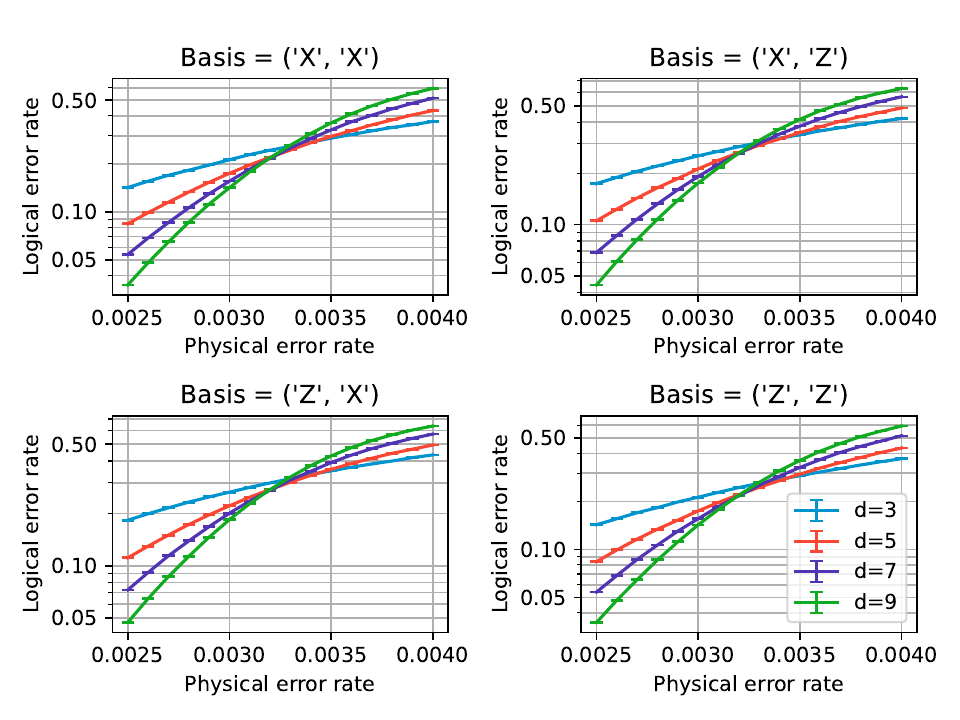}
        \caption{}
    \end{subfigure}
    \begin{subfigure}{0.49\linewidth}
        \includegraphics[width=\linewidth]{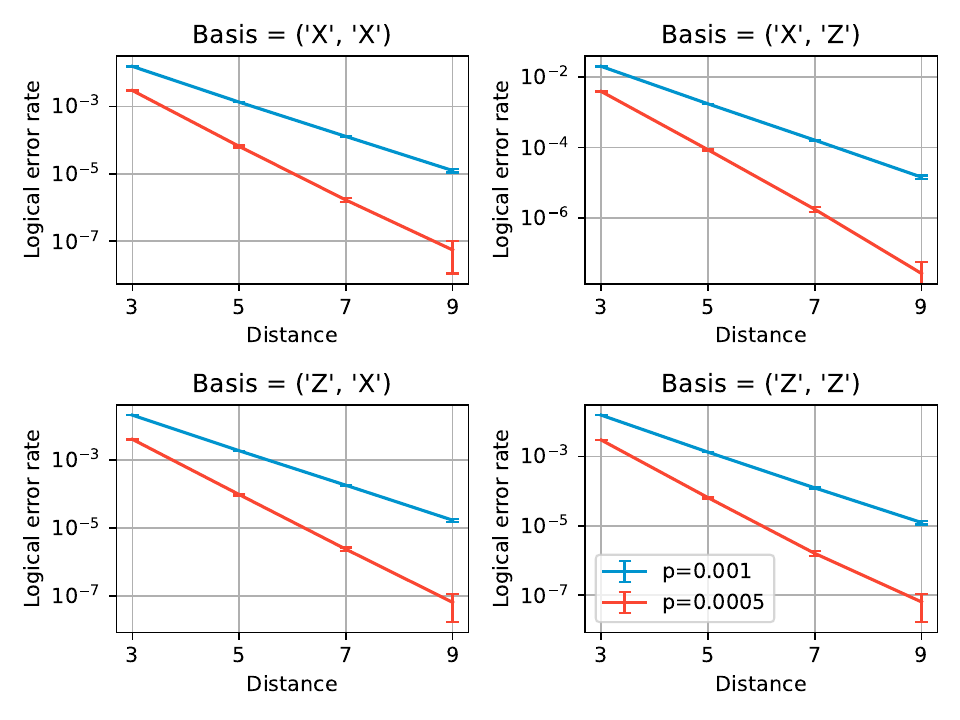}
        \caption{}
    \end{subfigure}
    \caption{Benchmarking a $(2d+1)$-round quantum memory on the CSS honeycomb Floquet code when decoded with PyMatching \cite{Higgott2025SparseBlossom}. (a) We estimate a threshold of around $0.32\%$. (b) Sub-threshold analysis for physical error rates of $0.1\%$ and $0.05\%$ show exponential error suppression.}
    \label{fig:memory-2d-plus-1-pymatching-full}
\end{figure}

\begin{figure}
    \centering
    \begin{subfigure}{0.49\linewidth}
        \includegraphics[width=\linewidth]{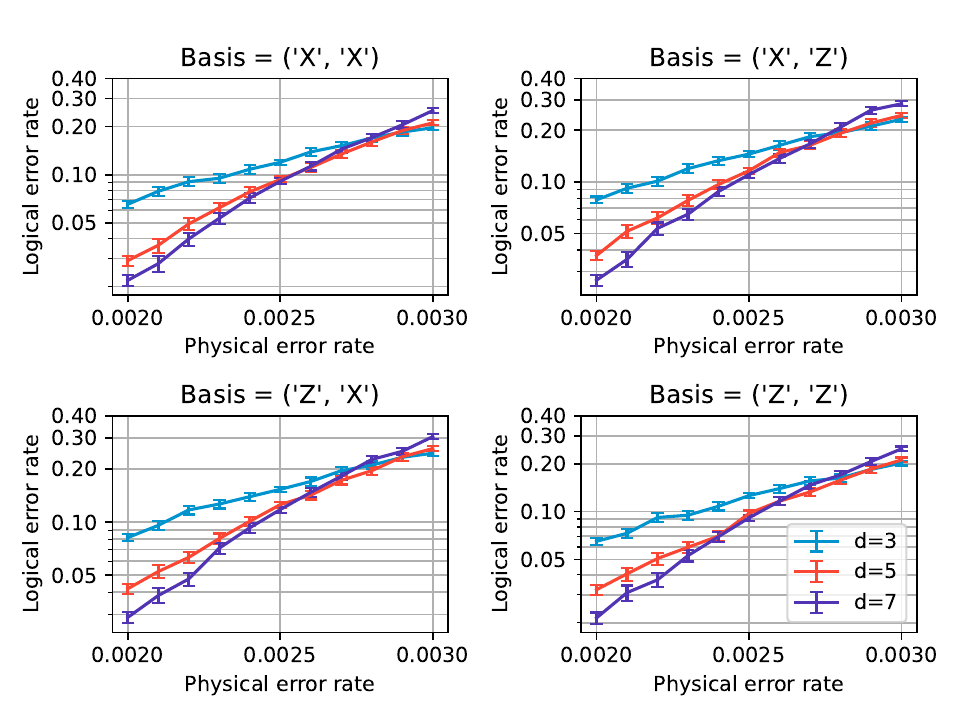}
        \caption{}
    \end{subfigure}
    \begin{subfigure}{0.49\linewidth}
        \includegraphics[width=\linewidth]{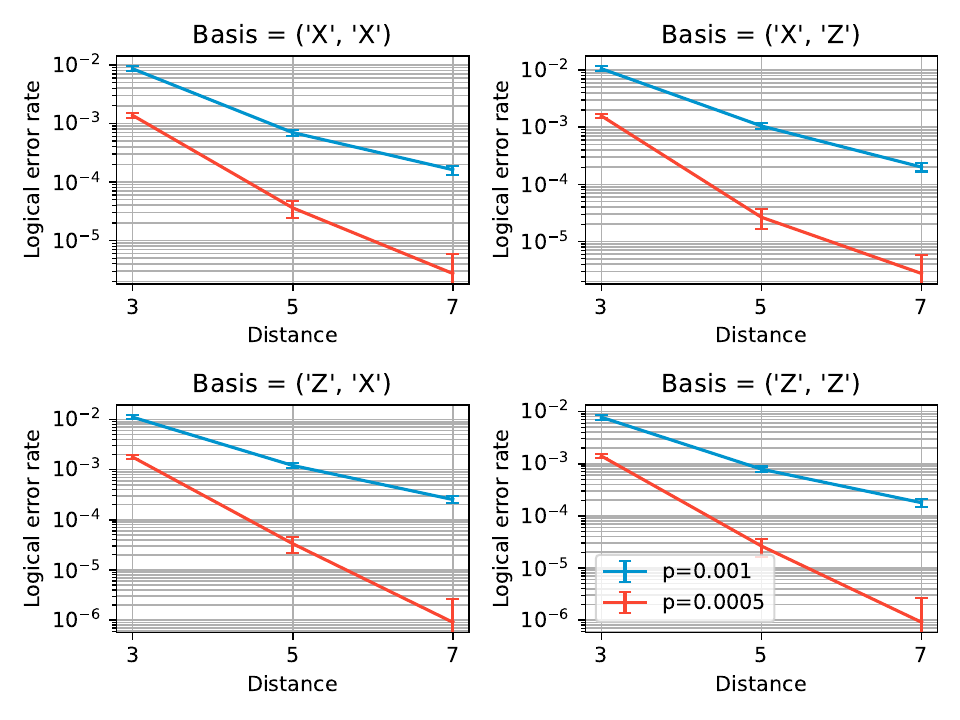}
        \caption{}
    \end{subfigure}
    \caption{Benchmarking a $(2d+1)$-round quantum memory on the CSS honeycomb Floquet code when decoded with BP+LSD-0 \cite{Hillmann2025BPLSD}. (a) We estimate a threshold of around $0.25$-$0.3\%$. (b) Sub-threshold analysis for physical error rates of $0.1\%$ and $0.05\%$. We attribute the flattening of the error suppression at higher distances to sub-optimal performance of BP+LSD-0.}
    \label{fig:memory-2d-plus-1-bp-lsd-full}
\end{figure}

\begin{figure}
    \centering
    \begin{subfigure}{0.49\linewidth}
        \includegraphics[width=\linewidth]{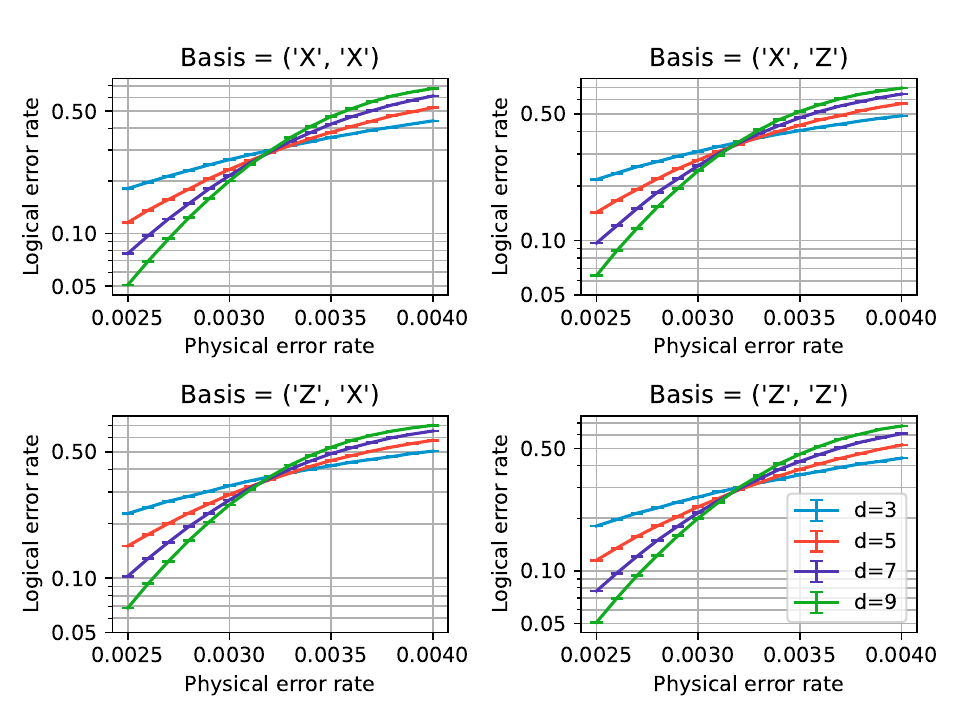}
        \caption{}
    \end{subfigure}
    \begin{subfigure}{0.49\linewidth}
        \includegraphics[width=\linewidth]{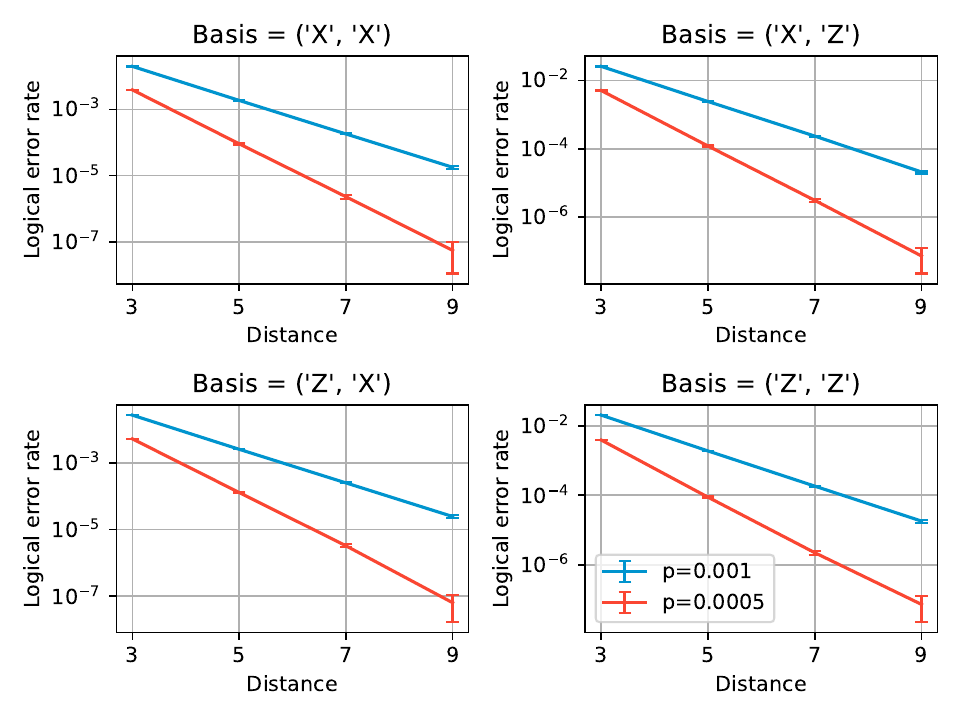}
        \caption{}
    \end{subfigure}
    \caption{Benchmarking a $3d$-round quantum memory on the CSS honeycomb Floquet code. (a) We estimate a threshold of around $0.32\%$. (b) Sub-threshold analysis for physical error rates of $0.1\%$ and $0.05\%$ shows exponential error suppression.}
    \label{fig:memory-3d-full}
\end{figure}

\begin{figure}
    \centering
    \begin{subfigure}{0.49\linewidth}
        \includegraphics[width=\linewidth]{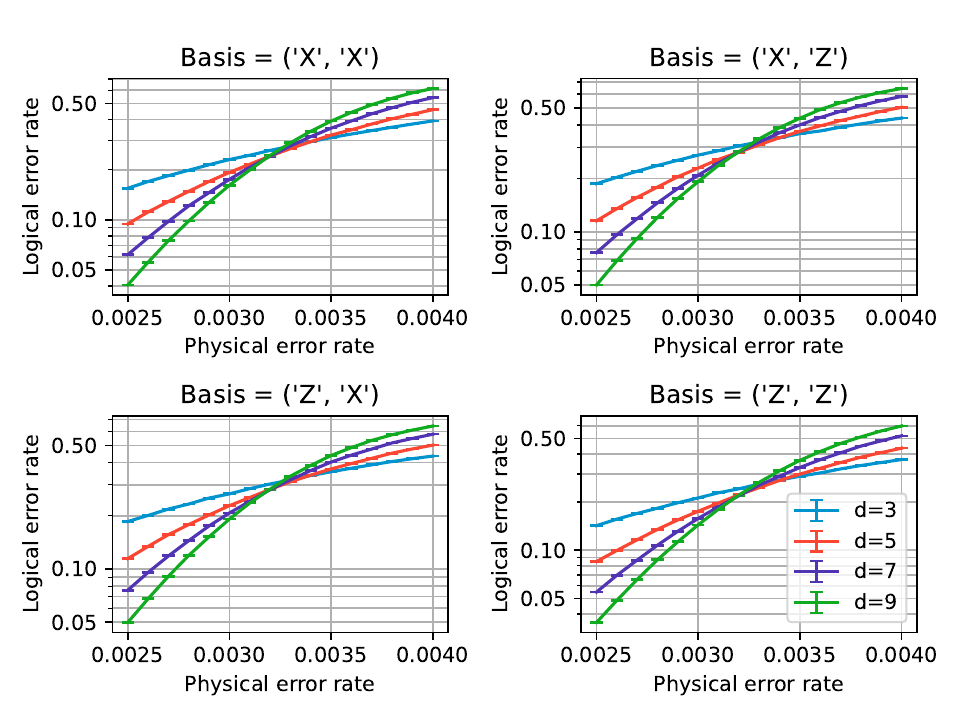}
        \caption{}
    \end{subfigure}
    \begin{subfigure}{0.49\linewidth}
        \includegraphics[width=\linewidth]{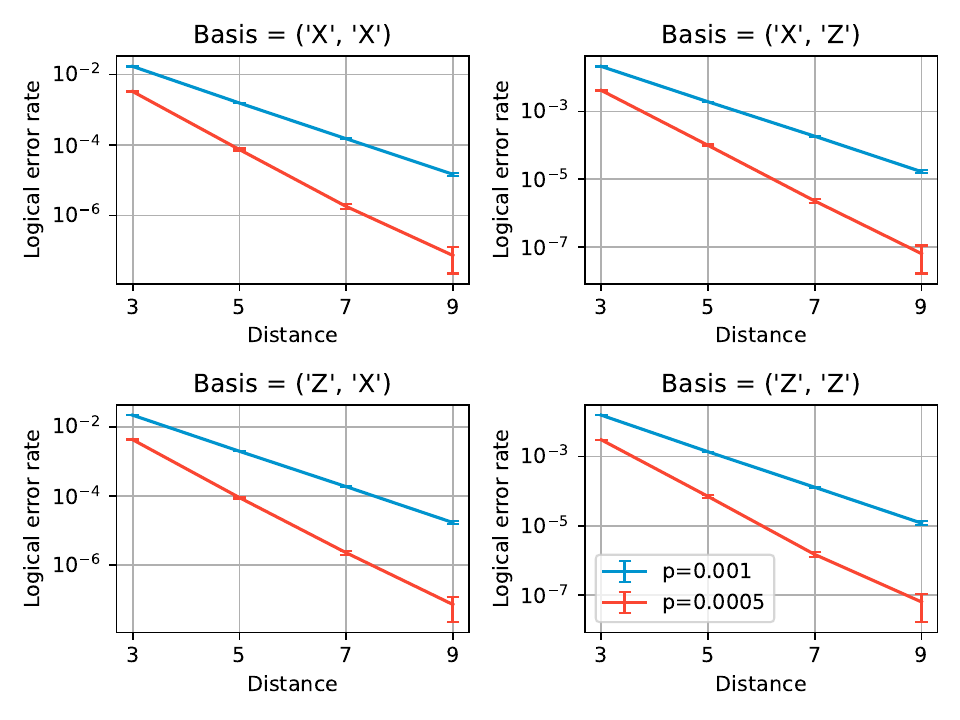}
        \caption{}
    \end{subfigure}
    \caption{Benchmarking a logical $\operatorname{H_0}\otimes \operatorname{H_1}$ gate on the CSS honeycomb Floquet code. (a)  We estimate a threshold of around $0.32\%$. (b) Sub-threshold analysis for physical error rates of $0.1\%$ and $0.05\%$ shows exponential error suppression.}
    \label{fig:hadamard-full}
\end{figure}

\begin{figure}
    \centering
    \begin{subfigure}{0.49\linewidth}
        \includegraphics[width=\linewidth]{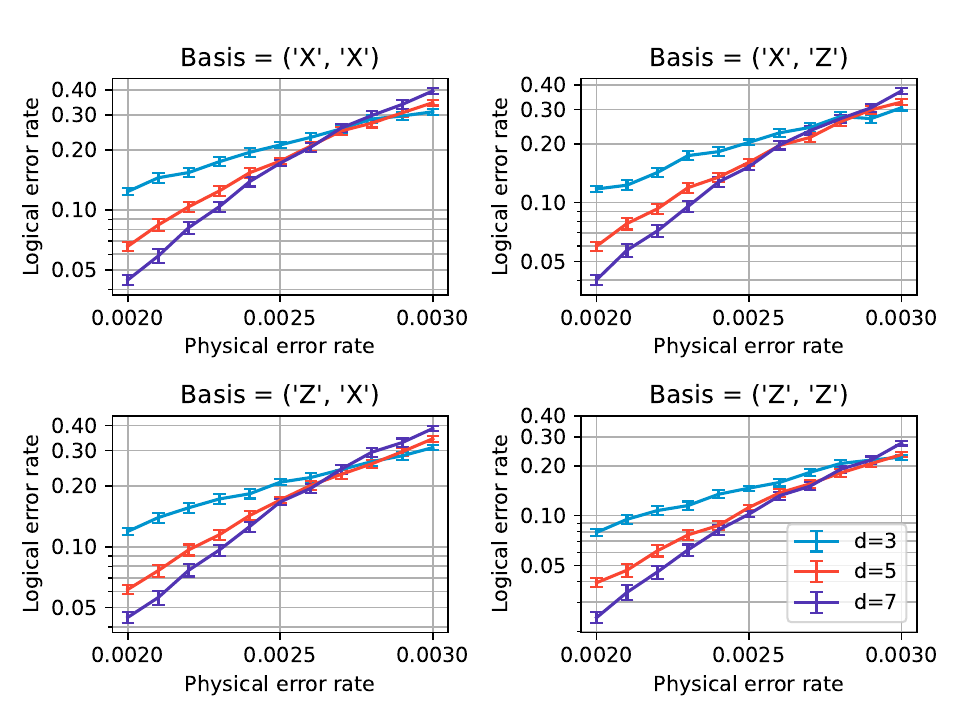}
        \caption{}
    \end{subfigure}
    \begin{subfigure}{0.49\linewidth}
        \includegraphics[width=\linewidth]{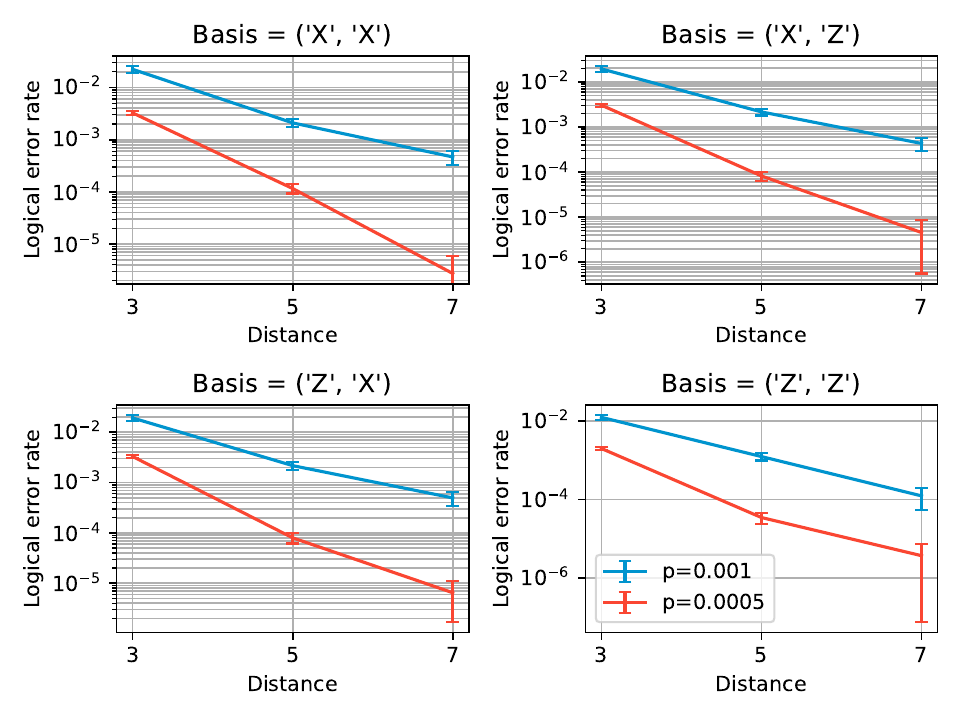}
        \caption{}
    \end{subfigure}
    \caption{Benchmarking a logical $\operatorname{S_0}\otimes \operatorname{S_1}$ gate on the CSS honeycomb Floquet code. (a)  We estimate a threshold of around $0.25$-$0.3\%$. (b) Sub-threshold analysis for physical error rates of $0.1\%$ and $0.05\%$. We attribute the flattening of the error suppression at higher distances to sub-optimality of BP+LSD-0.}
    \label{fig:s-gate-full}
\end{figure}

\begin{figure}
    \centering
    \begin{subfigure}{0.49\linewidth}
        \includegraphics[width=\linewidth]{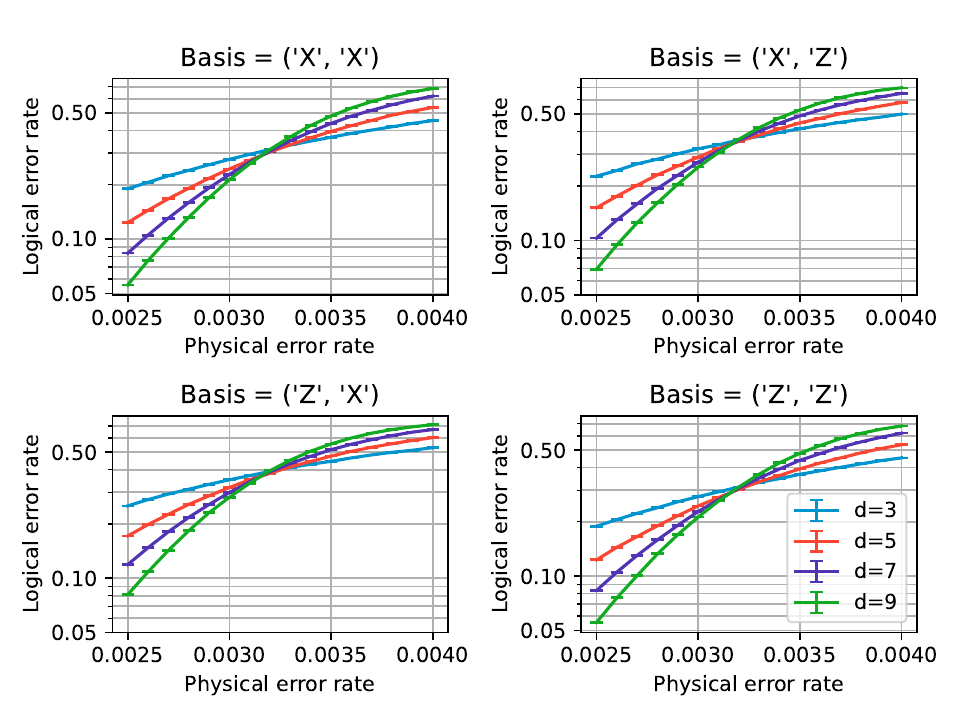}
        \caption{}
    \end{subfigure}
    \begin{subfigure}{0.49\linewidth}
        \includegraphics[width=\linewidth]{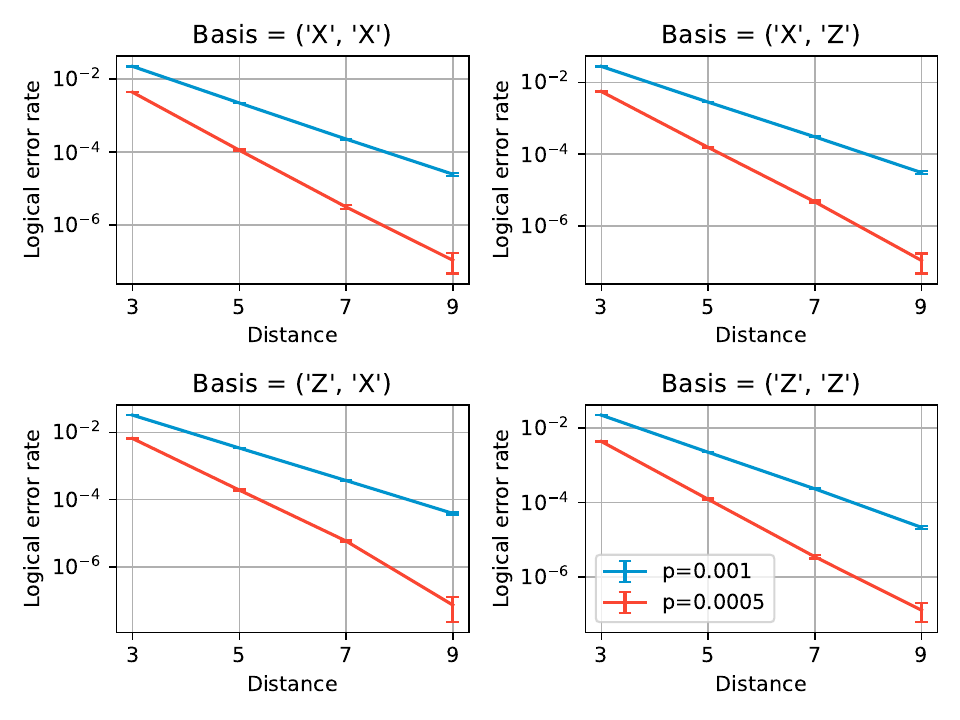}
        \caption{}
    \end{subfigure}
    \caption{Benchmarking a $d$-round horizontal Dehn twist on the CSS honeycomb Floquet code. (a)  We estimate a threshold of around $0.32\%$. (b) Sub-threshold analysis for physical error rates of $0.1\%$ and $0.05\%$ show exponential error suppression.}
    \label{fig:linear-horizontal-dehn-twist-full}
\end{figure}

\begin{figure}
    \centering
    \begin{subfigure}{0.49\linewidth}
        \includegraphics[width=\linewidth]{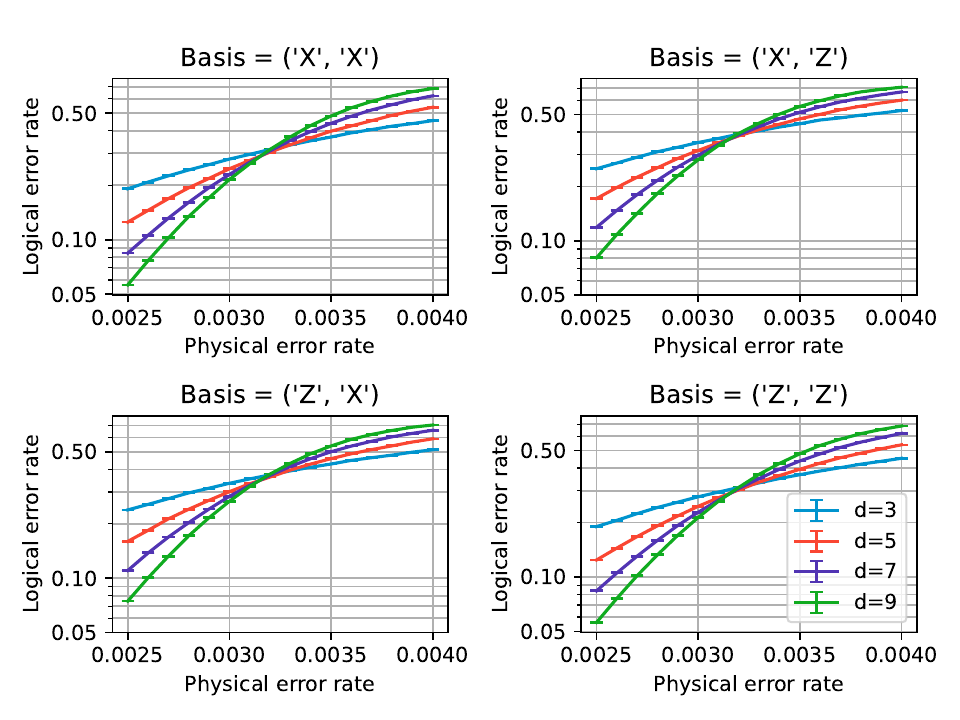}
        \caption{}
    \end{subfigure}
    \begin{subfigure}{0.49\linewidth}
        \includegraphics[width=\linewidth]{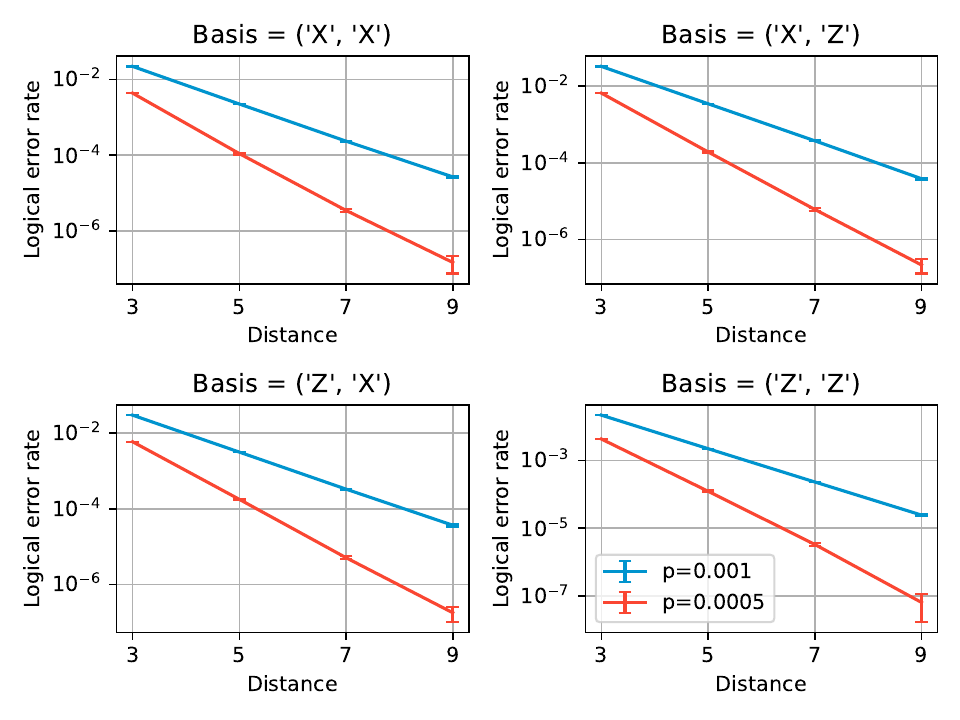}
        \caption{}
    \end{subfigure}
    \caption{Benchmarking a $d$-round vertical Dehn twist on the CSS honeycomb Floquet code. (a)  We estimate a threshold of around $0.32\%$. (b) Sub-threshold analysis for physical error rates of $0.1\%$ and $0.05\%$ show exponential error suppression.}
    \label{fig:linear-vertical-dehn-twist-full}
\end{figure}

In this appendix we present results for numerical benchmarks across all Pauli bases. These are presented in Figures \ref{fig:memory-2d-plus-1-pymatching-full}-\ref{fig:linear-vertical-dehn-twist-full}.

It is interesting to note that the performance of the logical $\operatorname{S}$-type gate is basis-dependent. In particular, the logical error rate when the logical qubits are prepared in the $\operatorname{X_0}\otimes \operatorname{X_1}$ basis is worse than when the logical qubits are prepared in $\operatorname{X_0}\otimes \operatorname{Z_1}$ or $\operatorname{Z_0}\otimes \operatorname{X_1}$, both of which are worse than when the logical qubits are prepared in $\operatorname{Z_0}\otimes \operatorname{Z_1}$. This is an effect of the logical $\operatorname{S_0}\otimes \operatorname{S_1}$ gate rather than this specific technique. To see this, note that the logical $\operatorname{S_0}\otimes \operatorname{S_1}$ gate maps the logical $\operatorname{X}$ basis to the logical $\operatorname{Y}$ basis. A logical qubit in the $\operatorname{Y}$ basis is sensitive to both $\operatorname{X}$ \textit{and} $\operatorname{Z}$ errors, whereas a logical qubit in the $\operatorname{X}$ (resp.\ $\operatorname{Z}$) basis is only sensitive to $\operatorname{Z}$ (resp.\ $\operatorname{X}$) errors. This effectively doubles the number of possible logical errors.

\begin{figure}
    \centering
    \begin{subfigure}{0.49\linewidth}
        \includegraphics[width=\linewidth]{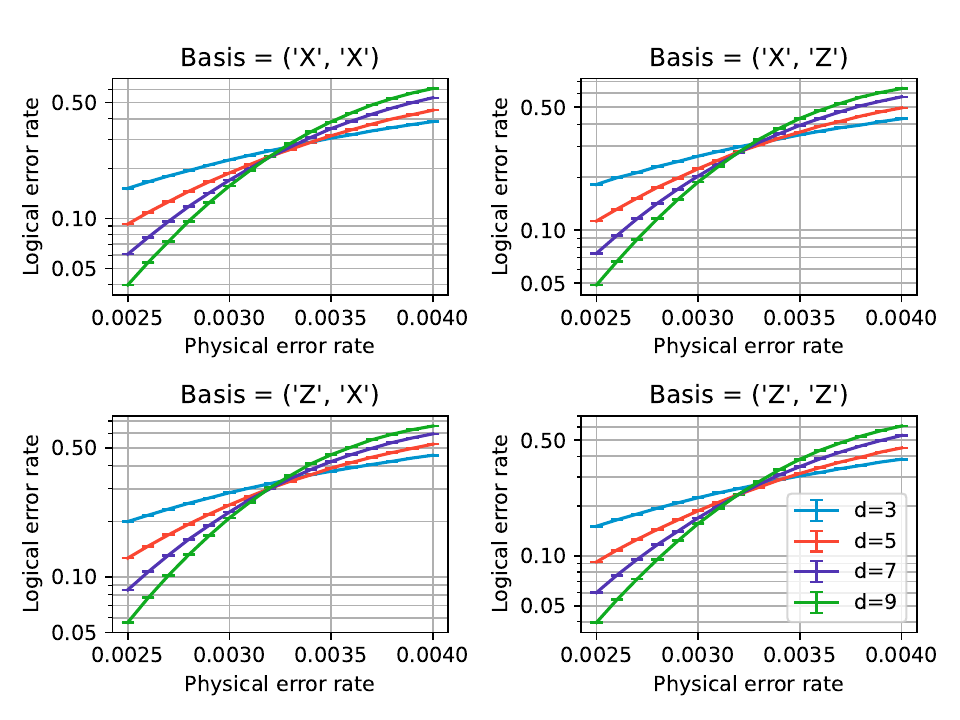}
        \caption{}
    \end{subfigure}
    \begin{subfigure}{0.49\linewidth}
        \includegraphics[width=\linewidth]{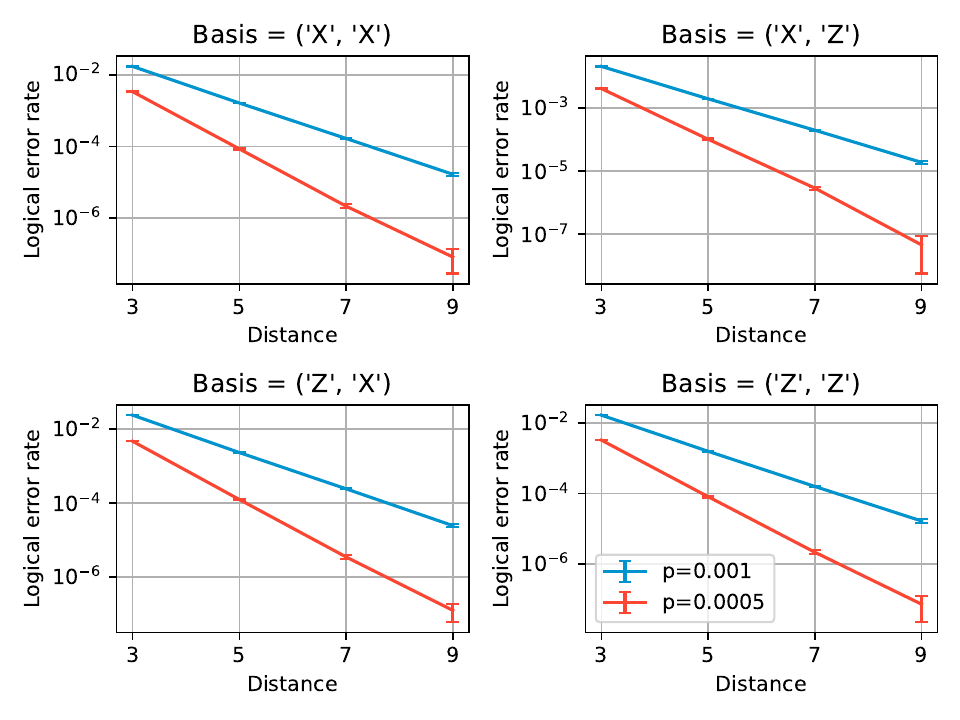}
        \caption{}
    \end{subfigure}
    \caption{Benchmarking an instantaneous horizontal Dehn twist on the CSS honeycomb Floquet code. (a) We estimate a threshold of around $0.32\%$. (b) Sub-threshold analysis for physical error rates of $0.1\%$ and $0.05\%$ shows exponential error suppression.}
    \label{fig:instantaneous-horizontal-dehn-twist-full}
\end{figure}

\begin{figure}
    \centering
    \begin{subfigure}{0.49\linewidth}
        \includegraphics[width=\linewidth]{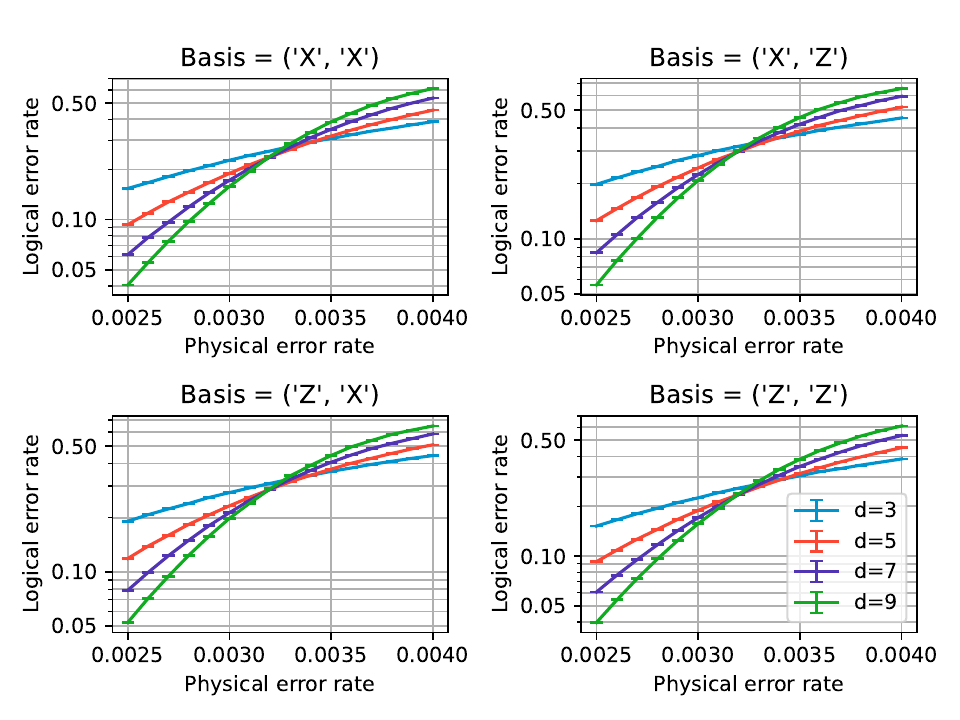}
        \caption{}
    \end{subfigure}
    \begin{subfigure}{0.49\linewidth}
        \includegraphics[width=\linewidth]{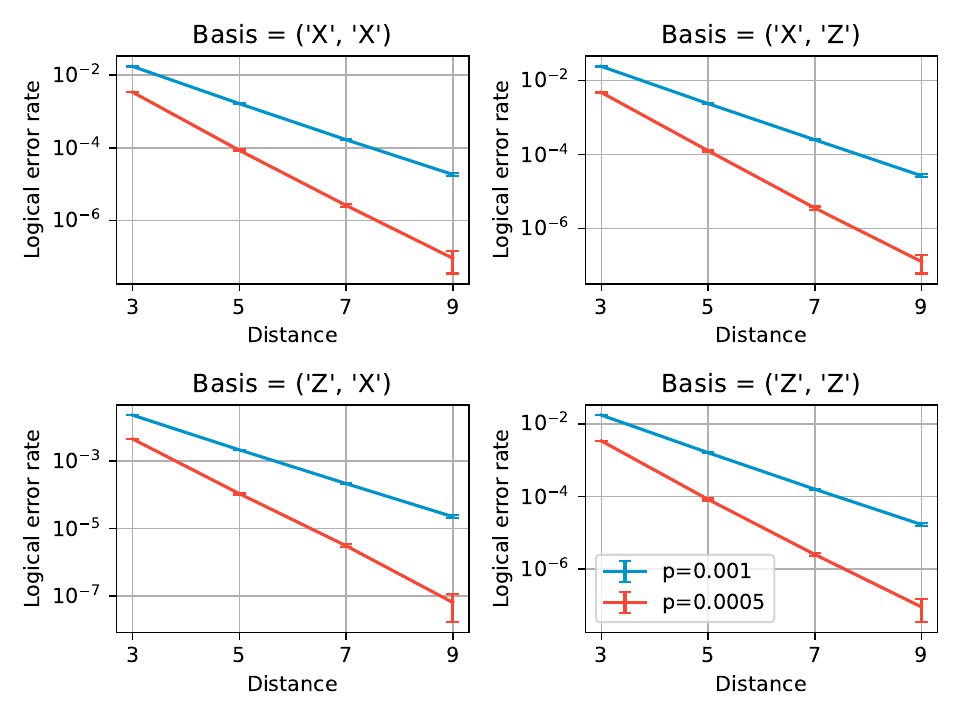}
        \caption{}
    \end{subfigure}
    \caption{Benchmarking an instantaneous vertical Dehn twist on the CSS honeycomb Floquet code. (a)  We estimate a threshold of around $0.32\%$. (b) Sub-threshold analysis for physical error rates of $0.1\%$ and $0.05\%$ shows exponential error suppression.}
    \label{fig:instantaneous-vertical-dehn-twist-full}
\end{figure}

\pagebreak
\bibliographystyle{quantum}
\bibliography{main}

\end{document}